\documentclass[aps,pra,twocolumn,showpacs,superscriptaddress,longbibliography]{revtex4-1}

\usepackage{blindtext}
\usepackage{graphicx}
\usepackage{verbatim}
\usepackage{color}
\usepackage{amsmath}
\usepackage{amsthm,amssymb}
\usepackage{microtype}
\usepackage{placeins}
\usepackage{bm}
\usepackage{bbold}
\usepackage{amsfonts}
\usepackage{amsbsy}
\usepackage{tabularx}
\usepackage[normalem]{ulem}
\usepackage{soul,wasysym}
\usepackage{upgreek}

\usepackage{hyperref}

\AtBeginDocument{%
    \newwrite\bibnotes
    \def\bibnotesext{Notes.bib}
    \immediate\openout\bibnotes=\jobname\bibnotesext
    \immediate\write\bibnotes{@CONTROL{REVTEX41Control}}
    \immediate\write\bibnotes{@CONTROL{%
    apsrev41Control,author="08",editor="1",pages="1",title="0",year="1"}}
     \if@filesw
     \immediate\write\@auxout{\string\citation{apsrev41Control}}%
    \fi
}%

\DeclareMathOperator{\Tr}{Tr}

\newcommand{\beq}{\begin{equation}}
\newcommand{\eeq}{\end{equation}}
\newcommand{\bqa}{\begin{eqnarray}}
\newcommand{\eqa}{\end{eqnarray}}
\newcommand{\bal}{\begin{equation}\begin{aligned}}
\newcommand{\eal}{\end{aligned}\end{equation}}
\newcommand{\ball}{\begin{align}}
\newcommand{\eall}{\end{align}}

\newcommand{\erf}[1]{Eq.~(\ref{#1})}

\newcommand{\erfs}[2]{Eqs.~(\ref{#1})--(\ref{#2})}

\newcommand{\arf}[1]{{Appendix}~\ref{#1}} 
\newcommand{\srf}[1]{Sec.~\ref{#1}}

\newcommand{\trf}[1]{Table ~\ref{#1}} 

\newcommand{\frf}[1]{Fig.~\ref{#1}}

\newcommand{\ie}{{\it i.e.}}
\newcommand{\dg}{^\dagger}

\newcommand{\bra}[1]{\left\langle #1\right|}
\newcommand{\ket}[1]{\left|#1\right\rangle}

\newcommand{\braketop}[3]{\left\langle #1\left|#2\right|#3\right\rangle}

\newcommand{\pref}[1]{(\ref{#1})}
\renewcommand{\eqref}[1]{Eq.~\pref{#1}}

\newcommand{\smallhalf}{{\textstyle\frac{1}{2}}}

\newcommand{\sch}{Schr\"odinger}

\newcommand{\ito}{It\^o}

\newcommand{\abs}[1]{\left| {#1} \right|}

\newcommand{\blk}{\color{black}}

\newcommand\stPW{\bgroup\markoverwith{\textcolor{red}{\rule[0.5ex]{2pt}{0.4pt}}}\ULon}

\newcommand{\eye}{{\bm 1}}

\begin{document}



\title{Solving quantum trajectories for systems with linear Heisenberg-picture dynamics and Gaussian measurement noise}


\date{\today}

\author{Prahlad~Warszawski}
 \affiliation{Centre for Engineered Quantum Systems, School of Physics,
The University of Sydney, Sydney, NSW 2006, Australia}
\author{Howard~M.~Wiseman}
 \affiliation{Centre for Quantum Computation and Communication Technology, 
Centre for Quantum Dynamics, Griffith University, Brisbane, Queensland 4111, Australia}
\author{Andrew~C.~Doherty}
 \affiliation{Centre for Engineered Quantum Systems, School of Physics,
The University of Sydney, Sydney, NSW 2006, Australia}


\begin{abstract}

We study solutions to the quantum trajectory evolution of $N$-mode open quantum systems possessing a time-independent Hamiltonian, linear Heisenberg-picture dynamics, and Gaussian measurement noise.  In terms of the mode annihilation and creation operators, a system will have linear Heisenberg-picture dynamics under two conditions.  First, the Hamiltonian must be quadratic.  Second, the Lindblad operators describing the coupling to the environment (including those corresponding to the measurement) must be linear.  In cases where we can solve the $2N$-degree polynomials that arise in our calculations, we provide an analytical solution for initial states that are arbitrary 
(\ie~they are not required to have Gaussian Wigner functions). The solution takes the form of an evolution operator, with the measurement-result dependence captured in $2N$ stochastic integrals over these classical random signals. The solutions also allow the POVM, which generates the probabilities of obtaining measurement outcomes, to be determined.  To illustrate our results, we solve some single-mode example systems, with the POVMs being of practical relevance to the inference of an initial state, via quantum state tomography.  Our key tool is the representation of mixed states of quantum mechanical oscillators as state vectors rather than state matrices (albeit in a larger Hilbert space).  Together with methods from Lie algebra, this allows a more straightforward manipulation of the exponential operators comprising the system evolution than is possible in the original Hilbert space.

\end{abstract}

\maketitle
  \section{Introduction}

The state of an open quantum system that is undergoing continuous measurement follows a quantum trajectory and is governed by a stochastic master equation (SME).
 Due to their importance to emerging quantum technologies, such systems have been studied extensively, both from a theoretical~\cite{belavkin1999measurement,belavkin1989quantum,CarQTraj,WisMil10,wisQTraj,herkommer1996localization,jacSteck,brun2002simple,caves1987quantum,diosi1988continuous,dalibard1992wave,daley2014quantum,sarlette2017deterministic} and, increasingly, experimental perspective~\cite{PhysRevLett.61.826,PhysRevLett.77.4887,peil1999observing,lu2003real,elzerman2004single,nagourney1986shelved,PhysRevLett.57.1696,PhysRevLett.57.1699,gleyzes2007quantum,neumann2010single,sayrin2011real,sun2014tracking,minev2019catch,campagne2016observing,vijay2011observation,murch2013observing,de2014reversing}.  
In this paper we focus on a particular class of Markovian open quantum systems: those with linear Heisenberg-picture (HP) dynamics and Gaussian measurement noise,
which are of wide practical importance, as well as being amenable to analytic techniques~\cite{huang2018smoothing,zhang2017prediction,braunstein2005quantum,weedbrook2012gaussian,PhysRevLett.94.070405,PhysRevLett.109.233906,wieczorek2015optimal,ockeloen2018stabilized,kohler2018negative,wade2015squeezing,laverick2019quantum,petersen2010quantum,james2008h}.  Physical systems that can be modeled in such a way include multimodal light fields, optical and optomechanical systems (including squeezing), microwave resonators and Bose-Einstein condensates.   Further motivation for their study arises due to recent interest in the control of such bosonic systems, potentially using feedback~\cite{ockeloen2018stabilized,mirrahimi2014dynamically,gough2010squeezing,WisMil10,combes2017slh}.  
Our specific research goal is to find an
evolution operator that can be applied to arbitrary (not necessarily Gaussian) initial states; in other words, to {\it solve} this class of SMEs.

In our paper, we refer to a `linear quantum system' as being one for which there exists a closed set of linear HP equations for a finite set of observables, in terms of which any system operator may be expressed.
A necessary requirement for this to be true is that the observables have the real line as their spectra and, consequently, describe bosonic modes.  A complete set of observables is provided by a canonically conjugate pair of position and momentum observables, one pair for each bosonic mode.  Equivalently, an annihilation and creation operator for each mode could be used instead.  

In the absence of monitoring (or by ignoring the measurement results) the dynamics of the system configuration will be linear given two restrictions.  Firstly, the Hamiltonian evolution must be at most quadratic in the bosonic annihilation and creation operators.  Although not necessary for the preservation of linearity, in our work we will take the quadratic terms as being time-independent (or made to be so by transformation to a new frame), so that analytic results are possible.  Secondly, the Lindblad operators describing the irreversible evolution must be linear in the annihilation and creation operators.  

When measurement of the environment is included, further restrictions must be placed to retain the linearity of the evolution when conditioning upon the measurement results. Specifically, the monitoring must be `diffusive', by which we mean that the measurement noise is Gaussian in nature, in contrast to jump-like trajectories. 
The jump class of trajectories arise when the measurement record is a point processes in which a detector `click' is accompanied by a finite change in the conditioned state matrix.  The diffusive class, by contrast, is one in which the stochasticity of the measurement results is described by a Wiener increment 
and the conditioned state evolves continuously (though non-differentiably) in time. 

The diffusive class of quantum trajectories is sufficient to describe 
all systems undergoing Markovian non-jumplike quantum evolution.  
The main examples of the diffusive class of unravellings, whether in the optical or microwave setting, are homo{\it dyne} and hetero{\it dyne} detection.  For example, in the optical regime, homodyne detection can be realized by coherently combining the light leaking out of an optical cavity with a very strong local oscillator before detection.  As almost all detection events are due to the local oscillator, the effect of each one on the system state becomes infinitesimal and a continuous description arises.  
This can be understood from the perspective of continuous state evolution occurring in the limit that the number of detection events is very large in a time period that is small compared to the system time scale.  In this limit, the Poissonian distributed photocount can be replaced by a Gaussian photocurrent~\cite{WisMil10, wiseman1993quantum}.
\blk
In this paper we will model the most general form of such {\it dyne} (diffusive) unravelings~\cite{WisDiosi}.

Given the restriction to diffusively monitored linear quantum systems with time-independent Hamiltonian, we will achieve our goal of solving the SME if certain polynomials of degree $2N$ can be solved, where $N$ is the number of physical modes.  In the case where the polynomials cannot be solved, our method of solution still provides a form amenable to efficient numerical simulation. Our theory applies to systems that include such features as squeezing, thermal or squeezed reservoirs and, very importantly, general forms of continuous (diffusive) measurement. 

When dealing with linear systems subject to diffusive monitorings it is common to assume that the initial system state is Gaussian. This is {\em not} assumed in our work.  We treat completely arbitrary initial states.
For initial Gaussian states, the system solution is well known, being governed by a Kalman filter.  Therefore, the extension our work provides is that of a more general solution to linear quantum systems undergoing diffusive measurement-induced evolution, being applicable to such initial non-Gaussian states as `cat' or Fock states.

The solution to a SME naturally involves classical random variables, as it represents the description of a particular quantum trajectory.  This is distinct from master equation (ME) solutions which are deterministic and provide a description that is inherently averaged over all possible trajectories.  By `analytical solution' of a SME, we therefore aim to find an expression for the system state at time $t$ in terms of a stochastic evolution operator that contains a finite number of stochastic integrals; this evolution operator will be independent of the initial state.  That is, rather than defining the evolved system state in terms of the infinity of numbers constituting the entire continuous measurement record, we will show that the final state is only dependent upon $2N$ complex-valued stochastic integrals.

The solution of the SME, given as a function of a finite number of stochastic integrals, 
has a number of uses, as we now discuss.  

A SME solution allows calculation of expectation values conditional upon the measurement results which are, in general, distinct from the values obtained from the average system behavior (described by the master equation).  Thus, a SME solution will be essential in {\it state}-based feedback control~\cite{DohJac}, by which knowledge of the system state is used for its accurate future control.

Possessing the SME solution also means that we can analyze what types of states are generated under measurement-induced evolution.  Notably, it will be found that the presence of measurement causes more than just phase-space displacements of the state. The SME solution will facilitate the engineering of desired dissipative dynamics and, in particular, conditional dynamics~\cite{verstraete2009quantum,toth2017dissipative}.   As an example, it could be investigated whether a desired Gaussian operation upon the state could be conditionally achieved~\cite{moore2017arbitrary}.

Another benefit of the SME solution is that it allows a characterization of the measurement, by defining the relevant POVM.  The POVM and related theoretical constructs, such as Bayesian inference, are of use in many contexts.  For example, they allow the optimal inference of the input system state via state tomography~\cite{six2016quantum,Chantasri2019,warszawski2019tomography}.  The motivation for solving the SME in \cite{warszawski2019tomography} was to know the POVMs relating to optomechanical position measurement with parametric amplification.  The method used there can be turned into a general method of solving SMEs, which is detailed in this paper.  To make the link more explicit to the previous work~\cite{warszawski2019tomography}, and to provide more detail regarding those calculations, we here consider the relevant optomechanical system as a specific single-mode example.

A related use of the POVM is that it allows a calculation of the probability density of obtaining a measurement sequence.  In combination with the system solution, we therefore have knowledge of the type of states obtainable under measurement and the probability distribution of such states.  This is extremely powerful: to simulate the system state at some specific future time one needs only to sample the state distribution, rather than integrating the SME.  We stress that this applies to non-Gaussian initial states that cannot be fully tracked by their first and second-order moments.  Potential specific applications include facilitating the investigation of the rate of decoherence of quantum superpositions~\cite{myatt2000decoherence} or entanglement dynamics~\cite{paz2008dynamics}.

Before closing this introduction, we briefly discuss the methods that we use to obtain SME solutions.  In order to make a solution tractable we use a {\it linear} SME~\cite{GoeGra,wisQTraj,jacLin}, in which some of the information concerning the probability of a measurement sequence occurring is contained in the norm (trace) of the density matrix.  It is important to note that the SME for the {\it normalized} quantum state is {\it nonlinear}, even when the system belongs to the class of diffusively monitored linear quantum systems which, by definition, possess linear quantum Langevin equations for the system configuration.  The use of a linear SME removes the measurement-induced nonlinearity and provides us with a pathway to calculate the POVM.  Our work in many regards generalizes that of Wiseman~\cite{wisQTraj}, and of 
Jacobs and co-workers~\cite{jacLin,jacSte}, which provided a general method of calculating the evolution operator for the stochastic Schr\"odinger equation (SSE).  
We extend the class of solutions to include arbitrary dyne measurements in systems requiring a mixed state description (that is, a SME rather than a SSE).  

Also influential is the application of group theory methods developed by such practitioners as Gilmore and Yuan~\cite{Gilmore,GilmoreSingle,Mufti,BanLie}.  Wilson and co-workers have obtained analytic solutions to master equations using Lie methods~\cite{galit2,galit1}.  Much of the problem of obtaining a practicable SME solution is contained in operator disentangling~\cite{disentanglingClarity} and re-ordering tasks, which are both a function of the operator commutation relations. Indeed, the ability to perform these tasks is what separates a bosonic system that is fully soluble (\ie~a linear quantum system with $N\leq 2$) using our methods from one that is not (a nonlinear, or $N>2$, system). As we will see, the dynamics of the SME imposes a structure upon the Lie group.  For the systems that we consider, this structure dictates the formation of subgroups that contain either deterministic or stochastic elements respectively. 

A necessary, but not sufficient, requirement for SME solvability is that the algebra defining the evolution closes under commutation, to form a finite dimensional Lie algebra. As an example, for the linear quantum systems that we consider, the algebra will always close. However, when we form a finite dimensional matrix representation of the operators, it will become crucial to solve polynomials of degree $2N$ in order to proceed with our method of solution.  Thus algebra closure, for $N>2$, does not imply that we can find an SME solution.  There is a considerable literature devoted to these topics, for example~\cite{Gilmore,GilmoreSingle,SchoWey,FANdis,FANdis2,VarMoy,Fer89,Wun01,Wil67,Das96,Twa93}.


The final method that will be mentioned here is that of the thermo-entangled state representation (sometimes called non-equilibrium thermo-field dynamics).  This key technique transforms the superoperators of the standard formulation of the SME into operators acting in a larger Hilbert space~\cite{ZhoFan,FANthermo,FANthermo2,Kosov,FANthermEta,UmeAri,Umebook,FanHu}.
We can then utilize powerful group theoretic tools to re-organize the infinite string of time slice evolutions.

It is well known~\cite{DruCor,3rdQuant,BazNag,ZhoFan} that in the absence of measurement the solution of linear quantum systems is possible via phase space methods, but there has been considerable interest in providing new methods of solution to the deterministic Gaussian master equation~\cite{FanHu,Ban,BazNag,3rdQuant}, so we note that our method of solution of the SME naturally subsumes non-stochastic systems and does so at the very general level described above.  

Our paper is organized as follows.  We begin by specifying, mathematically, the system of interest. Next, in \srf{sketch}, we sketch the steps that will be followed in order to solve the linear SME.  These steps are then carried out in \srf{SMEsol}.   In \srf{POVMsec}, the POVM pertaining to the compiled measurement of finite duration is obtained. The adjoint equation approach to finding the POVM is also discussed.  In \srf{recipe}, our calculational methods are condensed into a summary, for those wishing to apply them to their own systems.  In \srf{singleModeEgs}, we analyze some example single mode system to further illustrate our methods. The paper concludes with a discussion in \srf{conclusion}.  Many mathematical details are deferred to appendices, in order to improve the readability of the main text.

\section{System specification}
\label{sysSpecification}

An $N$-mode bosonic system undergoing linear Heisenberg-picture (HP) dynamics is subjected to an arbitrary number, $L$, of completely general dyne measurements \cite{WisDiosi}.  For illustrative purposes, we note that homodyne and heterodyne type measurements are two, experimentally prevalent, examples of `dyne' measurement.  We also reiterate that linear HP dynamics  is a completely distinct notion from that of the linearity, or otherwise, of the SME.  Linear HP dynamics will occur under two conditions.  Firstly, that the Hamiltonian be at most a quadratic function of the bosonic annihilation and creation operators.  Secondly, the $L$ Lindblad operators, which we write in column vector form as $\hat{\bm c}\equiv \left(\hat{c}_{1},\hat{c}_{2},...,\hat{c}_{L} \right)^{\rm T}$, are likewise limited to being arbitrary linear combinations of those operators.  

The nonlinear SME, describing the conditional evolution of the system density matrix in units where $\hbar=1$, is given by~\cite{chia2011complete}
\bqa
d\rho_{{\rm c}}(t) &=& -i\left[\hat{H},\rho(t) \right]dt+{\cal D}\left[\hat{{\bm c}} \right]\rho(t)  dt
\nonumber\\
&&+d{\bm w}^{{\rm T}}(t) 
{\cal H}\left[ {\bm M}^{\dag}\hat{{\bm c}}\right]\rho(t)  ,
\label{smeGenNonLin}
\eqa
where the superoperators are defined by
\beq
{\cal D}\left[\hat{{\bm c}} \right]\equiv \sum_{k=1}^{L}{\cal D}\left[\hat{c}_{k} \right],
\quad
{\cal D}\left[\hat{c} \right]\rho\equiv \hat{c}\rho\hat{c}^{\dag}-\smallhalf\hat{c}^{\dag}\hat{c}\rho
-\smallhalf \rho\hat{c}^{\dag}\hat{c}
\eeq
and
\beq
{\cal H}\left[\hat{{\bm c}}\right]\rho\equiv \hat{{\bm c}}\rho+\rho\hat{{\bm c}}^{\ddagger}
-\Tr \left[\hat{{\bm c}}\rho+\rho\hat{{\bm c}}^{\ddagger}\right]\rho,
\label{hSuper}
\eeq
with $\hat{{\bm c}}^{\ddagger}\equiv \left (\hat{{\bm c}}^{{\rm T}}\right)^{\dag}$ and $\hat{{\bm c}}^{\dagger}=\left(\hat{c}^{\dag}_{1},\hat{c}^{\dag}_{2},...,\hat{c}^{\dag}_{L} \right)$.
The subscript `c' of $d\rho_{{\rm c}}(t)$ is used to indicate conditioning on the set of measurement results at times up to and including $t$.  Of course, if measurement is ongoing then the density matrix on the RHS will also be a conditioned density matrix, but we omit a subscript there for simplicity of display.  The SME presented in \erf{smeGenNonLin} is nonlinear (due to the action of the superoperator ${\cal H}$), despite it describing linear HP dynamics.  This is necessary in order for $\rho_{{\rm c}}$ to remain normalized.

The nonlinear SME is written in terms of the measurement noise, $d{\bm w}^{{\rm T}}(t)$, which is related to the $2L\times 1$ column vector of measurement results as 
\beq
{\bm y}(t)dt=\langle {\bm M}^{\dag} \hat{{\bm c}}+{\bm M}^{{\rm T}}
 \hat{{\bm c}}^{\ddagger}\rangle dt+d{\bm w}(t),
 \label{measCurrent}
\eeq
where $\langle\cdots\rangle$ indicates a quantum expectation value. The length of ${\bm y}(t)$ is $2L$ because we allow for heterodyne-style measurement currents that can be decomposed into two real-valued components (we will often refer to ${\bm y}(t)$ as a measurement `current').  

The measurement noise, $d{\bm w}(t)$, is a vector of independent Wiener increments having statistics
\bqa
{\rm E}\left[d{\bm w}(t) \right]&=&{\bm 0},\label{dw}\\
 {\rm E}\left[d{\bm w}(t)d{\bm w}^{{\rm T}}(t) \right]&=&
 d{\bm w}(t)d{\bm w}^{{\rm T}}(t)= \eye_{2L}dt,\label{dw2}
\eqa
where ${\rm E}[\cdots]$ indicates a classical expectation value.  Note that \ito's rule allows the removal of the averaging in \erf{dw2}\cite{GarBook}.  From \erf{dw2}, it can be seen that $d{\bm w}(t)$ is typically of order $\sqrt{dt}$.

The statistics of the $L$ dyne measurement currents are also Gaussian, as follows from \erf{measCurrent} and \erfs{dw}{dw2}:
\bqa
{\rm E}\left[{\bm y}(t) \right]&=&\langle {\bm M}^{\dag} \hat{{\bm c}}+{\bm M}^{{\rm T}}
 \hat{{\bm c}}^{\ddagger}\rangle,\label{actExpY}\\
 dt{\rm E}\left[{\bm y}(t){\bm y}^{{\rm T}}(t) \right]&=&
 dt {\bm y}(t){\bm y}^{{\rm T}}(t)= \eye_{2L},\label{actExpY2}
\eqa
Note that ${\bm y}(t)dt$ is also typically of order $\sqrt{dt}$.

The complex, time-independent, matrix ${\bm M}$, of size $L\times 2L$, parameterizes the unraveling and defines the type of measurements being conducted.  It could be referred to as the measurement `setting'. One should not confuse the measurement setting, ${\bm M}$, with the measurement results themselves, ${\bm y}(t)$.  The set of allowed ${\bm M}$ is identified by the constraint ${\bm M}{\bm M}^{\dag}\in{\mathfrak H}$ where ${\mathfrak H}$ is
\beq
{\mathfrak H}=\left\{ {\bm H} ={\rm diag}({\bm \eta})|\,\forall k,\, \eta_k\in[0,1]\right\}.
\label{Mallowed}
\eeq
Note that ${\bm H}$ (capital $\eta$) is a diagonal matrix of detector efficiencies (not to be mistaken with the system Hamiltonian operator, $\hat{H}$).  The reader will observe that the {\it matrix} ${\bm M}$ generalizes the scalar detector efficiency factor $\sqrt{\eta}$ that would appear in a standard single channel homodyne nonlinear SME.  \erf{smeGenNonLin} is known as the $M$-representation of the nonlinear SME~\cite{chia2011complete}.

As previously stated, the Hamiltonian, $\hat{H}$, is quadratic at most (in the bosonic annihilation and creation operators), whilst the Lindblad operators, $\hat{{\bm c}}$, are linear.  It is standard procedure to write these operators in terms of pairs of canonically conjugate quadrature operators, with a single pair for each mode, $\hat{q}_{n},\hat{p}_{n}$, having commutation relation $\left[\hat{q}_{n},\hat{p}_{n} \right] =i$.  Thus, $\hat{q}_{n},\hat{p}_{n}$ are related to the annihilation and creation operator, $\hat{a}_{n},\hat{a}^{\dag}_{n}$, of each mode via
\bqa
 \begin{pmatrix}
  \hat{q}_{n}       \\
 \hat{p}_{n}      \\
\end{pmatrix}&=&
\frac{1}{\sqrt{2}}
 \begin{pmatrix}
 1&1\\
 -i&i\\
\end{pmatrix}
 \begin{pmatrix}
  \hat{a}      \\
 \hat{a}_{n}^{\dag}      \\
\end{pmatrix},
\label{canRel}
\eqa
with $\left[\hat{a}_{m},\hat{a}^{\dag}_{n} \right] =\delta_{mn}$. A vector of operators
\beq
\hat{{\bm x}}=\left(\hat{q}_{1},\hat{p}_{1},...,\hat{q}_{N},\hat{p}_{N}\right)^{{\rm T}}
\eeq
is defined, so that $\left[\hat{x}_{m},\hat{x}_{n} \right]=i{\bm \Sigma}_{mn}$, where the $2N\times 2N$ symplectic matrix is given by
\beq
{\bm \Sigma}=
\bigoplus^{N}_{n=1}
\begin{pmatrix}
 0&1       \\
 -1&0     \\
\end{pmatrix}.
\eeq
For later use, we also define the column vector of annihilation operators,
\beq
\hat{{\bm a}}=\left(\hat{a}_{1},\hat{a}_{2},...,\hat{a}_{N}\right)^{{\rm T}},
\label{aVec}
\eeq
from which follows the definition of $\hat{{\bm a}}^{\dagger}$.
Having laid the notational groundwork, we can then state the quadratic Hamiltonian as
\beq
\hat{H}=\frac{1}{2}\hat{{\bm x}}^{{\rm T}}{\bm G}\hat{{\bm x}}-
\hat{{\bm x}}^{{\rm T}}{\bm \Sigma}{\bm B}{\bm u},
\eeq
with the $2N\times 2N$ matrix ${\bm G}$ real and symmetric, ${\bm u}$ a classical drive, and a matrix, ${\bm B}$, that is also real.  To allow a formal analytic solution to be derived later, we have here assumed a time-independent Hamiltonian.  By making a canonical transformation, and then considering a shifted vacuum state, it is possible to remove the linear Hamiltonian term and also any constants in the Lindblad operators~\cite{3rdQuant}.  Consequently, without further loss of generality, the Hamiltonian is taken to be
\beq
\hat{H}=\frac{1}{2}\hat{{\bm x}}^{{\rm T}}{\bm G}\hat{{\bm x}}\label{hamiltonianG}
\eeq
and the vector of Lindblad operators is
\beq
\hat{{\bm c}}={\bm C}\hat{{\bm x}},\label{lindbladMatrix}
\eeq
for the $L\times 2N$ matrix ${\bm C}$.

The evolution described by \erf{smeGenNonLin}, with the specification of a quadratic Hamiltonian and linear Lindblad operators, is special in that it admits a Gaussian state as its solution.  That is, given an initial state possessing a Gaussian Wigner function, the system Wigner function will remain Gaussian at all future times. The evolution of the Gaussian state can be tracked just with the first and second order moments of the quadrature operators.  The equations governing these moments are jointly known as the generalized Kalman filter; the equation for the covariance matrix is of the form of a Riccati differential equation.  It is important to realize that in our work we go beyond this and treat arbitrary (that is, possibly non-Gaussian) initial states.

\section{Solution Sketch}
\label{sketch}

In this section, we describe the method used to solve for the evolution of the quantum system.  We focus upon the conceptual steps involved, with more technical details deferred, where appropriate, until later sections.  For further clarity, \frf{flowChart} schematically illustrates the process.

\begin{figure}
	\centering
 \includegraphics[width=1\hsize]{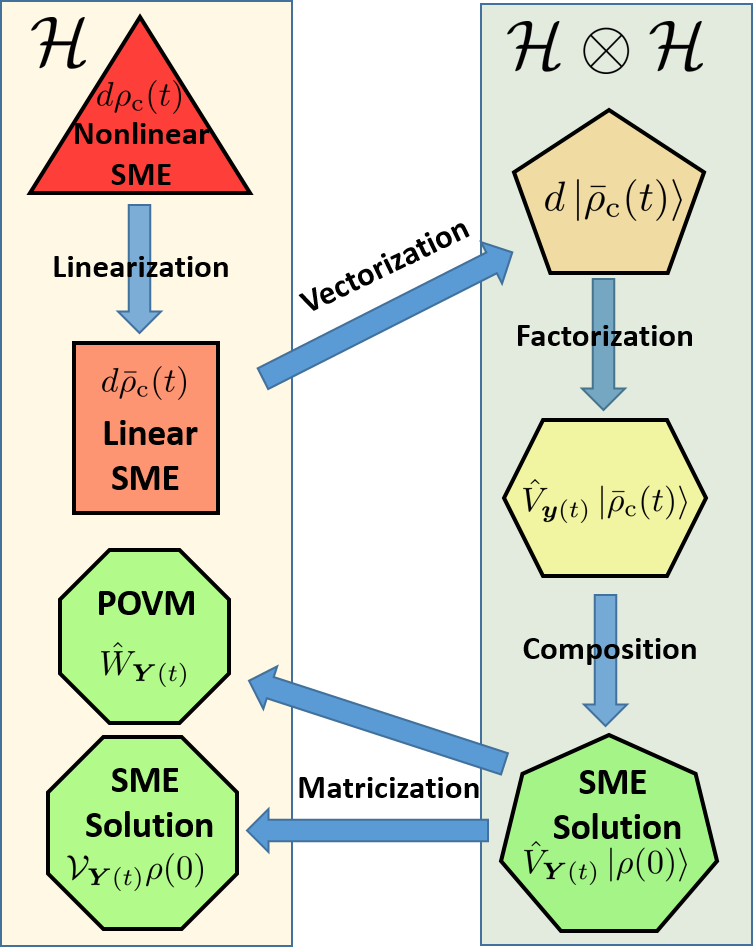}
	\caption{The sequence of steps involved in solving the SME and finding the POVM is schematically shown.  Within the smaller polygons, the mathematical object that most closely matches the outcome of the process arrow labels is shown.  Increasing side number of the polygon (together with shape color progression) represents sequential process steps, starting with the nonlinear SME and moving towards the twin goals of SME solution and obtaining the POVM.  The two larger, enveloping, rectangles indicate whether the contained objects reside in the original Hilbert space, ${\cal H}$, or the enlarged Hilbert space, ${\cal H}\otimes{\cal H}.$
	}
	\label{flowChart}
\end{figure}

Given the task of solving the SME, one might initially begin by hoping to obtain the dynamical mapping, in the form of an evolution superoperator ${\cal N}_{{\bm Y}(t)}$, that evolves the initial state of the system, $\rho(0)$, to the final state, $\rho(t)$.  The use of the subscript ${\bm Y}(t)$ is to indicate the set of measurement results obtained over the finite interval $[0,t]$, rather than the instantaneous results obtained at time $t$ (denoted by ${\bm y}(t)$).  That is, we want to find
\beq
{\cal N}_{{\bm Y}(t)}:\rho_{{\rm c}}(t)={\cal N}_{{\bm Y}(t)}\rho(0).
\eeq
The mapping ${\cal N}_{{\bm Y}(t)}$ will be stochastic, due to its dependence upon the measurement results.  In order for $\rho_{{\rm c}}(t)$ to be a normalized density matrix, the mapping ${\cal N}_{{\bm Y}(t)}$ is, in general, nonlinear.  This makes the task of directly finding it intractable.  

To avoid the nonlinearity imposed by \erf{smeGenNonLin}, it is necessary to consider the {\it linear} SME~\cite{GoeGra,wisQTraj,jacLin}.  A linear formulation being possible is a general feature of quantum measurement theory, thus extending the linearity of the unconditioned ME. The `cost' of formulating measurement dynamics in this manner is that the system state becomes unnormalised.  The state norm has meaning, as will be specified later in this section.  We term moving from the nonlinear to linear equation as `linearization' of the SME, and associate with it the traceful incremental evolution $d\bar{\rho}_{{\rm c}}(t)$.  
To be clear, linearization is not an approximation in any sense; the linearized SME is as accurate as the nonlinear SME. The normalised density matrix is, of course, found from $\rho_{{\rm c}}(t)=\bar{\rho}_{{\rm c}}(t)/\Tr\left[\bar{\rho}_{{\rm c}}(t)\right]$.  The overbar indicates an unnormalized state.

In order to deal with ordinary operators, rather than superoperators, step two of our solution method is to transform the linear SME for the unnormalized density matrix, $\bar{\rho}(t)$, into a linear equation for a state vector, $\ket{\bar{\rho}(t)}$ in a {\it larger Hilbert space}.  This `vectorized' equation in the larger Hilbert space would consist of left-only matrix multiplication of the state vector. For a finite dimensional basis, this can be achieved simply by column stacking the elements of $\rho$ to form a vector.  That it can be achieved for infinite dimensional bosonic modes, via the thermo-entangled state representation~\cite{FANthermEta,UmeAri,Umebook} in a larger Hilbert space ${\cal H}\otimes{\cal H}$, will be discussed when the detailed solution method is provided in the next section.  For the moment, we note that it is possible to recast the evolution $d\bar{\rho}_{{\rm c}}(t)$, via what we term `vectorization', into $d\ket{\bar{\rho}_{{\rm c}}(t)}$.

Our goal is to obtain the mapping from $\ket{\rho(0)}$ to $\ket{\bar{\rho}_{{\rm c}}(t)}$.  To achieve this, we first need to obtain the mapping corresponding to an infinitesimal time slice, $d t$,
\bqa
 \ket{\bar{\rho}_{{\rm c}}(t+d t)}&=&d \ket{\bar{\rho}_{{\rm c}}(t)}+\ket{\bar{\rho}(t)}\label{incrementState}\\
&=& \hat{V}_{{\bm y}(t)}\ket{\bar{\rho}(t)},\label{smallMap}
\eqa
where $\hat{V}_{{\bm y}(t)}$ represents nonunitary evolution.  That $ \hat{V}_{{\bm y}(t)}$ can be obtained from \erf{incrementState} is clear, given the linear form of the SME.  However, it is also useful to put $\hat{V}_{{\bm y}(t)}$ in an exponential form,
\beq
\hat{V}_{{\bm y}(t)}= e^{\hat{v}_{{\bm y}(t)}},
\label{expMap}
\eeq 
for some operator $\hat{v}_{{\bm y}(t)}$, as this allows contact with techniques from Lie algebra.  We refer to finding $\hat{V}_{{\bm y}(t)}$, and obtaining its desired form, as a process of `factorization' (see \frf{flowChart}).

The next step in obtaining the evolution operator for a finite time interval begins with the division of $t$ into a very large number, $J$, of time slices of length $dt$.  The finite evolution is given by
\bqa
\ket{\bar{\rho}_{{\rm c}}(t)}&=&\hat{V}_{{\bm y}(Jd t)}\cdots \hat{V}_{{\bm y}(2d t)}\hat{V}_{{\bm y}(d t)}\ket{\rho (0)}
\label{evoProd}\\
&\equiv&\hat{V}_{{\bm Y}(t)}\ket{\rho (0)}.
\label{evoOp}
\eqa
By finding the nonunitary evolution operator, $\hat{V}_{{\bm Y}(t)}$, which depends on the set, ${\bm Y}(t)$, of all measurement results over the interval $[0,t]$, the evolution is solved.  
In the limit $d t\rightarrow 0$, $\hat{V}_{{\bm Y}(t)}$ will contain integrals over the measurement record. The process of finding $\hat{V}_{{\bm Y}(t)}$, from the string of operators representing infinitesimal evolution, is termed `composition'.   

If desired, the vectorization to form $\ket{\bar{\rho}_{{\rm c}} (t)}$ can be unwound to write the solution in terms of $\bar{\rho}_{{\rm c}}(t)$. We refer to this unwinding as `matricization' as we are moving from the state vector, in the larger Hilbert space, back to the state matrix, in the original Hilbert space. The density matrix SME solution, as opposed to the state vector, is written in terms of a evolution {\it superoperator}, ${\cal V}_{{\bm Y}(t)}$, analogous to the evolution {\it operator}:
\beq
\bar{\rho}_{{\rm c}}(t)={\cal V}_{{\bm Y}(t)}\rho (0).
\eeq

Having sketched how to solve the SME, we now extend a little further and indicate how the POVM is subsequently obtained.  The POVM, which characterizes a quantum measurement,
is defined as the set of positive operators, $\{\hat{W}_{{\bm Y}(t)}:{\bm Y}(t)\}$, such that for all (normalized) $\rho(0)$
\beq
\wp_{{\bm Y}(t)} =\Tr[\hat{W}_{{\bm Y}(t)}\rho (0) ],
\label{POVMdef}
\eeq
where $\wp_{{\bm Y}(t)}$ is the probability density for the measurement record ${\bm Y}(t)$ and $\hat{W}_{{\bm Y}(t)}$ is known as the `effect' operator.

As mentioned earlier, the norm of the system state vector, $\left\Vert \ket{\bar{\rho}_{{\rm c}} (t)}\right\Vert$, has meaning when an initially normalized state is evolved using the linear SME.  Specifically, the norm is related to the probability of the measurement sequence, upon which the state is conditioned, as per  
\beq
\wp_{{\bm Y}(t)}=\left\Vert \ket{\bar{\rho}_{{\rm c}} (t)}\right\Vert
\wp_{{\rm ost}}({\bm Y}(t)).\label{normOst}
\eeq
There is some flexibility in the linear SME, relating to the choice of normalization.  However, 
$\wp_{{\bm Y}(t)}$ is a fixed quantity, so variations in $\left\Vert \ket{\bar{\rho}_{{\rm c}} (t)}\right\Vert$ must be compensated by the form of the `ostensible distribution', $\wp_{{\rm ost}}({\bm Y}(t))$~\cite{WisMil10}.  In this solution sketch, the specification of the particular form of the linear SME and ostensible distribution is not crucial, apart from noting that an analytic form can be found for both, and is left until later sections.  The reader is invited to read \arf{ostensible} for a more detailed discussion of the ostensible distribution.

The form of $\hat{W}_{{\bm Y}(t)}$ is fixed by ensuring that the two expressions for $\wp_{{\bm Y}(t)}$, given in \erf{POVMdef} and \erf{normOst}, are equal.  The effect operator is an operator in the original sized Hilbert space, so it is necessary to revert from the larger Hilbert space, via `matricization', when determining $\hat{W}_{{\bm Y}(t)}$.  As indicated by \erf{normOst}, and shown in \frf{flowChart}, we will find the POVM from $\ket{\bar{\rho}_{{\rm c}}(t)}$, the state vector form of the SME solution.  However, this is a matter of convenience; the POVM can, of course, be found from the SME density matrix solution.  The class of systems that we consider (those with linear HP dynamics and Gaussian measurement noise) will be found to have Gaussian effect operators.  A Gaussian operator is one which has a Gaussian Wigner function.

\section{Solving the stochastic master equation}

\label{SMEsol}

\subsection{Linearization}
In \srf{sketch}, a sequence of steps, that leads to the solution of a SME, was identified.  We now carry out these steps, beginning with the linearizing of the nonlinear SME.                                                                                                                                                                                                    
The nonlinear SME, for the system class of interest, has already been provided in \erf{smeGenNonLin}.  
As explained in \arf{ostensible}, the nonlinearity of the SME only affects the normalization of the density matrix so that 
one can faithfully propagate an unnormalized system state, $\bar{\rho}(t)$ using only linear terms, with what is known as a linear SME~\cite{chia2011complete}:
\bqa
d\bar{\rho}(t) &=& -i\left[\hat{H},\bar{\rho}(t) \right]dt+{\cal D}\left[\hat{{\bm c}} \right]\bar{\rho}(t)  dt
\nonumber\\
&&+{\bm y}^{{\rm T}}(t)dt 
\bar{{\cal H}}\left[ {\bm M}^{\dag}\hat{{\bm c}}\right]\bar{\rho}(t)  ,
\label{smeGen}
\eqa
where a linear form of the superoperator, ${\cal H}$, has been used
\beq
\bar{{\cal H}}\left[\hat{{\bm c}}\right]\rho\equiv \hat{{\bm c}}\rho+\rho\hat{{\bm c}}^{\ddagger}.\label{Hsuper}
\eeq
The bar on $d\bar{\rho}(t) $ is used to indicate a measurement-conditioned, unnormalized state, with the subscript `c' dropped for simplicity.  Note that the linear SME is written directly in terms of the measurement results ${\bm y}(t)$, which have the Gaussian statistics indicated  
by \erfs{actExpY}{actExpY2}.

\subsection{Vectorization}
The next step in the solution process is comprised of transforming the linear SME for the unnormalized density matrix into a vector form, in a larger Hilbert space.  The thermo-entangled state representation is now briefly reviewed, before being applied to the linear SME.


\subsubsection{A brief introduction to the thermo-entangled state representation}
To work in the thermo-entangled state representation, for a system consisting of $N$ physical modes described by a Hilbert space, ${\cal H}$, an ancillary Hilbert space of equal dimension is introduced, $\tilde{{\cal H}}$, which houses unphysical modes.  
This representation is based on prior work by Takahashi and Umezawa, relating to thermo field dynamics~\cite{UmeAri,Umebook}, in which a fictitious field is introduced in order to convert ensemble average calculations into equivalent pure state expressions.
Here, we focus on just the results that we require, and the interested reader is referred to the available literature for more detail~\cite{ZhoFan,FANthermo,FANthermo2,Kosov,FANthermEta,UmeAri,Umebook,FanHu,Ban}.

For an arbitrary operator $\hat{A}$ acting on vectors of the Hilbert space ${\cal H}$, there is a `tilde conjugate' operator $\tilde{A}$ that acts identically on vectors of the Hilbert space $\tilde{{\cal H}}$.  Without loss of generality, we can define the relationship between tilde and non-tilde operators as~\cite{Kosov}
\beq
\hat{A}=A(\hat{{\bm a}},\hat{{\bm a}}\dg), \quad \tilde{A}=A^*\left(\tilde{{\bm a}},\tilde{{\bm a}}\dg\right).
\label{tildeConjDef}
\eeq
Note that $\tilde{{\bm a}}=\left(\tilde{a}_{1},\tilde{a}_{2},...,\tilde{a}_{N}\right)^{\rm T\blk}$ has been introduced, and that taking the tilde conjugate of a matrix does not alter its dimensions.  This generalizes to the case where the object to be tilde-conjugated is itself a matrix of operators, $\hat{{\bm A}}={\bm A}(\hat{{\bm a}},\hat{{\bm a}}\dg)$, and the the matrix dimensions are left unaltered:
\beq
\tilde{{\bm A}}={\bm A}\blk ^{*}(\tilde{{\bm a}},\tilde{{\bm a}}\dg).
\label{tildeMat}
\eeq
The tilde and non-tilde annihilation and creation operators obey standard bosonic commutation relations:
\bqa
\left[\hat{a}_{m},\hat{a}_{n} \right]&=&\left[\tilde{a}_{m},\tilde{a}_{n} \right]=0,\\ \left[\hat{a}_{m},\hat{a}_{n} ^{\dg}\right]&=&\left[\tilde{a}_{m},\tilde{a}_{n}^{\dg}\right]=\delta_{mn},\\
\left[\hat{a}_{m},\tilde{a}_{n} \right]&=&\left[\hat{a}_{m},\tilde{a}_{n}^{\dg} \right]=0,
\label{commRelations}
\eqa
for $m,n\in\{1,\ldots,N\}$.

From \erf{tildeConjDef}, and the requirement that $ (\tilde{A})^{\widetilde{}}=\hat{A}$, the following `tilde-conjugation' rules may be inferred
\bal
&\left(z_1 \hat{A}_1+z_2 \hat{A}_2\right)^{\widetilde{}}=z_{1}^{*}\tilde{A}_1+z_{2}^{*}\tilde{A}_2,\\
&\left(\hat{A}_{1}\hat{A}_{2}\right)^{\widetilde{}}=\tilde{A_{1}}\tilde{A_{2}},\\
&(\hat{A}\dg)^{\widetilde{}}=(\tilde{A})\dg,
\label{tildeConjEqn}
\eal
for complex numbers $z_{1},z_{2}$. 
\blk

Similarly to multimode coherent states, the `mixed'-mode operator 
\beq
\hat{{\bm \beta}} = \hat{{\bm a}}-\tilde{{\bm a}}^{\ddagger}
\label{betaOpMixMode}
\eeq
(recall our previous definition, applying equally to tilde operators, that $\hat{{\bm a}}^{\ddagger}\equiv \left (\hat{{\bm a}}^{{\rm T}}\right)^{\dag}$) may be defined, which has eigenstates:
\begin{equation}
\hat{{\bm \beta}}\ket{{\bm \beta}}={\bm \beta}\ket{{\bm \beta}},\quad \hat{{\bm \beta}}^\dagger\ket{{\bm \beta}}={\bm \beta}^{\dag}\ket{{\bm \beta}}, \quad {\bm \beta}\in  \mathbb{C}^{N}. \label{mixMode}
\end{equation}
Note the use of the non-operator expression ${\bm \beta}^{\dag}=\left({\bm \beta}^{*}\right)^{{\rm T}}$, where ${\bm \beta}$ is a vector of complex numbers. Like position or momentum eigenstates, the states $\ket{{\bm \beta}}$ are not normalizable. 
From \erf{mixMode} we deduce that 
\beq
\hat{{\bm a}}\ket{{\bm \beta}={\bm 0}}=\tilde{{\bm a}}^{\ddagger}\ket{{\bm \beta}={\bm 0}},\quad \hat{{\bm a}}^{\ddagger}\ket{{\bm \beta}={\bm 0}}=\tilde{{\bm a}}\ket{{\bm \beta}={\bm 0}}.  
\label{thermoAnn}
\eeq
Going forward, the cumbersome notation of $\ket{{\bm \beta}={\bm 0}}$, indicating the zero eigenvector of $\hat{{\bm \beta}}$, will be dropped and merely written as $ \ket{{\bm 0}}\equiv \ket{{\bm \beta}={\bm 0}}$. To distinguish this zero-valued eigenvector of $\hat{{\bm \beta}}$ --- which we will often refer to as the thermo-entangled state vacuum --- from the $2N$-mode vacuum state of the Hilbert space ${\cal H}\otimes\tilde{{\cal H}}$, we will denote the latter by $\ket{ \mathbf{VAC}}$.

A useful consequence of \erf{thermoAnn} is that for the arbitrary, non-tilde, operator $\hat{{\bm A}}={\bm A}(\hat{{\bm a}},\hat{{\bm a}}\dg)$ (which acts as the identity on tilde modes), 
\beq
\hat{{\bm A}}\ket{{\bm 0}}=\tilde{{\bm A}}^{\ddagger}\ket{{\bm 0}}.
\eeq
For the special case in which $\hat{{\bm A}}$ is Hermitian, this relation simplifies further, to 
\beq
\hat{{\bm A}}\ket{{\bm 0}}=\tilde{{\bm A}}\ket{{\bm 0}}.
\label{HermVect}
\eeq

It is shown in~\cite{FANthermEta} that the thermo-entangled states are given by:
\begin{equation}
\ket{{\bm \beta}}=\exp\left(-\frac{\left| {\bm \beta}\right|^{2}}{2}+\hat{{\bm a}}^\dagger{\bm \beta}-{\bm \beta}^{\dag}\tilde{{\bm a}}^{\ddagger}+\hat{{\bm a}}^\dagger \tilde{{\bm a}}^{\ddagger}\right)
\ket{ \mathbf{VAC}}.
\label{tESVac}
\end{equation}
For later use, we note that the thermo-entangled state vacuum is given in terms of the Hilbert space vacuum by
\beq
\ket{{\bm 0}}\equiv \ket{{\bm \beta}={\bm 0}}=
\exp\left(
\hat{{\bm a}}^\dagger \tilde{{\bm a}}^{\ddagger}
\right)\ket{ \mathbf{VAC}},\label{hilbVacThermoVac}
\eeq
which follows directly from \erf{tESVac}.  As an aside, the thermo-entangled states are precisely the common eigenstates of the relative co-ordinate and total momentum of two particles, which are central to the original Einstein, Podolsky and Rosen (EPR) scheme~\cite{PhysRevA.49.704,PhysRev.47.777}. 

The thermo-entangled states allow us to represent an $N$-mode system density matrix, $\rho$, by a vector, $\ket{\rho}$, in the larger $2N$-mode Hilbert-space, via 
\beq
\ket{\rho}\equiv \rho\otimes\hat{\eye}\ket{{\bm 0}},
\eeq 
where the $N$-mode identity is acting on the unphysical modes. 

\subsubsection{Application of the thermo-entangled state representation}

As previously indicated, we will work in the thermo-entangled state representation, whereby the state matrix is vectorized according to $\ket{\bar{\rho}}\equiv \bar{\rho}\otimes\hat{\eye}\ket{{\bm 0}}$.  That is, \erf{smeGen}
is simply right multiplied by $\ket{{\bm 0}}$ and then the relations of \erfs{tildeConjDef}{tildeMat} and \erfs{thermoAnn}{HermVect} are used.  It is found that 
\bqa
d\ket{\bar{\rho}(t)} &=&\bigg[
\hat{S}\left[-i\hat{H}\right]dt
+\hat{D}\left[\hat{{\bm c}}\right]dt\nonumber\\
&&\quad+{\bm y}^{{\rm T}}(t)dt\hat{S}\left[{\bm M}^{\dag}\hat{{\bm c}}\right]
\bigg]\ket{\bar{\rho}(t)}.
\label{vecSME}
\eqa
The operator $\hat{D}\left[\hat{{\bm c}}\right]$ is defined to represent the unconditional decoherence terms
\beq
\hat{D}\left[\hat{{\bm c}}\right]=\tilde{{\bm c}}^{{\rm T}}\hat{{\bm c}}-\smallhalf
\hat{{\bm c}}^{\dag}\hat{{\bm c}}-\smallhalf
\left(\hat{{\bm c}}^{\dag}\hat{{\bm c}}\right)^{\sim},\label{DOPER}
\eeq
and we have also introduced notation for the sum of an operator and its tilde conjugate:
\beq
\hat{S}\left[{\bm M}^{\dag}\hat{{\bm c}}\right]=
{\bm M}^{\dag}\hat{{\bm c}}+\left({\bm M}^{\dag}\hat{{\bm c}}\right)^{\sim}.\label{SOPER}
\eeq
The RHS of \erf{vecSME}'s being invariant under tilde conjugation ensures that the hermiticity of the density matrix is preserved.  The vectorization step of the SME solution is now complete.

\subsection{Factorization}
\label{factorization}
 
The next step is to factorize the evolution.  From \erf{vecSME}, an expression for $\hat{V}_{{\bm y}(t)}$, such that
\beq
\ket{\bar{\rho}(t+dt)}=\hat{V}_{{\bm y}(t)}\ket{\bar{\rho}(t)},
\label{dMap}
\eeq
is trivially given by
\beq
\hat{V}_{{\bm y}(t)}=\hat{\eye}+
\hat{S}\left[-i\hat{H}\right]dt
+\hat{D}\left[\hat{{\bm c}}\right]dt+{\bm y}^{{\rm T}}(t)dt\hat{S}\left[{\bm M}^{\dag}\hat{{\bm c}}\right],
\label{factor1}
\eeq
with the $2N$-mode identity operator acting on all modes of the doubled Hilbert space.  

In order to calculate an evolution operator using the techniques of Lie algebra, an exponential operator form, $\hat{V}_{{\bm y}(t)}=\exp(\hat{v}_{{\bm y}(t)})$, is required.  An expression for $\hat{v}_{{\bm y}(t)}$, that provides a $\hat{V}_{{\bm y}(t)}$ accurate to $\mathcal{O}(dt)$, is formally achieved with 
\bqa
\hat{v}_{{\bm y}(t)}
&=&
\hat{S}\left[-i\hat{H}\right]dt+\hat{D}\left[\hat{{\bm c}}\right]dt
-\frac{1}{2}\left({\bm y}^{{\rm T}}(t)dt
\hat{S}\left[{\bm M}^{\dag}\hat{{\bm c}}\right]\right)^{2}
\nonumber\\
&&+{\bm y}^{{\rm T}}(t)dt
\hat{S}\left[{\bm M}^{\dag}\hat{{\bm c}}\right].
\label{vExp}
\eqa
To verify \eqref{vExp}, the exponential form of $\hat{V}_{{\bm y}(t)}$ should be expanded to first order in $dt$ and compared with \erf{factor1}.  This requires the inclusion of the second order contribution from ${\bm y}^{{\rm T}}(t)dt\hat{S}\left[{\bm M}^{\dag}\hat{{\bm c}}\right]$, as it is of order $\sqrt{dt}$ due to \ito's rule (see \erf{actExpY2}). The same
reasoning implies that $\left({\bm y}^{{\rm T}}(t)dt\hat{S}\left[{\bm M}^{\dag}\hat{{\bm c}}\right]\right)^{2}$ is actually deterministic and equal to $\left(dt\hat{S}\left[{\bm M}^{\dag}\hat{{\bm c}}\right]^{{\rm T}}\hat{S}\left[{\bm M}^{\dag}\hat{{\bm c}}\right]\right)$, to first order in $dt$.  Despite being deterministic, this term results from the presence of measurement. 

The first three terms of $\hat{v}_{{\bm y}(t)}$, written on the first line of \erf{vExp}, are deterministic, as well as being quadratic in the annihilation and creation operators.  The final term of \erf{vExp} is stochastic. It is linear in the measurement results, and linear in the annihilation and creation operators.  This motivates the simplifying expression
\beq
\hat{v}_{{\bm y}(t)}=\hat{Q}dt+d\hat{L}(t),
\label{littlev}
\eeq
where $\hat{Q}$ contains terms that are time independent and quadratic, while $d\hat{L}$ contains stochastic, linear terms. In summary, the dynamical map of \erf{dMap} is achieved by the nonunitary evolution operator 
\beq
\hat{V}_{{\bm y}(t)}=\exp\left[ \hat{Q}dt+d\hat{L}(t) \right],\label{evoOpdt}
\eeq 
with 
\bqa
\hat{Q}&=&
\hat{S}\left[-i\hat{H}\right]+\hat{D}\left[\hat{{\bm c}}\right]-\smallhalf\hat{S}\left[{\bm M}^{\dag}\hat{{\bm c}}\right]^{{\rm T}}
\hat{S}\left[{\bm M}^{\dag}\hat{{\bm c}}\right],
 \\
d\hat{L}(t)&=&
{\bm y}^{{\rm T}}(t)dt\hat{S}\left[{\bm M}^{\dag}\hat{{\bm c}}\right].
\eqa

\subsection{Composition}

Given the evolution operator, $\hat{V}_{{\bm y}(t)}$, which evolves the state vector forward a single time slice, the next task is to consider a sequence of them that evolves the initial state forward a finite duration:
\bqa
\ket{\bar{\rho}(t)}&=&\hat{V}_{{\bm y}(Jd t)}\cdots \hat{V}_{{\bm y}(2d t)}\hat{V}_{{\bm y}(d t)}\ket{\rho (0)},\label{evoOp2}\\
&=&\prod\limits_{j=1}^{J}\exp\left[\hat{Q} dt+d\hat{L}(j dt)\right]\ket{\rho (0)}\label{inftyRho}\\
&=&\hat{V}_{{\bm Y}(t)}\ket{\rho (0)}.\label{SMEsolVY}
\eqa
In the second expression, \erf{inftyRho}, in which the form of \erf{littlev} has been used, $Jdt=t$ and the product is enumerated with $j$ increasing from right to left.  Note that a normalized initial state has been assumed. The goal of the `composition' step is to find the finite evolution operator, $\hat{V}_{{\bm Y}(t)}$.  This expression for the state {\it matrix} (albeit in vectorized form) is analogous to that for the state {\it vector} obtained in~\cite{jacLin}, thus allowing a similar fundamental approach (but requiring different techniques).  

To derive a practicable expression for $\hat{V}_{{\bm Y}(t)}$, the long sequence of $J$ exponential operators must be re-ordered and composed (which will involve integration for $dt\rightarrow 0$) so as to form a small number of terms. This is a task in bosonic algebra and methods of Lie groups are used.

\subsubsection{Lie algebra}
\label{lieAlg}

To make contact with Lie algebra, note that $\hat{v}_{{\bm y}(t)}$ is comprised of terms that 
are at most quadratic in the $2N$-mode annihilation and creation operators. This is ensured by the assumption of linear HP dynamics (see \srf{sysSpecification}). Here is a list of all $(4N+1)(2N+1)$ such operators:
\bqa
\{&&\hat{1},\hat{a}_{m},\hat{a}^{\dagger}_{m},
\tilde{a}_{m},\tilde{a}_{m}^{\dg},\nonumber\\
&&\hat{a}_{m}\hat{a}_{n},
\tilde{a}_{m}\tilde{a}_{n},
\hat{a}_{m}^{\dagger}\hat{a}_{n}^{\dagger},
\tilde{a}_{m}^{\dagger}\tilde{a}_{n}^{\dagger},\nonumber\\
&&\hat{a}_{m}\tilde{a}_{n},
\hat{a}_{m}^{\dagger}\tilde{a}_{n}^{\dagger},
\hat{a}_{m}\tilde{a}_{n}^{\dagger},
\hat{a}_{m}^{\dagger}\tilde{a}_{n},\nonumber\\
&&\hat{a}_{m}^{\dagger}\hat{a}_{n}+\smallhalf\delta_{mn},
\tilde{a}_{m}^{\dagger}\tilde{a}_{n}+\smallhalf\delta_{mn}
\},
\label{opAlgebraA}
\eqa 
for $m,n\in\{1,\ldots,N\}$.  It is important to emphasize that unphysical tilde modes (represented by operators $\tilde{{\bm a}},\tilde{{\bm a}}^{\dag}$)
are on the same footing as physical modes (represented by operators $\hat{{\bm a}},\hat{{\bm a}}^{\dag}$) in terms of bosonic algebra calculations.  

The span over $\mathbb{C}$ of the operators is a subalgebra of the symplectic algebra $\mathfrak{sp}(4N+2)$~\cite{Gilmore,GilmoreCoh}.  The operator commutator is the Lie bracket for the Lie algebra.
The Lie algebra, $\mathfrak{g}$, has an associated Lie group, $G$, that is formed by taking the exponential map of $\mathfrak{g}$,
\beq
g=\exp(X),
\eeq
for $X\in\mathfrak{g}$ and $g\in G$.  It can be seen that $\hat{v}_{{\bm y}(t)}\in\mathfrak{g}$ and it will become clear that $\hat{V}_{{\bm Y}(t)}\in G$.  It is important to be clear that $\mathfrak{g}$ consists of arbitrary linear combinations of the operators listed in \erf{opAlgebraA}.

Of central importance to us is that there are group-theoretic calculations that are: (1) only dependent upon the algebra's commutation relations, and (2) are independent of the representation that the calculations are carried out in, provided that the representation respects the algebra's commutation relations~\cite{GilmoreSingle}.  In particular, this holds for multiplication within the group, and the related tasks of exponential operator disentanglement and reordering.  As $\hat{V}_{{\bm Y}(t)}$ is formed through group multiplication, it follows that $\hat{V}_{{\bm Y}(t)}\in G$.

In the context of a Lie group, disentanglement refers to the splitting of an exponential operator, whose exponent consists of multiple terms, into a product of exponential operators.  That is, for $Z\in\mathfrak{g}$ we disentangle the exponential operator into a product of $n$ terms as per
\beq
e^{Z}=e^{X_{1}}\ldots e^{X_{n}},\label{disEqn}
\eeq 
for $X_{1},\ldots ,X_{n}\in\mathfrak{g}$. To give an illustrative example, a simple, but well-known, disentanglement is the normal-ordering of the single-mode displacement operator
\beq
e^{\alpha\hat{a}^{\dag}-\alpha^{*}\hat{a}}=
e^{\alpha\hat{a}^{\dag}}e^{-\alpha^{*}\hat{a}}e^{-\frac{1}{2}\abs{\alpha}^{2}},
\eeq
for $Z=\alpha\hat{a}^{\dag}-\alpha^{*}\hat{a}$.  This disentanglement follows directly from the Zassenhaus formula~\cite{Wil67}
\beq
e^{X+Y}=e^{X}e^{Y}e^{-\smallhalf [X,Y]}\cdots,\label{zass}
\eeq
with further terms in the product containing higher order commutators, for example $[Y,[X,Y]]$.
In the case of the displacement operator, the higher order commutators evaluate to zero.  It is important to note that algebra described by \erf{opAlgebraA} is significantly more complicated than that of the subalgebra, $\{\hat{1},\hat{a},\hat{a}^{\dag}\}$, relevant to the displacement operator, and we will not in general use the Zassenhaus formula to evaluate disentanglements.  Despite this, the Zassenhaus formula highlights that disentanglement is only dependent upon the commutation relations of the algebra.

The second group-theoretic calculation of importance to us is exponential operator reordering.  Given the operator $e^{X}e^{Y}$, we may wish to write an equivalent expression in which the $e^{Y}$ term appears to the left.  That is, 
\beq
e^{X}e^{Y}=e^{Y}e^{Z},\label{reUp}
\eeq
for known $X,Y\in\mathfrak{g}$ and a to-be-determined $Z=e^{-Y}e^{X}e^{Y}\in\mathfrak{g}$.  It is possible to give an expression for $Z$~\cite{hall2015lie},
\beq
Z=X-[Y,X]+\frac{1}{2!}[Y,[Y,X]]+\ldots,\label{Zreorder}
\eeq
which makes it clear that operator reordering is a function of the group commutation relations only.

Having established the central role that the commutator plays in operator reordering and disentanglement, we are motivated to look at the commutator structure of the Lie algebra, $\mathfrak{g}$.  To do so we define a number of partially overlapping subalgebras.  We define the subalgebra $\mathfrak{q}$ as containing all the quadratic (in annihilation and creation operators) elements of $\mathfrak{g}$, together with the identity operator. Similarly, the subalgebra $\mathfrak{l}$ is defined as containing all linear elements of $\mathfrak{g}$, together with the identity.  Finally, we define $\mathfrak{i}$ as the subalgebra consisting only of the identity operator.  Then, we note the following useful facts:
\beq
[\mathfrak{q},\mathfrak{q}]\in\mathfrak{q},\label{QuadSub}
\eeq
\beq
[\mathfrak{l},\mathfrak{g}]\in\mathfrak{l},\label{LinIdeal}
\eeq
\beq
[\mathfrak{l},\mathfrak{l}]\in\mathfrak{i},\label{linLin}
\eeq
\beq
[\mathfrak{l},[\mathfrak{l},\mathfrak{l}]]=0.\label{linLinLin}
\eeq
That is, $\mathfrak{q}$ forms a subalgebra of $\mathfrak{g}$, and $\mathfrak{l}$ is an ideal of $\mathfrak{g}$. \erf{linLinLin} follows from \erf{linLin} as the identity commutes with every algebra element (it is the center of the algebra), but we state it explicitly for the reader due to its frequent application. 

The consequences of the algebraic facts of \erfs{QuadSub}{LinIdeal} at the group level are the following. Given \erf{disEqn} for $Z\in\mathfrak{q}$, it is true that $X_{1},\ldots ,X_{n}\in\mathfrak{q}$.  That is, the disentanglement of the exponential of a quadratic operator is given by a product of exponentials that do not involve any linear exponents.  Additionally, given \erf{reUp} for $X\in\mathfrak{q}$ and $Y\in\mathfrak{l}$, then $Z\in\mathfrak{l}$.  That is, when an exponential operator with quadratic exponent is moved through an exponential with linear exponent, the quadratic exponential is unchanged.  The exponential with linear exponent retains a linear exponent only, but one that is changed according to \erf{Zreorder}.  

As mentioned, we will not use the Zassenhaus formula, or \erf{Zreorder}, to calculate operator disentanglements and reordering for general elements $g\in G$.  The algebra of $\mathfrak{g}$ is, in general, too complicated to make this feasible.  Instead, we use the fact that a representation of the algebra that upholds the algebra's commutation relations can be employed to perform group calculations, with the results then abstracted back to the level of the algebra.  To make the representation-independent calculations, it is beneficial to choose as simple as possible a representation that faithfully respects the algebra $\mathfrak{g}$.  The bosonic operators of $\mathfrak{g}$ are typically represented in the infinite dimensional Fock basis, but this is an unnecessary complication as far as group calculation is concerned.  It is a convenient fact that there exist faithful finite dimensional matrix representations of $\mathfrak{g}$.  All such faithful finite dimensional matrix representations are known, with the smallest using matrices of dimension $(4N+2)\times (4N+2)$ in order to represent the $2N$-mode algebra, $\mathfrak{g}$, of \erf{opAlgebraA}~\cite{gilmore2012lie,gilmore2012lie,Gilmore}.  As a reminder, this $2N$-mode algebra corresponds to $N$ physical modes as well as $N$ unphysical modes that were introduced to facilitate the vectorization of the SME. 

Let us now assume, as will be true in practice, that we have chosen the minimally sized $(4N+2)$ matrices to represent the algebra $\mathfrak{g}$.  This leads to the disentanglement and operator reordering equations, \erfs{disEqn}{reUp}, being represented as matrix equations.  In other words, to find the parameters which describe the disentanglement and reordering, a finite set of algebraic equations needs solving. To construct these equations requires the exponentiation of symbolic matrices; for our systems this involves solving polynomials of degree $2N$.  As no known general solution exists for polynomials of degree higher than quartic, this approach has some intrinsic limitations for greater than two physical modes. Despite this, the Lie algebra facts detailed in this section will provide useful information regarding the form of the solution of the SME for arbitrary $N$, as well as more explicit results for $N\leq 2$.  
Finite dimensional matrix representations, and how they can be utilized in our calculations, are discussed further in \arf{LieAlgebraSec}.  We prefer to defer explicit finite dimensional matrix calculations to appendices in order to improve the readability of the main text.

Having described the essential techniques of Lie algebra, we can now return to the composition of the nonunitary evolution operator, $\hat{V}_{{\bm Y}(t)}$.

\subsubsection{Evolution operator form}
\label{Evolution}

In the remainder of this section, we will be heuristic in our calculations.  The reason for this is that the details will likely not add significantly to the conceptual understanding gained by the reader.  Despite this, the practical implementation of our SME solutions is obviously important, so we provide a recipe for their use in \srf{recipe}, as well as examples in \srf{singleModeEgs}.  The reader will also be referred to appendices as appropriate.  In this subsection, we obtain the form of the evolution operator $\hat{V}_{{\bm Y}(t)}$.  

Before reordering the product of exponential terms in \erf{inftyRho}, we perform the simple disentanglement of splitting the quadratic and linear terms that belong to each time slice. That is,
\bqa
\hat{V}_{{\bm y}(t)}&=&\exp\left[ \hat{Q}dt+d\hat{L}(t) \right]\\
&=&\exp\left[d\hat{L}(t) \right]\exp\left[ \hat{Q}dt\right],\label{splitExp}
\eqa
which is correct up to order $dt$.  This can be seen from \erf{zass}, as the corrections involve 
commutators of the infinitesimal operators.  For example, $[\hat{Q}dt,d\hat{L}(t)]={\cal O}(dt^{3/2})$. 

The general strategy to find $\hat{V}_{{\bm Y}(t)}$ begins with moving the rightmost quadratic exponential through the linear exponential to its left~\cite{notationClean}.  After this is done, it is in contact with a second quadratic exponential, with which it can be combined,
\beq
\exp\left[ \hat{Q}dt\right]\exp\left[ \hat{Q}dt\right]=\exp\left[ 2\hat{Q}dt\right],
\eeq
as the exponents obviously commute for our time-independent $\hat{Q}$.  This combined quadratic exponential is then moved through the next linear exponential to the left and combined with a third quadratic exponential.  After this has been repeated $j-1$ times, the task of reordering the combined quadratic exponential with the next linear exponential is given by
\bqa
e^{d\hat{L}(jdt)}\prod\limits_{k=1}^{j}\exp\left[\hat{Q} dt\right]&=&e^{d\hat{L}(\tau)}e^{\hat{Q} \tau}\label{linComm2}\\
&=&e^{\hat{Q}\tau}e^{d\hat{L}^{\prime}(\tau)},
\label{linComm}
\eqa
with the interim time $jdt$ labeled as $\tau$. The reordering, performed in \erf{linComm}, is an example of the reordering shown in \erf{reUp} for $Y\in\mathfrak{q}$ and $X\in\mathfrak{l}$. As described in the previous subsection, the quadratic exponential is left unchanged while the linear exponential remains linear but is modified.  The use of the prime in \erf{linComm}, for the linear exponential, is to indicate this modification.  Following the movement of the quadratic exponentials through all the linear exponentials, we are left with 
\beq
\hat{V}_{{\bm Y}(t)}=e^{\hat{Q}t}\prod\limits_{j=1}^{J}\exp\left[d\hat{L}^{\prime}(j dt)\right].
\eeq

Next, we wish to combine all the linear exponentials.   We note that  $d\hat{L}^{\prime}(j dt)\in \mathfrak{l}$.  Thus, according to \erf{zass} and \erfs{linLin}{linLinLin}, we can write
\beq
\prod\limits_{j=1}^{J}\exp\left[d\hat{L}^{\prime}(j dt)\right]=
\exp\left[\delta_{{\bm Y}(t)}\hat{1}+\sum_{j=1}^{J}d\hat{L}^{\prime}(j dt)\right],
\label{tooLate}
\eeq
with $\delta_{{\bm Y}(t)}$ being a scalar that is a function of the stochastic measurement record.  From this point onward, the identity operator is not explicitly written, due to its trivial action. For convenience, we combine the linear term and that proportional to the identity into a single operator as per
\beq
\hat{L}^{\prime}(t)=\delta_{{\bm Y}(t)}+\sum_{j=1}{J}d\hat{L}^{\prime}(j dt),
\eeq
where $\hat{L}^{\prime}(t)$ is finite and will convert into a stochastic {\ito } integral in the limit $dt\rightarrow 0$.  The $\delta_{{\bm Y}(t)}$ term does not affect the system state, but is relevant to the probability of obtaining the measurement record (see \erf{normOst}).

We can now give the form of $\hat{V}_{{\bm Y}(t)}$ as
\beq
\hat{V}_{{\bm Y}(t)}=e^{\hat{Q}t}e^{\hat{L}^{\prime}(t)}.
\eeq
As desired, the evolution operator has been composed into a product of a finite number of exponentials (in this case, two).  The first is deterministic and contains quadratic operator terms.  The second is linear and is a function of the measurement record.
When $\hat{V}_{{\bm Y}(t)}$ is used in \erf{SMEsolVY}, the solution of the SME,
\beq
\ket{\bar{\rho} (t)}=e^{\hat{Q}t}e^{\hat{L}^{\prime}(t)}\ket{\rho (0)},
\label{solSumm}
\eeq
is obtained.  Note that the invariance of $\hat{V}_{{\bm Y}(t)}$ under tilde conjugation, $\hat{V}_{{\bm Y}(t)}=\tilde{V}_{{\bm Y}(t)}$, ensures that $\bar{\rho}(t)$ is Hermitian.
In the next two subsections we will provide more detail concerning $\hat{L}^{\prime}(t)$ and find a more convenient expression for $e^{\hat{Q}t}$.

\subsubsection{Investigating $e^{\hat{L}^{\prime}(t)}$}
\label{SMEsolPrimed}

In this subsection, we give expressions for $e^{\hat{L}^{\prime}(t)}$ that will be of later use, with calculational details deferred to appendices.  For notational convenience, and to emphasize that unphysical tilde modes are on the same footing as physical modes in terms of bosonic algebra calculations, the $\hat{{\bm a}}$ and $\tilde{{\bm a}}$ column vectors are placed into a single column vector of length $2N$:
\beq
\hat{{\bm b}}=\left(\hat{{\bm a}};\tilde{{\bm a}}\right)=\left(\hat{a}_{1},...,\hat{a}_{N},\tilde{a}_{1},...,\tilde{a}_{N}\right)^T.
\label{bVec}
\eeq
We can now write $\hat{Q}$ and $d\hat{L}(t)$ in terms of the bosonic vector $\hat{{\bm b}}$:
\begin{align}
\hat{Q}t=&\hat{{\bm b}}^{\dag}{\bm R}\hat{{\bm b}}^{\ddagger}+
\hat{{\bm b}}^{\dag}{\bm D}\hat{{\bm b}}+
\hat{{\bm b}}^{{\rm T}}{\bm L}\hat{{\bm b}}
\label{L}\\
d\hat{L}(t)=&\hat{{\bm b}}^{\dag}d{\bm r}+d{\bm l}\hat{{\bm b}},
\label{dS}
\end{align}
with the $2N\times2N$ matrices ${\bm R},{\bm D},{\bm L}$, the $2N\times 1$ column vector, $d{\bm r}$, and the $1\times 2N$ row vector, $d{\bm l}$, all constrained by Hermiticity preservation of the evolved state matrix. 
Rather than use $\left\{{\bm r},{\bm l}\right\}$ we have introduced $\left\{d{\bm r},d{\bm l}\right\}$, in the linear evolution containing the measurement noise, to emphasize that they are infinitesimal and are of $\mathcal{O}(\sqrt{dt})$.  
The relationship between the system description in \eqref{smeGen}, that is given in terms of $\{{\bm G},{\bm C},{\bm M}\}$, and the parameterization of $\{{\bm R},{\bm D},{\bm L},d{\bm l},d{\bm r}\}$ of \erfs{L}{dS} is provided in \arf{RDLdldrDescription}.  Note that $d\hat{L}^{\prime}(t)$ is defined analogously to \erf{dS}, in terms of $\{ d{\bm r}^{\prime},d{\bm l}^{\prime}\}$ (see \erf{dS2}).

The explicit expression for $\hat{L}^{\prime}(t)$ is not difficult to derive using a finite dimensional representation of the algebra $\mathfrak{sp}(4N+2)$ (see \arf{LieAlgebraSec} for details of this calculation, in particular \arf{reorderAppLin1} and \arf{reorderAppLin2}).  In this section we state it as
\beq
e^{\hat{L}^{\prime}(t)}=e^{h}e^{\hat{{\bm b}}^{\dag} {\bm r}^{\prime}}e^{{\bm l}^{\prime} \hat{{\bm b}}},
\label{linPiece}
\eeq
with
$h$ being a scalar non-Gaussian complex-valued stochastic integral and 
\bqa
{\bm r}^{\prime}(\tau)&=&\int\limits_{0}^{\tau}d{\bm r}^{\prime}(s),
\label{intNote}
\eqa
with equivalent notation for ${\bm l}^{\prime}(\tau)$.
The explicit time dependence of $\{{h,\bm l}^{\prime},{\bm r}^{\prime}\}$ in \erf{linPiece} has been suppressed for display purposes.  The relevance (or lack thereof) of the presence of the non-Gaussian random variable, $h$, will be discussed in detail in relation to the POVM, in \srf{POVMsec}.  For the moment we note that $h$ has no effect upon the system state, as the non-operator multiplicative factor, $e^{h}$, will be removed when the state is normalized.  In other words, rather than storing the entire measurement record, ${\bm Y}(t)$, it is sufficient to track only $\{{\bm l}^{\prime},{\bm r}^{\prime}\}$ in order to follow the system state.  
In closing this subsection, we remind the reader that $e^{\hat{L}^{\prime}(t)}$ is the only term in our SME solution impacted explicitly by the measurement record. 

\subsubsection{Disentangling $e^{\hat{Q}t}$}

In order to facilitate calculations, such as expectation values, it is often convenient to use a disentangled exponential operator, with an ordering chosen to suit the calculation.  In this subsection, we give the disentangled form of $e^{\hat{Q}t}$.
This should be understood in the context of \erf{disEqn} for 
$Z,X_{1},\ldots ,X_{n}\in\mathfrak{q}$.   That is, we split $e^{\hat{Q}t}$ into a product of exponentials with quadratic (and no linear) exponents. 
Once again, only the heuristic form is provided in this section as the explicit results are obtained from the finite dimensional representation of $\mathfrak{sp}(4N+2)$ (see~\arf{LieAlgebraSec}, together with~\srf{recipe} for details and examples).  Using the form of $\hat{Q}$ given in~\erf{L} we state
\begin{align}
e^{\hat{Q} t}=&\exp\left[\hat{{\bm b}}^{\dag}{\bm R}\hat{{\bm b}}^{\ddagger}+
\hat{{\bm b}}^{\dag}{\bm D}\hat{{\bm b}}+
\hat{{\bm b}}^{{\rm T}}{\bm L}\hat{{\bm b}}
 \right]
\label{gen}\\
=&e^{\delta^{\prime}}\exp\left[\hat{{\bm b}}^{\dag}{\bm R}^{\prime}\hat{{\bm b}}^{\ddagger}\right]
\exp\left[\hat{{\bm b}}^{\dag}\underline{{\bm D}}\hat{{\bm b}}
\right]
\exp\left[\hat{{\bm b}}^{{\rm T}}{\bm L}^{\prime}\hat{{\bm b}} \right]
\label{disG}\\
=& e^{\delta^{\prime}}\exp\left[\hat{{\bm b}}^{\dag}{\bm R}^{\prime}\hat{{\bm b}}^{\ddagger}\right]
\left(
:\exp\left[\hat{{\bm b}}^{\dag}{\bm D}^{\prime}\hat{{\bm b}} \right]:
\right)
\exp\left[\hat{{\bm b}}^{{\rm T}}{\bm L}^{\prime}\hat{{\bm b}} \right],
\label{disGnorm2}
\end{align}
\blk
with the primes and underline indicating different functions of system parameters.  Also,
${\bm D}^{\prime}=e^{\underline{\bm D}}-\eye_{2N}$ by a standard operator identity for normal ordering~\cite{FAMmulti}.
Note the appearance of the scalar $\delta^{\prime}$, due to the disentanglement involving the commutation of quadratic terms.

We have now solved the linear SME in the enlarged Hilbert space, ${\cal H}\otimes\tilde{{\cal H}}$.  The solution is comprised of \erf{solSumm}, \erf{linPiece} and \erf{disG}, together with the explicit expressions for primed variables contained in \arf{LieAlgebraSec}.
Some readers will object to the presence of the unphysical tilde modes, but we note that there are advantages in the thermo-entangled state representation of the solution.  For example, to find expectation values of a system operator $A(\hat{{\bm a}},\hat{{\bm a}}\dg)$ we use~\cite{FANthermo2}
\beq
\Tr[A(\hat{{\bm a}},\hat{{\bm a}}\dg)\bar{\rho}(t)]=\braketop{{\bm 0}}{A(\hat{{\bm a}},\hat{{\bm a}}\dg)}{\bar{\rho}(t)}.
\label{expect}
\eeq
As the system state is mixed in general (that is, impure) we cannot separately factorize the exponentials of physical and unphysical mode operators.  The conversion of the state vector, $\ket{\bar{\rho}(t)}$, to the state matrix, $\bar{\rho}(t)$, will necessitate the use of superoperators instead of operators or, alternatively, the power series expansion of exponentials.  Indeed, when finding the POVM representing the composite measurement, we find it more simple to use the $\ket{\bar{\rho}(t)}$ solution.  However, it is of clear relevance to show that we can find $\bar{\rho}(t)$, so this is performed in the next subsection.

For completeness, we perform one final operator reordering, being that of normal ordering the full evolution operator.  That is
\bqa
\hat{V}_{{\bm Y}(t)}&=&e^{\hat{Q}t}e^{\hat{L}^{\prime}(t)}\label{preoper}\\
&=&e^{h+\delta^{\prime}}e^{\hat{{\bm b}}^{\dag}\underline{ {\bm r}}+\hat{{\bm b}}^{\dag}{\bm R}^{\prime}\hat{{\bm b}}^{\ddagger}}
\left(:e^{\hat{{\bm b}}^{\dag}{\bm D}^{\prime}\hat{{\bm b}}}:\right)
e^{\hat{{\bm b}}^{{\rm T}}{\bm L}^{\prime}\hat{{\bm b}} +\underline{{\bm l}} \hat{{\bm b}}},\label{fullevoopeqn}
\eqa
where the reordering has lead to modification of the linear terms from $\{{\bm l}^{\prime},{\bm r}^{\prime}\}$ to $\{\underline{{\bm l}},\underline{{\bm r}}\}$, as detailed in \arf{acute}.

\subsection{Matricization}

The solution $\ket{\bar{\rho}(t)}$ is a vector in ${\cal H}\otimes\tilde{{\cal H}}$ and contains
the unphysical mode operators $\tilde{{\bm a}}$ and $\tilde{{\bm a}}\dg$ (within the mixed mode operators $\hat{{\bm b}}$ and $\hat{{\bm b}}^{\dag}$).   These can be removed in the following way.  Recall that $\ket{\rho(0)}\equiv\rho(0)\otimes\hat{\eye} \ket{\bm{0}}$, where $\rho(0)$ is the physical mode density matrix.  All tilde mode operators commute through $\rho(0)$ to act on $\ket{\bm{0}}$, with the conversion to physical mode operators as per \erf{thermoAnn}.

To identify the tilde mode operators, the compactifying notation of the the $\hat{{\bm b}}$ operators is unwound.  As should be evident already, there are an endless number of operator orderings that can be chosen, each with a parameterization.  To avoid having to repeat the definition, we note that for all $\{{\bm R},{\bm L},{\bm D},{\bm r},{\bm l}\}$ matrices (including $\{\underline{\bm R},\underline{\bm L},\underline{\bm D},\underline{\bm r},\underline{\bm l}\}$ and $\{{\bm R}^{\prime},{\bm L}^{\prime},{\bm D}^{\prime},{\bm r}^{\prime},{\bm l}^{\prime}\}$ or any other disentanglement parameters), the following block form will hold 
\bqa
{\bm L}&=&\left[
\begin{array}{cc}
    {\bf L}      & \breve{{\bf L}} \\
   \breve{{\bf L}}	 &  {\bf L}^{ *}
\end{array}\right]\label{LBlockeqn}\\
{\bm R}&=&\left[
\begin{array}{cc}
    {\bf R}      & \breve{{\bf R}} \\
   \breve{{\bf R}}	 &  {\bf R}^{ *}
\end{array}\right],\\
{\bm D}&=&\left[
\begin{array}{cc}
     {\bf D}     & \breve{ {\bf D}} \\
   \breve{ {\bf D}}^{ *}	 &   {\bf D}^{*}
\end{array}\right],
\label{allMatrixBlock}
\eqa
with $\breve{{\bf R}}=\breve{{\bf R}}^{\dg}=\breve{{\bf R}}^{{\rm T}}$ and ${\bf R}={\bf R}^{{\rm T}}$.  The same respective properties hold for $\{{\bf L},\breve{{\bf L}}\}$, but not necessarily for the block matrices of ${\bm D}$.
The vectors $\{{\bm l},{\bm r}\}$ (and operator reordered variations) all have the block form
\bqa
{\bm l}&=&\left[
\begin{array}{cc}
    { {\bf l}}    &  {{\bf l}}^{ *}
\end{array}\right],\\
{\bm r}&=&\left[
\begin{array}{c}
    { {\bf r}  }  \\
 { {\bf r}}^{ *}
\end{array}\right].
\label{allVecBlock}
\eqa
The requirement that evolution preserves state matrix hermiticity has been enforced for all parameters.  Note that Roman font has been used to distinguish the $N\times N$ block matrices from the full matrix (which is $2N\times 2N$), and similarly for the block vectors of length $N$ rather than $2N$.  To give an example to illustrate our notation, we have implied that the block form of $\{{\bm R}^{\prime}, {\bm D}^{\prime}\}$ is
\bqa
{\bm R}^{\prime}&=&\left[
\begin{array}{cc}
    {\bf R}^{\prime}      & \breve{{\bf R}}^{\prime} \\
   \breve{{\bf R}}^{\prime }	 &  {\bf R}^{\prime *}
\end{array}\right],\\
{\bm D}^{\prime}&=&\left[
\begin{array}{cc}
     {\bf D}^{\prime}     & \breve{ {\bf D}}^{\prime} \\
   \breve{ {\bf D}}^{\prime *}	 &   {\bf D}^{\prime *}
\end{array}\right],
\label{allMatrixBlock2}
\eqa
with $\breve{{\bf R}^{\prime}}=\breve{{\bf R}}^{\prime \dg}=\breve{{\bf R}}^{\prime{\rm T}}$ and ${\bf R}^{\prime}={\bf R}^{\prime{\rm T}}$.


By substituting the block form of \erfs{LBlockeqn}{allVecBlock} into \erf{fullevoopeqn}, one can obtain an expression for $\hat{V}_{{\bm Y}(t)}$ in terms of $\{\hat{{\bm a}},\hat{{\bm a}}^{\dg},\tilde{{\bm a}}^{\dag},\tilde{{\bm a}}^{\dag}\}$, rather than $\{\hat{{\bm b}},\hat{{\bm b}}^{\dag}\}$.  This gives a lengthy, but useful, aid
for moving to the state matrix solution of the linear SME:
\bqa
&&\hat{V}_{{\bm Y}(t)}=\nonumber\\
&&\exp\left[
\hat{{\bm a}}^{\dg}\underline{\bf r} +\tilde{{\bm a}}^{\dg}\underline{\bf r}^{*}+
\hat{{\bm a}}^{\dg}{\bf R}^{\prime}\hat{{\bm a}}^{\ddagger}+
\tilde{{\bm a}}^{\dg}{\bf R}^{\prime *}\tilde{{\bm a}}^{\ddagger}+
2\hat{{\bm a}}^{\dg}\breve{{\bf R}}^{\prime}\tilde{{\bm a}}^{\ddagger}
\right]\nonumber\\
&&\times e^{
\hat{{\bm a}}^{\dg}\underline{\breve{{\bf D}}}\tilde{{\bm a}}+
\tilde{{\bm a}}^{\dg}\underline{\breve{{\bf D}}}^{*}\hat{{\bm a}}
}
e^{
\hat{{\bm a}}^{\dg}\underline{\bf D}\hat{{\bm a}}}
e^{
\tilde{{\bm a}}^{\dg}\underline{\bf D}^{*}\tilde{{\bm a}}}\times\nonumber\\
&&\exp\left[
\hat{{\bm a}}^{{\rm T}}{\bf L}^{\prime}\hat{{\bm a}}+
\tilde{{\bm a}}^{{\rm T}}{\bf L}^{\prime *}\tilde{{\bm a}}+
2\hat{{\bm a}}^{{\rm T}}\breve{{\bf L}}^{\prime}\tilde{{\bm a}}+
\underline{\bf l}\hat{{\bm a}} +\underline{\bf l}^{*}\tilde{{\bm a}}
\right],\label{vectorizedEvoOp}
\eqa
which manifestly preserves state matrix Hermiticity.  Note that we choose to use the underlined operator ordering $e^{\hat{{\bm b}}^{\dag}\underline{\bm D}\hat{{\bm b}}}$ rather than the primed normal odering $:e^{\hat{{\bm b}}^{\dag}{\bm D}^{\prime}\hat{{\bm b}}}:$ for this purpose.
After the tilde operators are moved through $\rho(0)$ to act on $\ket{\bm{0}}$, they become right-multiplying operators onto $\rho(0)$.  

Our first option for representing the state matrix solution, $\bar{\rho}(t)$, is to place it in superoperator form
\beq
\bar{\rho}(t)={\cal V}_{{\bm Y}(t)}\rho(0).
\eeq
We find it necessary to introduce new superoperators, as well as clarifying how superoperator matrix-multiplication functions.  Our new superoperators, for an arbitrary vector of operators, $\hat{{\bm c}}$, and square matrix ${\bm M}$ (of the same length as $\hat{{\bm c}}$), are
\bqa
{\cal J}[\hat{{\bm c}}^{{\rm T}}]{\bm M}\rho &=&\hat{{\bm c}}_{i}^{{\rm T}}{\bm M}_{ij}\rho\hat{{\bm c}}_{j}^{\dg}\\
{\cal K}[\hat{{\bm c}}^{{\rm T}}]{\bm M}\rho &=&\hat{{\bm c}}_{i}^{{\rm T}}{\bm M}_{ij}\rho\hat{{\bm c}}_{j}.
\eqa
Thus, although ${\cal J}[\hat{{\bm c}}^{{\rm T}}]{\bm M}$ appears to be a row-vector, when acted upon an operator we define it so that it produces a scalar operator.  This convention is used for all superoperators and shows how the right-multiplying operator indices are summed with the left-multiplying portion.  We also use the superoperator $\bar{\cal H}$ from \erf{Hsuper}.
We can then write
\bqa
{\cal V}_{{\bm Y}(t)}&=&\exp\left[
\bar{{\cal H}}\left[\hat{{\bm a}}^{\dg}\underline{\bf r}+
\hat{{\bm a}}^{\dg}{\bf R}^{\prime}\hat{{\bm a}}^{\ddagger}
\right]+
2{\cal J}\left[\hat{{\bm a}}^{\dg}\right]\breve{{\bf R}}^{\prime}
\right]\times\nonumber\\
&&\exp\left[
{\cal K}\left[\hat{{\bm a}}^{\dg}\right]\underline{\breve{{\bf D}}}+
{\cal K}\left[\hat{{\bm a}}^{{\rm T}}\right]\underline{\breve{{\bf D}}}^{\dg}
\right]
\exp\left[
\bar{\cal H}\left[\hat{{\bm a}}^{\dg}\underline{{\bf D}}\hat{{\bm a}}\right]
\right]\times\nonumber\\
&&\exp\left[
\bar{\cal H}\left[\underline{\bf l}\hat{{\bm a}}+
\hat{{\bm a}}^{{\rm T}}{\bf L}^{\prime}\hat{{\bm a}}
\right]+
2{\cal J}\left[\hat{{\bm a}}^{{\rm T}}\right]\breve{{\bf L}}^{\prime}
\right].
\label{superOpEvoFull}
\eqa


Our second option, for writing $\bar{\rho}(t)$, is to use power series expansions of the exponentials instead of superoperators.  The notation to represent the general multi-mode case in this way is extremely cumbersome as one must link left-multiplying and right-multipling factors with summations that are associated with matrix multiplication. To avoid providing an expression which is too complicated to be of value, we limit ourselves to the single mode case, for which the vector notation is not needed (no bold font required).  Using the explicitly normal ordered form of \erf{disGnorm2}, we obtain
\bqa
\bar{\rho}(t)&=&
\sum_{jkmn=0}^{\infty}
\frac{(2\breve{{\rm L}})^{\prime j}{\breve{{\rm D}}}^{\prime *k}
{\breve{{\rm D}}}^{\prime m}(2\breve{{\rm R}}^{\prime })^{n}}
{j!k!m!n!}
\hat{a}^{\dg n+k}
e^{
\underline{\rm r}\hat{a}^{\dg} +
{\rm R}^{\prime }\hat{a}^{\dg 2}
}
\times \nonumber\\
&&
\left(:e^{
{{\rm D}}^{\prime }\hat{a}^{\dg}\hat{a}
}:\right)
\hat{a}^{j+m}
e^{
{\rm L}^{\prime }\hat{a}^{2}+
\underline{\rm l}\hat{a} 
}
\rho(0)
e^{\underline{\rm l}^{*}\hat{a}^{\dg}+
{\rm L}^{\prime *}\hat{a}^{\dg 2}
}
\hat{a}^{\dg j+k}\nonumber\\
&&\times\left( :e^{
{{\rm D}}^{\prime *}\hat{a}^{\dg}\hat{a}
}:\right)
e^{
{\rm R}^{\prime *}\hat{a}^{2}+
\underline{\rm r}^{*}\hat{a}
}
\hat{a}^{n+m},\label{powerexpand}
\eqa
which is also fully normally ordered.  The lack of explicit Hermiticity is superficial; this can seen by writing $\bar{\rho}(t)$ with the dummy indices $k\leftrightarrow m$ reversed and then halving the sum of both expressions.

As a reminder, the feasibility of our method of solution method depends upon being able to find the disentanglement and reordering parameters, $\{{\bf R},{\bf L},{\bf D},\breve{{\bf R}},\breve{{\bf L}},\breve{{\bf D}},{\bf r},{\bf l}\}$.  To do so, $2N\times 2N$ matrices must be characterized and manipulated, as explained further in \arf{LieAlgebraSec} which, for our systems, turns out to involve solving polynomials of degree $2N$.  This becomes insurmountable, in general, beyond $N=2$.  Later, in \srf{singleModeEgs}, we will solve some example single-mode systems.

The SME solution, found in either \erf{fullevoopeqn}, \erf{superOpEvoFull} or \erf{powerexpand} (as well as the differently operator ordered forms) represents an important result of our paper.  We now compliment it by finding the probability density associated with the measurement sequence upon which the system is conditioned.  That is, we find the POVM.

\section{Finding the POVM}

\label{POVMsec}

Given a normalized initial state, $\ket{\rho (0)}$ (that can be non-Gaussian), we have shown how to find $\ket{\bar{\rho} (t)}$. 
It has been seen that there is a deterministic factor, $e^{\hat{Q}t}$, and, in addition, terms involving the stochastic integrals $\{h,{\bm l}^{\prime}, {\bm r}^{\prime}\}$.  Given recorded measurement currents, the experimentalist can therefore follow the system state, which, when normalized, will be independent of $h$.  However, it is of interest to perform calculations as to the {\it expected} characteristics of the system evolution; for this, one requires the probability density of the state at time $t$, given by $\wp(h,{\bm l}^{\prime}, {\bm r}^{\prime})$.  For an arbitrary normalized initial state, a POVM, $\{\hat{W}_{h,{\bm l}^{\prime}, {\bm r}^{\prime}}:h,{\bm l}^{\prime}, {\bm r}^{\prime}\}$, achieves this via
\bqa
\wp(h, {\bm l}^{\prime}, {\bm r}^{\prime}|\rho(0))&=&\Tr\left[\hat{W}_{h,{\bm l}^{\prime}, {\bm r}^{\prime}}\rho(0)\right]\label{trW}\\
&=&\bra{{\bm 0}}e^{\hat{Q}t}e^{\hat{L}^{\prime}(t)}\ket{\rho(0)}\nonumber\\
&&\times\wp_{{\rm ost}}(h,{\bm l}^{\prime}, {\bm r}^{\prime}),\label{evoOst}\\
&=&
\bra{{\bm 0}}e^{\hat{Q}t}e^{\hat{{\bm b}}^{\dagger} {\bm r}^{\prime} }e^{{\bm l}^{\prime} \hat{{\bm b}}}
\ket{\rho(0)}e^{h}\nonumber\\
&&\times \wp_{{\rm ost}}(h, {\bm l}^{\prime}, {\bm r}^{\prime}),\label{evoOst}
\eqa
where \erf{expect} has been used, with $A(\hat{{\bm a}},\hat{{\bm a}}\dg)=\hat{1}$, and also \erf{normOst} and \erf{linPiece}.  For clarity, we remind the reader that the thermo-entangled state vacuum, $\ket{{\bm 0}}$, is the zero-valued eigenvector of $\hat{{\bm \beta}}$ (see \erf{betaOpMixMode}), which differs from the multimode ground state, $\ket{\mathbf{VAC}}$.  It is important to note that the POVM, $\{\hat{W}_{h,{\bm l}^{\prime}, {\bm r}^{\prime}}:h,{\bm l}^{\prime}, {\bm r}^{\prime}\}$, represents the {\it compiled measurement} up until a time $t$.  The reason for working with the thermo-entangled state representation of the SME solution when finding the POVM is that it allows us to use the powerful Lie algebra methods described in \srf{lieAlg}.  In contrast, \erf{powerexpand} involves non-exponentiated operators that are not elements of the Lie group $G$.

The reader will note that in writing \erf{evoOst} we have moved from considering the entire measurement record, ${\bm Y}(t)$, in \erf{normOst}, to only the relevant stochastic integrals, $\{h,{\bm l}^{\prime}, {\bm r}^{\prime}\}$, in \erf{trW}.  That is, the information obtained relating to the system at the initial time, $t=0$, can be fully summarized in terms of a finite number of integrals over the continuous measurement record. Commensurate with this observation, a change of variables has been performed from ${\bm Y}(t)$ to $\{h,{\bm l}^{\prime}, {\bm r}^{\prime}\}$.
However, despite $\wp_{{\rm ost}}({\bm Y}(t))$ having a known analytic form, being that of Gaussian white noise~\cite{WisMil10} (see \arf{ostensible}),
the calculation of $\wp_{{\rm ost}}(h, {\bm l}^{\prime}, {\bm r}^{\prime})$ is made difficult due to $h$ being a non-Gaussian random variable.  This is addressed in the next subsection, where it will be shown that the POVM can be made independent of $h$, without loss of predictive (or retrodictive) power with regards to measurement outcomes. Detailed comments relating to an alternative method of calculating the POVM independent of $h$, via the adjoint equation~\cite{PhysRevLett.111.160401}, will also be given later.

For now, we proceed with the direct calculation of the POVM, $\{\hat{W}_{h,{\bm l}^{\prime}, {\bm r}^{\prime}}:h,{\bm l}^{\prime}, {\bm r}^{\prime}\}$, in the thermo-entangled state representation (which has been used elsewhere with regards to retrodiction of the quantum state, see \cite{ban2007quantum}).  To obtain the POVM from \erfs{trW}{evoOst}, the operator $e^{\hat{Q}t}e^{\hat{L}^{\prime}(t)}$ needs to be converted into one that only contains physical mode operators.  This can achieved by acting the tilde mode operators backwards onto $\bra{{\bm 0}}$, via \erf{thermoAnn}: $\bra{{\bm 0}}\tilde{{\bm a}}=\bra{{\bm 0}}\hat{{\bm a}}^{\ddagger}$ and $\bra{{\bm 0}}\tilde{{\bm a}}^{\ddagger}=\bra{{\bm 0}}\hat{{\bm a}}$. However, it is a nontrivial task to move all the tilde operators into direct contact with $\bra{{\bm 0}}$, due to the non-commutativity of the terms containing tilde mode operators.
Additionally, once all the tilde mode operators are converted we will re-order the POVM towards normal order.  These two processes can lead to apparently very complex expressions for the disentangling parameters that can be difficult to simplify.



To proceed with greater simplicity and elegance to the POVM, we use the following calculational trick.  Instead of disentangling $e^{\hat{Q}t}$ as per \erf{disG}, we note that $\bra{{\bm 0}}e^{\hat{Q}t}$ is required for the POVM.   As per \erf{hilbVacThermoVac}, the thermo-vacuum can be expanded in terms of the standard coherent vacuum (for which $\hat{{\bm a}}\ket{ \mathbf{VAC}}=\tilde{{\bm a}}^{{\rm T}}\ket{ \mathbf{VAC}}={\bm {\bm 0}}$) giving~\cite{FANthermEta}
\begin{equation}
\bra{{\bm 0}}e^{\hat{Q}t}=\bra{ \mathbf{VAC}}e^{\tilde{{\bm a}}^{{\rm T}}\hat{{\bm a}}}e^{\hat{Q}t}.
\label{coh2}
\end{equation}

It is then clear that the disentanglement required is of $e^{\tilde{{\bm a}}^{{\rm T}}\hat{{\bm a}}}e^{\hat{Q}t}$, which we choose to be ordered as
\bqa
e^{\tilde{{\bm a}}^{{\rm T}}\hat{{\bm a}}}e^{\hat{Q}t}&=&
\exp\left[\hat{{\bm b}}^{\dag}{\bm R}^{\prime\prime}\hat{{\bm b}}^{\ddagger}\right]
\exp\left[\hat{{\bm b}}^{\dag}{\bm D}^{\prime\prime}\hat{{\bm b}}
+\delta^{\prime\prime}\right]\nonumber\\
&&\times
\exp\left[\tilde{{\bm a}}^{{\rm T}}\hat{{\bm a}}\right]
\exp\left[\hat{{\bm b}}^{{\rm T}}{\bm L}^{\prime\prime}\hat{{\bm b}}\right],
\label{povmDis}
\eqa
Double primes indicate that a different form (from the single prime matrices of \erf{disG}) is expected due to the inclusion and embedding of $e^{\tilde{{\bm a}}^{{\rm T}}\hat{{\bm a}}}$ (appearing as the second last exponential).  As a reminder, the vector $\hat{{\bm b}}$ contains both tilde and non-tilde operators, whilst $\hat{{\bm a}}$ and $\tilde{{\bm a}}$ are individually `unmixed'.  The expression, on the RHS of \erf{povmDis}, achieves a simultaneous disentangling and re-ordering of the product of the two exponentials on the LHS of \erf{povmDis}.  The explicit form of it is once again found by using the finite dimensional representation of $\mathfrak{sp}(4N+2)$ (see~\arf{reorderAppLin4}).  The advantage of this particular disentanglement is that most of the terms annihilate against the multimode coherent vacuum that appears in \eqref{coh2}.  This simplifies the disentanglement procedure, as we only need to solve for one of the parameters (${\bm L}^{\prime\prime}$).  It also removes the need for the tedious and complexifying re-ordering of exponentials when finding the POVM.   Up to a constant factor we are thus left with
\begin{equation}
\bra{\mathbf{VAC}}
\exp\left[\tilde{{\bm a}}^{{\rm T}}\hat{{\bm a}} \right]
\exp\left[\hat{{\bm b}}^{{\rm T}}{\bm L}^{\prime\prime}\hat{{\bm b}}\right]
=\bra{{\bm 0}}
\exp\left[\hat{{\bm b}}^{{\rm T}}{\bm L}^{\prime\prime}\hat{{\bm b}}\right],
\end{equation}
where the thermo-entangled vacuum has been reconstituted on the RHS.  

In order to facilitate the POVM being written in terms of physical mode operators only, ${\bm L}^{\prime\prime}$ is now written in block form, separating the physical and unphysical modes, as per the convention of \erf{LBlockeqn}.
This allows us to write 
\beq
\hat{{\bm b}}^{{\rm T}}{\bm L}^{\prime\prime}\hat{{\bm b}}=
\hat{{\bm a}}^{{\rm T}}{\mathbf L}^{\prime\prime}\hat{{\bm a}}+
2\tilde{{\bm a}}^{{\rm T}} \breve{{\bf L}}^{\prime\prime}\hat{{\bm a}}+
\tilde{{\bm a}}^{{\rm T}}{\mathbf L}^{\prime\prime *}\tilde{{\bm a}},
\label{tildConjPres}
\eeq
so that we can act the tilde exponentials onto $\bra{{\bm 0}}$ and convert them to physical mode operators. 
The ordering of the (commuting) exponential terms \blk is chosen so that the disentanglement parameter matrices $\{{\bf L}^{\prime\prime},\breve{{\bf L}}^{\prime\prime}\}$ are altered as little as possible when then conversion takes place:
\bqa
&&\bra{{\bm 0}}
e^{2\tilde{{\bm a}}^{{\rm T}} \breve{{\bf L}}^{\prime\prime}\hat{{\bm a}}}
e^{\hat{{\bm a}}^{{\rm T}}{\mathbf L}^{\prime\prime}\hat{{\bm a}}}
e^{\tilde{{\bm a}}^{{\rm T}}{\mathbf L}^{\prime\prime *}\tilde{{\bm a}}}\nonumber\\
&&=\bra{{\bm 0}}
e^{\hat{{\bm a}}^{\dag}{\mathbf L}^{\prime\prime*}\hat{{\bm a}}^{\ddagger}}
\left(
:e^{2\hat{{\bm a}}^{\dag}\breve{{\bf L}}^{\prime\prime}\hat{{\bm a}}}:
\right)
e^{\hat{{\bm a}}^{{\rm T}}{\mathbf L}^{\prime\prime}\hat{{\bm a}}}\\
&&=
\bra{{\bm 0}}
e^{\hat{{\bm a}}^{\dag}{\mathbf L}^{\prime\prime *}\hat{{\bm a}}^{\ddagger}}
e^{\hat{{\bm a}}^{\dag}\ln \left(\eye_{N}+2\breve{{\bf L}}^{\prime\prime}\right)\hat{{\bm a}} }
e^{\hat{{\bm a}}^{{\rm T}}{\mathbf L}^{\prime\prime}\hat{{\bm a}}},
\eqa
with the last line obtained by application of the operator identity, \erf{opId}.

The final task is to consider the linear component, $e^{\hat{L}^{\prime}(t)}$, of the POVM, that is detailed in \erf{linPiece}.  First, $\{{\bm r}^{\prime},{\bm l}^{\prime}\}$ are broken into their non-tilde and tilde components, according to ${\bm l}^{\prime}=\left({\mathbf  l}^{\prime},{\mathbf  l}^{*\prime}\right)$ and ${\bm r}^{\prime}=\left({\mathbf  r}^{\prime};{\mathbf  r}^{*\prime}\right)$, which leads to
\bqa
\hat{{\bm b}}^{\dag}{\bm r}^{\prime}&=&\hat{{\bm a}}^{\dag}{\mathbf  r}^{\prime}
+\tilde{{\bm a}}^{\ddagger}{\mathbf  r}^{\prime *}\label{r1}\\
{\bm l}^{\prime}\hat{{\bm b}}&=&{\mathbf  l}^{\prime }\hat{{\bm a}}
+{\mathbf  l}^{\prime *}\tilde{{\bm a}}.\label{l1}
\eqa
Then the tilde terms must be brought into contact with $\bra{{\bm 0}}$ (or $\ket{{\bm 0}}$, as they could be moved through the physical mode density operator) and converted.  
Finally, we perform a normal re-ordering of the linear pieces using the finite dimensional matrix representation.

Finally, we arrive at the Hermitian POVM element:
\bqa
\hat{W}_{h,{\mathbf l}^{\prime}, {\mathbf r}^{\prime}}&=&e^{\Delta}
e^{\hat{{\bm a}}^{\dag}{\bf d}+\hat{{\bm a}}^{\dag}{\mathbf L}^{\prime\prime\dg}\hat{{\bm a}}^{\ddagger}}
e^{\hat{{\bm a}}^{\dag}\ln \left(\eye_{N}+2\breve{{\bf L}}^{\prime\prime}\right)\hat{{\bm a}} }
e^{\hat{{\bm a}}^{{\rm T}}{\mathbf L}^{\prime\prime}\hat{{\bm a}}+{\bf d}^{\dag}\hat{{\bm a}}}
e^{h}\nonumber\\
&&\times\wp_{{\rm ost}}(h,{\mathbf l}^{\prime}, {\mathbf r}^{\prime}),
\label{pom}
\eqa
where $e^{\Delta}$ collects all the constant terms (non-stochastic, non-operator) that have been picked up along the path of our derivation.  We have also introduced the stochastic vector, ${\bf d}$, defined by
\beq
{\bf d}={\mathbf l}^{\prime \dagger}+
2{\mathbf L}^{\prime\prime\dag}{\mathbf r}^{\prime *}+
\left(\eye_{N}+2\breve{{\bf L}}{}^{\prime\prime}\right){\mathbf r}^{\prime}.
\label{R}
\eeq 
The form of ${\bf d}$ is found using the finite dimensional matrix representation, which is explained further in \arf{reorderAppLin4}.
Examples of explicit POVMs will be given in~\srf{singleModeEgs}.

There are two significant issues with the expression for the POVM in \erf{pom}.  Firstly, the joint ostensible distribution for $\wp_{{\rm ost}}(h, {\mathbf l}^{\prime}, {\mathbf r}^{\prime})$ will be difficult to determine analytically due to the non-Gaussian nature of $h$.  Secondly, all calculations should actually be independent of $h$ as it does not affect the system state.  As an example, the system state is {\it retrodicted} via the application of Bayes' rule,
\beq
\wp(\rho(0)|h,{\mathbf l}^{\prime}, {\mathbf r}^{\prime})=
\frac{\wp(h,{\mathbf l}^{\prime}, {\mathbf r}^{\prime}|\rho(0))\wp(\rho(0))}{\wp(h,{\mathbf l}^{\prime}, {\mathbf r}^{\prime})},
\label{bayes}
\eeq
by which the scalar factor $e^{h}$ cancels from numerator and denominator, so that the RHS in independent of $h$.  In fact, we see that the dependence of the RHS upon $\{{\mathbf l}^{\prime}, {\mathbf r}^{\prime} \}$ is contained in the single stochastic integral, ${\bf d}$.  In \cite{jacLin}, the authors suggest resorting to numeric calculation of the ostensible statistics of $h$.  Rather than take that approach, we now try to analytically determine the effect operator averaged over $h$, which represents a minimalistic POVM that, nonetheless, contains all relevant statistics.

\subsection{A simplified POVM, \texorpdfstring{$\hat{W}_{{\bf d}}$}{TEXT}}
\label{simplePOVM}
The POVM with $h$ absent is straightforward to formally write:
\bqa
\hat{W}_{{\bf d}}&=&\int\hat{W}_{h,{\bf d}}d^{2}h\\
&=&e^{\Delta}
e^{\hat{{\bm a}}^{\dag}{\bf d}+\hat{{\bm a}}^{\dag}{\mathbf L}^{\prime\prime *}\hat{{\bm a}}^{\ddagger}}
e^{\hat{{\bm a}}^{\dag}\ln \left(\eye_{N}+2\breve{{\bf L}}^{\prime\prime}\right)\hat{{\bm a}} }
e^{\hat{{\bm a}}^{{\rm T}}{\mathbf L}^{\prime\prime}\hat{{\bm a}}+{\bf d}^{\dag}\hat{{\bm a}}}
\nonumber\\
&&\times \int
e^{h}\wp_{{\rm ost}}(h, {\bf d})d^{2}h,
\label{pNoA}
\eqa
however, the non-Gaussianity of $h$ makes this difficult to evaluate.  (Note that in writing \erf{pNoA} in terms of ${\bf d}$ rather than $\{{\mathbf l}^{\prime}, {\mathbf r}^{\prime} \}$ we have performed another change of variables.)
It is also far from obvious that the above POVM, when viewed only in a mathematical sense, is capable of giving Gaussian statistics for ${\bf d}$.  Despite this, given an initial Gaussian state, it would be highly surprising if $\wp({\bf d}|\rho(0))$ was not Gaussian, as the SME maps Gaussian states to Gaussian states.  To explain further, ${\bf d}$ is a linear integral of the measurement currents, as is clear from \erf{R}.  In turn, the statistics of the measurement currents are given by \erfs{actExpY}{actExpY2}, which have white noise added to linear functions of the first order moments.  It is a well-known property of the Kalman filter that these currents will have Gaussian statistics for initial Gaussian states.  Given conviction from physical arguments, in \arf{intAOut} we show mathematically how $\int
e^{h}\wp_{{\rm ost}}(h, {\bf d})d^{2}h$ does, in fact, provide Gaussian statistics.
The essence of the argument is that when the integrals that define $h$ are discretized, it can be seen that $h$ is a linear combination of chi-squared random variables.  The integral in \erf{pNoA} then reduces to Gaussian integrals, which of course provide a Gaussian outcome.  In what follows, an explicit expression for the POVM will be found, without having to calculate the integral over $h$ directly. \blk

In order to determine $\hat{W}_{{\bf d}}$, we will find its $Q$-function, $\bra{{\bm \alpha}}\hat{W}_{{\bf d}}\ket{{\bm \alpha}}=\wp( {\bf d}|{\bm \alpha})$.  Here, $\ket{{\bm \alpha}}$ is the $N$-mode coherent state of amplitude ${\bm \alpha}$ (not to be confused with  a $2N$-mode thermo-entangled state).   As the $Q$-function of an Hermitian operator is unique, finding it will be sufficient to specify $\hat{W}_{{\bf d}}$.  From \erf{pNoA}, the ${\bm \alpha}$-dependent factors of $\bra{{\bm \alpha}}\hat{W}_{{\bf d}}\ket{{\bm \alpha}}$ are simple to to find.  However, there is ${\bf d}$-dependence arising from the integral over $h$, and this prevents us immediately inferring the precise Gaussian form of $\bra{{\bm \alpha}}\hat{W}_{{\bf d}}\ket{{\bm \alpha}}$.  Instead, we will use the fact that the ${\bm \alpha}$-dependent factors are known in order to first find $\wp({\bm \alpha}|{\bf d} )$.  In turn, this will fix $\wp( {\bf d}|{\bm \alpha})$ via an application of Bayes' theorem,
\beq
\wp( {\bf d}|{\bm \alpha})=\frac{\wp({\bm \alpha}|{\bf d} )\wp({\bf d} )}{\wp({\bm \alpha} )}.
\label{bayesNoh}
\eeq
To be clear, in this section we are not ultimately interested in performing retrodiction.  The utilization of Bayes' theorem is as a mathematical tool to step from a quantity that we can more easily to determine, to the quantity that is desired.

To infer $\wp({\bm \alpha}| {\bf d})$, a normalized Gaussian distribution for ${\bm \alpha}$, all that is needed is the mean, $\langle{\bm \alpha}\rangle$, covariance, ${\bm \Gamma}$, and pseudo-covariance, ${\bm \Upsilon}$.  These can be determined by equating the ${\bm \alpha}$-dependent pieces of $\bra{{\bm \alpha}}\hat{W}_{{\bf d}}\ket{{\bm \alpha}}$ with the general form of a multidimensional complex normal distribution~\cite{539051}
\bqa
\wp({\bm \alpha}|{\bf d})&=&
{\cal N}_{1}\exp\left[-({\bm \alpha}-\langle{\bm \alpha}\rangle_{{\bf d}})^{\dag} ({\bm P}^{-1})^{*}
({\bm \alpha}-\langle{\bm \alpha}\rangle_{{\bf d}})\right.\nonumber\\
&&+\left.
{\rm Re}\left(({\bm \alpha}-\langle{\bm \alpha}\rangle_{{\bf d}})^{{\rm T}} {\bm Q}^{{\rm T}} ({\bm P}^{-1})^{*}
({\bm \alpha}-\langle{\bm \alpha}\rangle_{{\bf d}}) \right)
   \right],\nonumber\\
   \label{genPDF}
\eqa
with ${\bm Q}={\bm \Upsilon}{\bm \Gamma}^{-1}$, ${\bm P}={\bm \Gamma}^{*}-{\bm Q}{\bm \Upsilon}$ and $\langle{\bm \alpha}\rangle_{{\bf d}}$ being a function of ${\bf d}$.  ${\cal N}_{1}$ is a normalization that depends only on ${\bm Q},{\bm \Upsilon}$.

The ${\bm \alpha}$-dependent pieces of $\bra{{\bm \alpha}}\hat{W}_{{\bf d}}\ket{{\bm \alpha}}$ are easily determined, given that a normally ordered form is given in \erf{pNoA}.
Comparing \erf{genPDF} with $Q$-function of the POVM in \erf{pNoA}, we find the relations for the distribution parameters
\bqa
 \smallhalf({\bm P}^{-1})^{*}&=& \breve{{\bf L}}^{\prime\prime}\label{a}\\
\smallhalf{\bm Q}^{{\rm T}} ({\bm P}^{-1})^{*}&=&{\mathbf L}^{\prime\prime}\label{b}\\
  2 \breve{{\bf L}}^{\prime\prime}\langle{\bm \alpha}\rangle_{{\bf d}}-\left({\mathbf L}^{\prime\prime\dag}+{\mathbf L}^{\prime\prime}\right)
    \langle{\bm \alpha}\rangle_{{\bf d}}^{*} &=&{\bf d}
    \label{alphaAve}
\eqa
and 
\beq
   {\cal N}_{1}=
    (2/\pi)^{N}\sqrt{
    {\rm det}\left(\breve{{\bf L}}^{\prime\prime}- \breve{{\bf L}}^{\prime\prime -1}{\mathbf L}^{\prime\prime\dag}{\mathbf L}^{\prime\prime}\right)
{\rm det}\left( \breve{{\bf L}}^{\prime\prime}\right)},\label{d}
\eeq
where \erfs{alphaAve}{d} have been simplified with the use \erfs{a}{b}.
As per the comments below \erf{expect}, the feasibility of obtaining explicit analytic expressions depends on the ability of characterizing matrices; in this case it is necessary to invert matrices to find $\langle{\bm \alpha}\rangle_{{\bf d}}$ from \erf{alphaAve}.  If $\langle{\bm \alpha}\rangle_{{\bf d}}$ can be found, an analytic expression for $\wp({\bm \alpha}|{\bf d})$ results.  For a single mode, the matrices $\{{\bf L}^{\prime\prime},\breve{{\bf L}}^{\prime\prime}\}$ are, of course, just scalars.

Having found $\wp({\bm \alpha}|{\bf d})$, we now consider the two remaining factors in \erf{bayesNoh} that are required to fix $\wp( {\bf d}|{\bm \alpha})$: $\wp({\bm \alpha})$ and $\wp({\bf d})$.  

For simplicity, we assume no knowledge of ${\bm \alpha}$ exists before measurement begins; the prior distribution for ${\bm \alpha}$ is flat and can be represented by a (multi-dimensional) Gaussian of infinite variance.  In \erf{bayesNoh}, $\wp({\bm \alpha})$ can consequently be treated as a constant factor independent of ${\bm \alpha}$.

The expression for $\wp({\bf d})$ (the final factor on the RHS of \erf{bayesNoh} yet to be determined) is given by
\beq
\wp( {\bf d})=\int \wp({\bf d}|{\bm \alpha})
\wp({\bm \alpha})d^{2}{\bm \alpha},
\eeq
where both $\wp({\bf d}|{\bm \alpha})$ and $\wp({\bm \alpha})$ are Gaussian distributions, as we have argued.  As such, the integral over the $N$-dimensional complex plane can be carried out analytically.  The only aspect of it that we need is that the resulting Gaussian, for $\wp({\bf d})$, will have an infinite variance.  This follows because if the mean of ${\bf d}$ is linearly dependent upon ${\bm \alpha}$, and ${\bm \alpha}$ has a flat distribution, then ${\bf d}$ itself will also have a flat distribution.  This linear relationship is inferred from \erf{alphaAve}.  Consequently, $\wp({\bf d})$ can be treated as a constant factor independent of ${\bf d}$.

Using \erf{bayesNoh}, we can draw together our knowledge of $\wp({\bm \alpha}|{\bf d})$, $\wp({\bm \alpha})$ and $\wp({\bf d})$ to write
\beq
\wp( {\bf d}|{\bm \alpha})={\cal N}_{2}\wp({\bm \alpha}|{\bf d} ),
\eeq
with $\wp({\bm \alpha}|{\bf d} )$ given by \erf{genPDF} and $\langle{\bm \alpha}\rangle_{{\bf d}}$ expressed as a function of ${\bf d}$ through \erf{alphaAve}. ${\cal N}_{2}$ is a new normalization constant independent of $\{{\bm \alpha},{\bf d}\}$.

Having determined the $Q$-function, $\bra{{\bm \alpha}}\hat{W}_{{\bf d}}\ket{{\bm \alpha}}$, the operator form can be inferred by inspection of \erf{genPDF}.  To explain further, the operator dependence is determined from the ${\bm \alpha}$ terms, while the terms without powers of ${\bm \alpha}$ provide the scalar factors.  We deduce that
\bqa
\hat{W}_{{\bf d}}&=&{\cal N}
\exp\left[ 
-
2\langle{\bm \alpha}\rangle_{{\bf d}}^{\dag}\breve{{\bf L}}^{\prime\prime}\langle{\bm \alpha}\rangle_{{\bf d}}
+\langle{\bm \alpha}\rangle_{{\bf d}}^{{\rm T}}{\mathbf L}^{\prime\prime}\langle{\bm \alpha}\rangle_{{\bf d}}\right.\nonumber\\
&&\left.\quad\quad\quad+
\langle{\bm \alpha}\rangle_{{\bf d}}^{\dag}{\mathbf L}^{\prime\prime\dag}\langle{\bm \alpha}\rangle_{{\bf d}}^{*}
\right]\exp\left[
\hat{{\bm a}}^{\dag}{\bf d}+
\hat{{\bm a}}^{\dag}{\mathbf L}^{\prime\prime}\hat{{\bm a}}^{\ddagger}
\right]\times
\nonumber\\
&&
\left(
 :\exp\left[
2\hat{{\bm a}}^{\dag}\breve{{\bf L}}^{\prime\prime}\hat{{\bm a}}
\right]:
\right)
\exp\left[
\hat{{\bm a}}^{{\rm T}}{\mathbf L}^{\prime\prime\dag}\hat{{\bm a}}
+{\bf d}^{\dag}\hat{{\bm a}}
\right],\label{finalPOVM}
\eqa 
with ${\cal N}$ a function of $\{{\mathbf L}^{\prime\prime},\breve{{\bf L}}^{\prime\prime} \}$ that is fixed by normalization of $\wp( {\bf d}|{\bm \alpha})$ with respect to ${\bf d}$.
The normal ordering could be removed via \erf{opId}.  Note the similarity of \erf{finalPOVM} to \erf{pom}; the operator factors have remained the same.  However, more than just the normalization has been found (from the perspective of ${\bf d}$) as there is quadratic dependence upon ${\bf d}$ contained in the exponent of the first exponential.  This piece originates from the integral over the non-Gaussian variable, $\int
e^{h}\wp_{{\rm ost}}(h, {\mathbf l}^{\prime}, {\mathbf r}^{\prime})d^{2}h$, that we have avoided evaluating directly.  If only the operator dependence of $\hat{W}_{{\bf d}}$ had been found, then the correct statistics of ${\bf d}$ would remain unknown. 

An interesting aspect of \erf{finalPOVM} is that it implies that the probability of obtaining a particular measurement record depends only on the one stochastic integral ${\bf d}$.  In contrast, the values of $\{{\mathbf l}^{\prime}, {\mathbf r}^{\prime}\}$, must both be known in order to determine the system state.  
When examples are provided, we will see that there do exist cases for which one of the two stochastic integrals, $\{{\mathbf l}^{\prime}, {\mathbf r}^{\prime}\}$, is strictly zero.  Thus, there is a natural classification of systems subjected to dyne measurement: whether or not ${\bf d}$ and, consequently, the POVM, is sufficient to determine the evolution operator.

We conclude this subsection by noting that the POVM is of a Gaussian form, yielding Gaussian statistics for ${{\bf d}}$.  Given that a Gaussian operator can be characterized by a vector of means and a covariance matrix, it is natural to wonder whether the result could have been obtained by a more direct approach.  In the next subsection, we describe an existing method to obtain the POVM that proceeds in such a manner.

\subsection{Finding the POVM via the adjoint equation} 
\label{adjointSec}

In solving the linear SME, we worked in the {\sch} picture, whereby the system state evolves in time.  An equivalent pathway is to consider the evolution of an arbitrary operator as defined by the adjoint to the stochastic master equation.  To explain this approach, let us consider the evolution superoperator that evolves the state forward by $dt$
\beq
{\cal V}_{{\bm y}(t)}:\bar{\rho}(t+dt)={\cal V}_{{\bm y}(t)}\bar{\rho}(t),
\eeq
with the mapping, ${\cal V}_{{\bm y}(t)}$, determined by the linear SME.  As we are interested in obtaining the POVM, we choose to illustrate the adjoint master equation for the effect operator.  It is defined by the adjoint mapping, ${\cal V}_{{\bm y}(t)}^{\dg}$ such that~\cite{infDim},
\beq
\Tr [\hat{W}{\cal V}_{{\bm y}(t)}[\bar{\rho}(t) ]]=\Tr [{\cal V}_{{\bm y}(t)}^{\dg}[\hat{W} ]\bar{\rho}(t)].
\label{traceAd}
\eeq
That is, the adjoint equation to the linear SME evolves the effect operator (or any observable) and represents a quasi-HP.  It is not a true HP as the resultant operator equations of motion will not preserve the operator algebra of the system~\cite{WisMil10}.  So, for example, in general ${\cal V}_{{\bm y}(t)}^{\dg}[\hat{W}^{2} ]\neq ({\cal V}_{{\bm y}(t)}^{\dg}[\hat{W} ])^{2}$.  To calculate functions of an operator, care must be taken to re-integrate the quasi-HP equation for the new operator.  
Nevertheless, when this quasi-HP equation is integrated we still obtain correct expectation values of that particular operator (by definition, from \erf{traceAd}).  To calculate functions of an operator, care must be taken to re-integrate the quasi-HP equation for the new operator.  

The two pictures ({\sch} and quasi-Heisenberg) represent two different ways to proceed; in particular, we could have solved for the time evolution of the effect operator in the quasi-HP in order to obtain the POVM.  
In the literature, the quasi-HP approach to the POVM has been partially explored using methods different to those of this paper~\cite{zhang2017prediction,huang2018smoothing,lammers2018state,laverick2019quantum} and we now compare and contrast it with our current work.

\subsubsection{Adjoint equation for the effect operator}

A feature of the adjoint equation is that it leads to backwards-in-time evolution of operators --- this can be seen by considering two successive updates in \erf{traceAd} and then using the cyclic properties of the trace operation to obtain the adjoint mapping.  Using the backwards increment notation 
\beq
d\hat{W}(t)=\hat{W}(t-dt)-\hat{W}(t),
\eeq
with $dt$ positive, it is straightforward to derive an adjoint equation to the linear SME for the effect operator~\cite{PhysRevLett.111.160401}
\bqa
d\hat{W}(t)&=& i\left[\hat{H},\hat{W}(t) \right]dt+{\cal D}^{\dag}\left[\hat{{\bm c}} \right]\hat{W}(t)  dt\nonumber\\
&&+{\bm y}^{{\rm T}}(t-dt)dt{\bar {\cal H}}\left[{\bm M}^{{\rm T}}\hat{{\bm c}}^{\ddagger} \right]
\hat{W}(t).
\label{adjEqn}
\eqa
This generalizes the adjoint equation contained in \cite{PhysRevLett.111.160401} by allowing for completely general measurement parameterizations, ${\bm M}$.  To obtain the POVM element applicable to the entire measurement record, the effect operator needs to be integrated from the final (absolute) measurement time backwards to the time at which measurement is physically turned on, $t_{0}$.   (To clarify, in our work we take the `final condition' for a backwards-in-time differential equation as referencing the starting point for the integration and, consequently, the latest {\it absolute} time).  The superoperator ${\cal D}^{\dag}$ is defined by
\beq
{\cal D}^{\dag}\left[\hat{{\bm c}} \right]\equiv \sum_{k=1}^{L}{\cal D}^{\dag}\left[\hat{c}_{k} \right],
\quad
{\cal D}^{\dag}\left[\hat{c} \right]\rho\equiv \hat{c}^{\dag}\rho\hat{c}-\smallhalf\hat{c}^{\dag}\hat{c}\rho
-\smallhalf \rho\hat{c}^{\dag}\hat{c}.
\eeq

Note that the adjoint equation for $\hat{W}$, \erf{adjEqn}, is a linear equation that preserves Hermiticity. That is, it is of the same general form as the linear SME, \erf{smeGen}.  Consequently, the methods we have described in our paper for solving the linear SME could be applied to solve the adjoint equation.  Specifically, by using a thermo-entangled state representation and then applying techniques from Lie algebra.  However, other authors~\cite{zhang2017prediction,huang2018smoothing,lammers2018state,laverick2019quantum} have taken a different approach to solving the adjoint equation that is designed to avoid much of the complication that we have considered.  Namely, they choose to apply a phase space representation of the effect operator, which is characterized by its first and second order moments.  For these moments to be calculated correctly, a normalized effect operator is required.  As such, \erf{adjEqn} can be put in an explicitly trace preserving form:
\bqa
d\hat{W}(t)&=& i\left[\hat{H},\hat{W}(t) \right]dt\nonumber\\
&&+\left(
{\cal D}^{\dag}\left[\hat{{\bm c}} \right]\hat{W}(t) 
+{\rm Tr}\left[\left(\hat{{\bm c}}^{\dag}\hat{{\bm c}}-\hat{{\bm c}}^{{\rm T}}\hat{{\bm c}}^{\ddagger}
\right)\hat{W} \right]\hat{W} 
\right) dt\nonumber\\
&&
+d{\bm w}^{{\rm T}}(t-dt)dt {\cal H}\left[{\bm M}^{{\rm T}}\hat{{\bm c}}^{\ddagger} \right]
\hat{W}(t),
\label{adjEqnNorm}
\eqa
where the nonlinear superoperator ${\cal H}$, from \erf{hSuper}, appears, as does the measurement noise, $d{\bm w}$.

Working with a normalized effect operator marks a significant departure from our methods: it disregards the scalar factor that depends on the measurement record.  We know this factor exists due to its presence in \erf{finalPOVM}.  In order to retrieve it, the procedure described in \srf{simplePOVM} could be undertaken. If it is not retrieved then the correct statistics are lost. This is not quite as dire as it sounds; because it is only a scalar factor, there is no problem created for retrodiction of the prior quantum state as the scalar norm of the effect operator is removed by renormalization (see the discussion below \erf{bayes}).

Tracking the effect operator by only its first and second order moments is possible if the effect operator is Gaussian.  The duality between \erf{adjEqnNorm} and the nonlinear SME, 
means that it {\it does} map Gaussian effect operators to Gaussian effect operators~\cite{PhysRevLett.111.160401}. Additionally, given that the adjoint equation is backwards in time, the appropriate final condition for $\hat{W}$ is the identity operator, $\hat{1}$, as no measurement data has yet been used in a mathematical sense.  
The identity operator can be viewed as a Gaussian operator of infinite variance, meaning that the evolution induced by \erf{adjEqnNorm} can be tracked simply by following the mean, ${\bm x}=\Tr\left[\hat{{\bm x}}\hat{W} \right]$, and variance, ${\bm V}_{ij}=\smallhalf\Tr\left[\left(\hat{{\bm x}}_{i}\hat{{\bm x}}_{j}+\hat{{\bm x}}_{j}\hat{{\bm x}}_{i}\right)\hat{W} \right]-
\Tr\left[\hat{{\bm x}}_{i}\hat{W} \right]\Tr\left[\hat{{\bm x}}_{j}\hat{W} \right]$, of $\hat{W}$.  

Using \erf{adjEqnNorm}, it is straightforward to find dynamical equations for ${\bm x}$ and ${\bm V}$:
\bqa
d{\bm x}&\equiv &{\bm x}(t-dt)-{\bm x}(t)\nonumber\\
&=&-{\bm A}{\bm x}dt+\left(2{\bm V}{\bm B}^{{\rm T}}+{\bm S}^{{\rm T}}\right)d{\bm w}\label{meanW}\\
\frac{d{\bm V}}{dt}&\equiv &\frac{{\bm V}(t-dt)-{\bm V}(t)}{dt}\nonumber\\
&=&
-{\bm A}{\bm V}-{\bm V}{\bm A}^{{\rm T}}+{\bm E}\nonumber\\
&&-
\left(2{\bm V}{\bm B}^{{\rm T}}+{\bm S}^{{\rm T}}  \right)
\left(2{\bm V}{\bm B}^{{\rm T}}+{\bm S}^{{\rm T}}  \right)^{{\rm T}},
\label{varW}
\eqa
for matrices ${\bm A}={\bm \Sigma}\left({\bm G}+{\rm Im}\left\{{\bm C}^{\dag}{\bm C} \right\}\right)$, 
${\bm B}={\rm Re}\left\{{\bm M}^{\dag}{\bm C} \right\}$, ${\bm S}={\rm Im}\left\{{\bm M}^{\dag}{\bm C} \right\}{\bm \Sigma}^{{\rm T}}$ and ${\bm E}={\bm \Sigma}{\rm Re}\left\{{\bm C}^{\dag}{\bm C} \right\}{\bm \Sigma}^{{\rm T}}$.  
\erfs{meanW}{varW} once again represent a generalization of \cite{PhysRevLett.111.160401} to account for arbitrary `dyne' unravelings. Specifically, a diagonal matrix of detector efficiencies has been replaced by the matrix ${\bm M}$, the allowable form of which is dictated by \erf{Mallowed}.  This generalization also appears in \cite{laverick2019quantum}, albeit using the U-representation~\cite{chia2011complete} of diffusive monitorings.  The U-representation is alternative, but equivalent, to the M-representation that is used in this paper (see discussion around \erf{Mallowed}).

Given that we wish to assign a value of infinity to the variance as a starting point for the backwards-in-time integration, it is worth noting that this will lead to infinite `kicks' to the mean.  In order to perform the backwards integration accurately~\cite{laverick2019quantum,huang2018smoothing} it is, therefore, necessary to introduce a new variable, ${\bm z}={\bm \Lambda}{\bm x}$ for ${\bm \Lambda}={\bm V}^{-1}$.  Both ${\bm z}$ and ${\bm \Lambda}$ have a starting value of zero~\cite{fraser1969optimum}.  The backwards-in-time equation for ${\bm z}$ is
\bqa
d{\bm z}&=&\left[{\bm A}+2{\bm S}^{{\rm T}}{\bm B} - \left({\bm E}-{\bm S}^{{\rm T}}{\bm S}\right){\bm \Lambda}\right]^{{\rm T}}{\bm z}dt\nonumber\\
&&+\left(2{\bm B}^{{\rm T}}+ {\bm \Lambda}{\bm S}^{{\rm T}}\right){\bm y}dt.
\label{meanw}
\eqa
For simplicity, it has been written in terms of ${\bm y}dt$ rather than $d{\bm w}$. 
In \srf{singleModeEgs}, \erf{meanw} will be used to solve the adjoint equation when investigating a single mode example.
\blk

\erfs{meanW}{varW} (or more conveniently \erf{meanw}) are Kalman filter equations and are amenable to analytic solution provided sufficiently small system dimension or other simplifying features.  
It is clear that the derivation here is much simpler than that which we used to obtain \erf{finalPOVM}.  Consequently, we now wish to discuss whether we have arrived at exactly the same object (that is, the POVM for the compiled measurement) and, if not, whether it is of the same utility.

Firstly, to reiterate, it lacks the scalar factor that depends on the measurement record in \erf{finalPOVM}. This could be retrieved, using the same procedure as \srf{simplePOVM}, to obtain the complete POVM.  Secondly, although the adjoint equation is dual to the linear SME, such that it contains all the information of the linear SME, some of this generality is lost when the final condition of the identity operator is used.  That is, even if the scalar stochastic factor dependent upon ${\bf d}$ is regained, it is not, in general, possible to infer the system state.  This is due to there not being a one-to-one correspondence between the POVM and the final system state.  In contrast to the identity being used as the final condition, the reader is reminded that when we solved the linear SME, nothing was assumed about the initial state.


In summary, finding the POVM via a phase space representation of the adjoint equation provides a simple way to obtain a great deal of information, 
but it does not complete the state-inclusive analysis of the compiled measurement. 

\section{A black box approach to solving the SME and finding the POVM}
\label{recipe}

In this section, we provide the reader with a minimalistic description for finding both the SME solution and the POVM, applicable to a compiled measurement up to a time $t$.  
In prior sections, the methodology was firstly sketched, and then carried out in detail, in order to provide a pedagogic pathway.  Here, our purpose is to summarize a `black box' type recipe for arriving at the end result. For example, it is desired that a SME solution could be obtained by the practitioner using straightforward algebraic methods, without knowledge (or at least very little knowledge) of the thermo-entangled state representation or Lie groups.

To define our recipe, we need to clearly state what the input and output from the black box will be, as there is some flexibility to this.  Firstly, let us consider the description of the system.  The system and dynamics are specified by the nonlinear SME in \erf{smeGenNonLin}.  To complete this description, the Hamiltonian, $H$, Lindblad operators, $\hat{{\bm c}}$ and measurement setting ${\bm M}$ must be provided. The matrix ${\bm G}$ of \erf{hamiltonianG} parameterizes the Hamiltonian, the matrix ${\bm C}$ of \erf{lindbladMatrix} parameterizes the Lindblad terms, while the matrix ${\bm M}$ itself defines the measurement setting. It is common for a SME to be given in terms of $\{\hat{{\bm a}},\hat{{\bm a}}^{\dg}\}$ instead of 
$\hat{{\bm x}}$.  That is, 
\beq
\hat{H}=\frac{1}{2}\left(\hat{a}_{1}^{\dg},\hat{a}_{1},...,\hat{a}_{N}^{\dg},\hat{a}_{N}\right){\bm F}\left(\hat{a}_{1},\hat{a}_{1}^{\dg},...,\hat{a}_{N},\hat{a}_{N}^{\dg}\right)^{{\rm T}},\label{hamiltonianF}
\eeq
for Hermitian ${\bm F}$, and 
\beq
\hat{{\bm c}}={\bm Z}\left(\hat{a}_{1},\hat{a}_{1}^{\dg},...,\hat{a}_{N},\hat{a}_{N}^{\dg}\right)^{{\rm T}}.
\eeq
The conversion between the two representations can be simply performed:
\bqa
{\bm G}&=&{\bm X}{\bm F}{\bm X}^{\dg}\label{convertG}\\
{\bm C}&=&{\bm Z}{\bm X}^{\dg}\label{convertA}
\eqa
for 
\beq
{\bm X}=\bigoplus^{N}_{n=1}\frac{1}{\sqrt{2}}\left[
\begin{array}{cc}
    1      &1 \\
   -i & i
\end{array}\right].\label{XMat}
\eeq
In summary, the first step in obtaining an SME solution and POVM is to write the matrices $\{{\bm G},{\bm C},{\bm M}\}$, using \erfs{convertG}{convertA} if necessary.

Having defined the system input, we now define the product of our recipe.  This is a solution to the linear SME, specified either (depending on preference) in the enlarged Hilbert space, ${\cal H}\otimes\tilde{{\cal H}}$, appropriate for the vectorized solution, 
\beq
\ket{\bar{\rho} (t)}=\hat{V}_{{\bm Y}(t)}\ket{\rho (0)},
\label{solSummRecipe}
\eeq
or the original Hilbert space, ${\cal H}$, appropriate for the density matrix solution
\beq
\bar{\rho} (t)={\cal V}_{{\bm Y}(t)}\rho (0),
\eeq
with ${\bm Y}(t)$ representing the set of all measurement results and $\bar{\rho} (t)$ being an unnormalized state. Obviously, the practitioner not interested in the thermo-entangled state representation would use the density matrix solution.
Both the evolution operator, $\hat{V}_{{\bm Y}(t)}$, defined in \erf{vectorizedEvoOp} and evolution superoperator, ${\cal V}_{{\bm Y}(t)}$, defined in \erf{superOpEvoFull}, are parameterized by the $N\times N$ matrices $\{{\bf R}^{\prime},\breve{{\bf R}}^{\prime},{\bf L}^{\prime},\breve{{\bf L}}^{\prime},\underline{\bf D},\underline{\breve{{\bf D}}}\}$ and length $N$ stochastic vectors $\{\underline{\bf r},\underline{\bf l}\}$.
Additionally, the POVM describing the composite measurement is found with our methods.  It was found, in \srf{POVMsec}, that the probability of obtainig a sequence of measurement results was only dependent upon the stochastic integral ${\bf d}$.  The effect operator for the measurement result ${\bf d}$, $\hat{W}_{{\bf d}}$, was defined in \erf{finalPOVM}.  It is parameterized by $\{{\bf L}^{\prime\prime},\breve{{\bf L}}^{\prime\prime},{\bf d},\langle{\bm \alpha}\rangle_{{\bf d}}\}$, where $\langle{\bm \alpha}\rangle_{{\bf d}}$ is defined implicitly, in \erf{alphaAve}, through the other parameters.

Consequently, our recipe for finding $\{\ket{\bar{\rho} (t)},\bar{\rho} (t),\hat{W}_{{\bf d}}\}$ must describe how to obtain the parameters 
\beq
\{{\bf R}^{\prime},\breve{{\bf R}}^{\prime},{\bf L}^{\prime},\breve{{\bf L}}^{\prime},\underline{\bf D},\underline{\breve{{\bf D}}},\underline{\bf r},\underline{\bf l},{\bf L}^{\prime\prime},\breve{{\bf L}}^{\prime\prime},{\bf d}\}\label{solParams}
\eeq
from the inputs $\{{\bm G},{\bm C},{\bm M}\}$ (or $\{{\bm F},{\bm Z},{\bm M}\}$).  Rather than give a series of extremely lengthy formulae for the solution parameters in terms of $\{{\bm G},{\bm C},{\bm M}\}$, we break the calculation into a few steps.

Given inputs in the appropriate form, $\{{\bm G},{\bm C},{\bm M}\}$, the first step is to calculate the $\{{\bm R},{\bm D},{\bm L},d{\bm l},d{\bm r}\}$ parameters.  Explicit formulae are given for these parameters in \arf{RDLdldrDescription}.

Next, the now known matrices $\{{\bm R},{\bm D},{\bm L}\}$ are used to calculate the finite dimensional matrix representation of $e^{\hat{Q}t}$.  This is done by substituting $\{{\bm R},{\bm D},{\bm L}\}$ into \erf{L} and then using the representation of $\{\hat{{\bm b}}^{\dag},\hat{{\bm b}}\}$ provided in \trf{GilmoreTable}.  The symbolic matrix exponentiation will need to be carried out by software, as it results in typically lengthy expressions for the block ${\bm N}$ matrices of the LHS of \erf{disMatrix}.
The expression for $e^{\hat{Q}t}$ can then be equated with a desired disentanglement (for example \erf{disG}) with the parameterized disentanglement also put into the finite dimensional representation in the same manner.  The parameters of the disentanglement, $\{{\bm R}^{\prime},\underline{{\bm D}},{\bm L}^{\prime}\}$ are found by solving these algebraic equations, with the obtained expressions shown in \erfs{Dunder}{Lprimeapp}.  From $\{{\bm R}^{\prime},\underline{{\bm D}},{\bm L}^{\prime}\}$, we obtain $\{{\bf R}^{\prime},\breve{{\bf R}}^{\prime},{\bf L}^{\prime},\breve{{\bf L}}^{\prime},\underline{\bf D},\underline{\breve{{\bf D}}}\}$ using the block form of \erfs{LBlockeqn}{allMatrixBlock}.

We now wish to calculate the linear parameters $\{\underline{\bf r},\underline{\bf l}\}$.  To do so, the parameters $\{d{\bm l}^{\prime},d{\bm r}^{\prime}\}$ must first be found from \erf{linPrime} and then integrated to give $\{{\bm l}^{\prime},{\bm r}^{\prime}\}$.  Using the ${\bm N}$ matrices and $\{{\bm l}^{\prime},{\bm r}^{\prime}\}$ in \erfs{normLittlel}{normLittlel2} gives $\{\underline{\bf r},\underline{\bf l}\}$.  We have now found all the parameters that define the solution to the SME.  They can be substituted into, for example, \erf{vectorizedEvoOp} in order to determine the evolution operator.

To find the POVM, an expression for ${\bf L}^{\prime\prime}$ is required.  After the transformation of \erf{transform} is made to \erf{Lprimeapp}, ${\bm L}^{\prime\prime}$ is found directly in terms of already known parameters.  Upon using the block form of \erf{LBlockeqn}, one obtains $\{{\bf L}^{\prime\prime},\breve{{\bf L}}^{\prime\prime}\}$.
The next POVM parameter, ${\bf d}$, is calculated from \erf{R}.  Finally, we solve \erf{alphaAve} for $\langle{\bm \alpha}\rangle_{{\bf d}}$.  Substituting $\{{\bf L}^{\prime\prime},\breve{{\bf L}}^{\prime\prime},{\bf d},\langle{\bm \alpha}\rangle_{{\bf d}}\}$ into \erf{finalPOVM} gives the POVM.

\section{Single mode examples}
\label{singleModeEgs}

In the case of a single mode, much of the notational complexity of our methods is removed, with the solution parameters, of \erf{solParams}, being scalars. This allows a heightened focus on important conceptual aspects.  Additionally, many important examples are single modes.  For these reasons, single mode examples will be used to illustrate the techniques of prior sections. The reader is also invited to peruse \arf{singleModeSec}, where the single mode case is further discussed.

\subsection{Homodyne detection in the presence of a thermal input bath}
\label{homdyneg}
A simple, but important, example is a single mode subjected to homodyne detection of the $x$-quadrature, with an input bath in a thermal state and no Hamiltonian evolution.  
We use the linear SME for this system given in \cite{WisMil10}:
\bqa
d\bar{\rho}(t)&=&\gamma(K+1){\cal D}[\hat{a}]\bar{\rho}(t)dt
+\gamma K{\cal D}[\hat{a}^{\dag}]\bar{\rho}(t)dt\nonumber\\
&&+y(t)dt\bar{{\cal H}}\left[
\frac{\sqrt{\gamma\eta}}{\sqrt{2K+1}}\left((K+1)\hat{a}-K\hat{a}\dg\right)
\right]{\rho}(t), \nonumber\\
\label{homeg}
\eqa
where $\eta$ is the detector efficiency, $\gamma$ is the system decay rate and $K$ describes the temperature of the thermal bath. 
This can be put in a form consistent with the $M$-representation given in \erf{smeGen} via a two-component column vector of Lindblad operators
\beq
\hat{{\bm c}}={\bm C}\hat{{\bm x}}={\bm Z}\left[
\begin{array}{c}
\hat{a} \\
\hat{a}^{\dag}
\end{array}\right],
\label{LVecHom}
\eeq
with ${\bm Z}$ being the $2\times 2$ matrix
\beq
{\bm Z}=\sqrt{\frac{\gamma}{2K+1}}
\left[
\begin{array}{cc}
K+1& -K\\
\sqrt{K(K+1)} & \sqrt{K(K+1)}
\end{array}\right].
\eeq
From ${\bm Z}$, it is straightforward to determine ${\bm C}={\bm Z}{\bm X}^{\dg}$, via \erf{XMat}. The form of the $2\times 4$ ${\bm M}$-matrix  also follows from \erf{homeg}.  We find
\beq
{\bm M}=
\left[
\begin{array}{cccc}
\sqrt{\eta}& 0&0&0\\
0&0&0&0
\end{array}\right].
\eeq
Note that the measurement current current, of \erf{smeGen}, is given by ${\bm y}^{{\rm T}}(t)=\left (y(t),0,0,0\right )$.  Also, ${\bm G}=0$, as there is no Hamiltonian.

Having defined the input parameters $\{{\bm G},{\bm C},{\bm M}\}$, we have the choice of utilizing a black box approach, in which we are only interested in obtaining the solution parameters of \erf{solParams}, or we can follow the steps outlined in \srf{sketch} in order to build understanding of the solution method.  
In this subsection, we take the time to explore some of the conceptual steps involved in solving the SME, in the context of our simple example.

The first step is to vectorize the linear SME.  This is done by right multiplying $d\bar{\rho}(t)$ by $\ket{0}$ and then using the usual identities $\hat{a}\ket{0}=\tilde{a}\dg\ket{0}$, $\hat{a}\dg\ket{0}=\tilde{a}\ket{0}$.  This gives
\bqa
d\ket{\bar{\rho}(t)}& =&\left(\gamma (K+1)dt\hat{D}[\hat{a}]+
\gamma Kdt\hat{D}[\hat{a}^{\dg}]+\right.\nonumber\\
&&\left. y(t)dt\sqrt{\frac{\gamma\eta}{2K+1}}\hat{S}[(K+1)\hat{a}-K\hat{a}\dg]\right)\rho(t),\nonumber\\
&&
\label{vecDRhoHom}
\eqa
where the operator functions $\hat{D}$ and $\hat{S}$ are defined in \erfs{DOPER}{SOPER}.

The next step, which we called `factorization' is to form the exponential evolution operator, $\hat{V}_{{\bm y}(t)}$, for the infinitesimal time slice.  This gives the updated density state vector as per $\ket{\bar{\rho}(t+dt)}=\hat{V}_{{\bm y}(t)}\ket{\bar{\rho}(t)}$.  We find
\beq
\hat{V}_{{\bm y}(t)}=\exp\left[ \hat{Q}dt+d\hat{L}(t) \right],
\eeq
with 
\bqa
\hat{Q}&=&\gamma (K+1)\hat{D}[\hat{a}]+
\gamma K\hat{D}[\hat{a}^{\dg}]\nonumber\\
&&-\frac{\gamma\eta}{2(2K+1)}\hat{S}[(K+1)\hat{a}-K\hat{a}\dg]^{2}\\
d\hat{L}(t)&=&
y(t)dt\sqrt{\frac{\gamma\eta}{2K+1}}\hat{S}\left[(K+1)\hat{a}-K\hat{a}\dg\right].
\label{vecRhoHom}
\eqa



By equating the above expressions for $\hat{Q}$ and $d\hat{L}(t)$ with \erfs{L}{dS} we can determine the matrices $\{{\bm R},{\bm D},{\bm L},d{\bm l},d{\bm r}\}$, such that we can express $\hat{V}_{{\bm y}(t)}$ in a standardized manner.  We find
\bqa
{\rm R}&=&-\frac{\gamma\eta K^{2}}{2(2K+1)}\label{Rrr}\\
\breve{{\rm R}}&=&\frac{\gamma K}{2(2K+1)}\left(2K+1-\eta K\right)\\
{\rm L}&=&-\frac{\gamma\eta( K+1)^{2}}{2(2K+1)}\\
\breve{{\rm L}}&=&\frac{\gamma (K+1)}{2(2K+1)}\left(2K+1-\eta (K+1)\right)\\
{\rm D}&=&\frac{\gamma }{2(2K+1)}\left(2\eta K(K+1)-(2K+1)^{2}\right)\\
\breve{{\rm D}}&=&\frac{\gamma\eta K(K+1)}{2K+1}\label{Dbreve}\\
d{\rm r}&=&-y(t)dtK\sqrt{\frac{\gamma\eta}{2K+1}}\\
d{\rm l}&=&y(t)dt(K+1)\sqrt{\frac{\gamma\eta}{2K+1}},
\eqa
which form $\{{\bm R},{\bm D},{\bm L},d{\bm l},d{\bm r}\}$ via \erfs{LBlockeqn}{allVecBlock}.  The parameters $\{{\bm R},{\bm D},{\bm L}\}$ are sufficient to form $e^{\hat{Q}t}$.

To compose the infinite string of infinitesimal evolution operators into a finite evolution operator, as per \erfs{evoOp2}{SMEsolVY}, all the quadratic exponential operators must be moved through the linear exponential terms, as described in \srf{SMEsolPrimed}.  To find a disentanglement of $e^{\hat{Q}t}$ and the parameters $\{{\rm r}^{\prime},{\rm l}^{\prime}\}$ that form $e^{\hat{L}^{\prime}(t)}$, the finite dimensional representation of $e^{\hat{Q}t}$ is required.  Its form is specified by \erf{LouiMat}, into which the expressions for $\{{\rm R},\breve{{\rm R}},{\rm L},\breve{{\rm L}},{\rm D},\breve{{\rm D}}\}$, from \erfs{Rrr}{Dbreve}, can be substituted.  Calculation of the exponential of the symbolic matrix is best done using software.  Using the notation of the RHS of \erf{LouiMat}, we obtain 
\bqa
q&=&e^{-\gamma  t /2}\frac{ 1-K (\eta  (K+1)-2 K-3)}{2 K+1}
\nonumber\\
&&+e^{\gamma  t /2}\frac{ K (\eta +(\eta -2) K-1)}{2 K+1}\\
s&=&\frac{2 \eta  K (K+1) \sinh \left(\gamma  t /2\right)}{2 K+1}\\
u&=&\frac{2 K (1-(\eta -2) K)}{2 K+1} \sinh \left(\frac{\gamma  t }{2}\right)\\
v&=&-\frac{2 \eta  K^2 }{2 K+1}\sinh \left(\frac{\gamma  t }{2}\right)\\
w&=&\frac{2 (K+1) (\eta +(\eta -2) K-1) }{2 K+1}\sinh \left(\frac{\gamma  t }{2}\right)\\
x&=&\frac{2 \eta  (K+1)^2 }{2 K+1}\sinh \left(\frac{\gamma  \tau }{2}\right)\\
y&=&e^{-\gamma  t/2}\frac{ K (\eta +(\eta -2) K-1)}{2 K+1}\nonumber\\
&&- e^{\gamma  t /2} \frac{(K+1)((\eta -2) K-1)}{2 K+1}\\
z&=&-\frac{2 \eta  K (K+1) }{2 K+1}\sinh \left(\frac{\gamma  t }{2}\right).
\eqa
These parameters are the single mode version of the ${\bm N}$ matrices found in \arf{disMatrix}.  Noting that the representation of $\hat{Q}$ is completely real, we can calculate the disentanglement parameters by substituting $\{q,s,u,v,w,x,y,z\}$ into \erfs{greekDis1}{greekDis}.  Given their straightforward calculation, we leave this task to the reader.  Similarly, the parameters $\{{\rm r}^{\prime},{\rm l}^{\prime}\}$ are obtained by integrating \erfs{linMove2}{linMove}.  We note that in the case of a zero-temperature ($K=0$) thermal bath ${\rm r}^{\prime}=0$ and knowledge of the stochastic integral ${\rm l}^{\prime}$, that determines the POVM effect, through $d$, is sufficient to specify the system state at the end of the measurement period (see \erf{linPiece}).   
\blk

Of interest are the POVM parameters $\{{\rm L}^{\prime\prime},\breve{ {\rm L}}^{\prime\prime},{\rm d}\}$.  They are calculated from \erfs{greekDis5}{greekDis} and \erf{R2}, together with the replacements of \erf{transform}.  We find that
\bqa
&&\breve{ {\rm L}}^{\prime\prime}={\rm L}^{\prime\prime}=-\frac{\left(1-e^{-\gamma t}\right) \eta }
{2+ 4K\left ( 1-\eta\left(1-   e^{-\gamma t}\right)\right)} \label{POVMHomParams2}\\
&&{\rm d}=\frac{ \sqrt{\gamma  \eta  (1+2K)}}{1+2 K \left(1-\eta  \left(1-e^{-\gamma  t}\right)\right)}
\int\limits_{0}^{t} e^{-\frac{\gamma  \tau}{2} }y(\tau)d\tau,
\label{POVMHomParams}
\eqa
which are to be substituted into \eqref{pomSingle} to obtain the POVM element.  We note that Ref.~\cite{Genoni} provides a general `dyne' POVM for the instantaneous measurement result, but here we form the composite measurement up to a finite time $t$.

The POVM defined by \erfs{POVMHomParams2}{POVMHomParams} can be compared with the literature~\cite{wisQTraj}  \blk
in the limit that an initial pure state is kept pure; that is, perfect efficiency detection and a zero temperature bath.  In this scenario, a stochastic Schr\"odinger equation (SSE) rather than SME is sufficient.  For $\eta=1$ and $K=0$, the POVM simplifies to
\bqa
\hat{W}_{{\rm pure}}&=&\exp\left[\sqrt{\gamma }\int\limits_{0}^{t} e^{-\frac{1}{2} \gamma  \tau }y(\tau)d\tau \hat{a}\dg-\frac{1}{2}(1-e^{-\gamma t}) \hat{a}^{\dagger 2}\right]\nonumber\\
&&\times\exp\left[-\gamma  \hat{a}^\dagger \hat{a} t\right]\nonumber\\
&&\times\exp\left[-\smallhalf(1-e^{-\gamma t}) \hat{a}^{2}+\sqrt{\gamma }\int\limits_{0}^{t} e^{-\smallhalf \gamma  \tau }y(\tau)d\tau \hat{a}\right]\nonumber\\
\label{purePOVM}
\eqa
which agrees with the result contained in~\cite{wisQTraj}.

We began this section by assuming that the $x$-quadrature was being measured, but it is easy to retrieve an arbitrary quadrature from our results.  Initially, a canonical transformation such as 
\bqa
\hat{a}\rightarrow \hat{a} e^{i\Phi}\quad{\rm and}\quad \hat{a}\dg\rightarrow \hat{a}\dg e^{-i\Phi},
\label{canon}
\eqa
together with the implied tilde conjugate transformation (based on \erf{tildeConjEqn}), could have been made to absorb the quadrature phase.  This preserves the commutation relations of $\hat{a},\hat{a}\dg,\tilde{a},\tilde{a}^{\dg}$.  As the disentanglement and re-ordering of the group elements is a function only of the commutation relations, this is permissible and is undone at the end of the calculation.  Thus, with the replacement $\hat{a}\rightarrow \hat{a} e^{i\Phi}$ and $\hat{a}\dg\rightarrow \hat{a}\dg e^{-i\Phi}$ in our POVM results (or SME solutions), measurement of any quadrature can be analysed.

\subsubsection{Adjoint equation}

The POVM can also be investigated via the adjoint equation, as per \srf{adjointSec}.  We now take this approach for the case of homodyne detection in the presence of a thermal input bath (that is, the linear SME specified in \erf{homeg}).  The matrices ${\bm A}$, ${\bm B}$, ${\bm S}$ and ${\bm E}$, that are defined below \erf{varW}, follow from knowledge of ${\bm C},{\bm M},{\bm G}$.  Using these matrices in \erfs{meanW}{varW}
provides the backwards-in-time equations of motion for the phase space representation of the effect operator:  
\bqa
\left[
\begin{array}{c}
dx \\
dp
\end{array}
\right]&=&
\frac{\gamma}{2}\left[
\begin{array}{c}
x \\
p
\end{array}
\right] d\tau \blk\nonumber\\
&&+
\sqrt{\frac{\gamma\eta}{2(1+2K)}}\left[
\begin{array}{c}
2V_{xx} +1+2K\\
2V_{px}
\end{array}
\right]dw(\tau) \blk\nonumber\\
\label{eomWmean}
\eqa
\bqa
\left[
\begin{array}{cc}
\dot{V}_{xx} & \dot{V}_{xp}\\
\dot{V}_{px} & \dot{V}_{pp}
\end{array}
\right]&=&\frac{\gamma(1+2K)}{2}\left[
\begin{array}{cc}
1-\eta &0\\
0 & 1
\end{array}
\right]\nonumber\\
&&+
\gamma\left[
\begin{array}{cc}
(1-2\eta)V_{xx} &V_{xp}-\eta V_{px} \\
(1-\eta) V_{px} & V_{pp}
\end{array}
\right]\nonumber\\
&&-
\frac{2\gamma}{1+2K}\left[
\begin{array}{cc}
\eta V_{xx}^{2} &\eta V_{px}V_{xx} \\
\eta V_{px}V_{xx} & \eta V_{px}^{2}
\end{array}
\right].\nonumber\\
\label{eomWvar}
\eqa
For clarity, we will use a backwards-in-time integration variable $\tau$, and a measurement completion time $t_{m}$ (so that $\tau\leq t_{m}$).   After the backwards-in-time integration has been performed we can set $\tau =0$ and $t_{m}=t$ to obtain the effect operator for a measurement compiled over the same time frame $[0,t]$ as \erf{POVMHomParams}. 

From \erf{eomWvar}, it can be seen that the equation for $\dot{V}_{xx}$ is decoupled. It can be solved using an infinite variance Gaussian, having no $x$-$p$ correlation, as the final condition (which is a suitable  approximation to the identity matrix for the effect operator).  We obtain 
\beq
V_{xx}=\smallhalf (1+2 K)\left(\frac{1}{\eta  \left(1-e^{-\gamma  (t_{m}-\tau) \blk}\right)}-1\right).
\label{varVxx}
\eeq
In contrast, $V_{pp}$ does not become finite when it is integrated backwards-in-time, which is consistent with homodyne detection of the $x$-quadrature providing no information about the $p$-quadrature. 
By inspection, we also observe that the matrix Riccati equation will not evolve the values of $\{V_{xp},V_{px}\}$ away from zero. 

The solution of \erf{eomWmean} for $x$ is made problematic due to the divergence of $V_{xx}$ at $\tau=t_{m}$.  Consequently, we use \erf{meanw} and instead work with $w_{x}=x v_{xx}^{-1}$.  We do not need to consider $w_{p}$ as it is decoupled from $w_{x}$, given our final condition together with \erf{eomWvar}.  Using \erf{meanw},
\bqa
dw_{x}&=&-\frac{\gamma}{2}\left[(1-2\eta)
+\Lambda_{xx}(1-\eta)(1+2K)
\right]w_{x}d\tau\nonumber\\
&&+
\sqrt{\frac{2\gamma\eta}{1+2K}}\left(1+\frac{\Lambda_{xx}}{2}\left(1+2K\right)\right)y d\tau,
\label{dwEqtn}
\eqa
where $\Lambda_{xx}=V_{xx}^{-1}$ and is determined by \erf{varVxx}.  After solving \erf{dwEqtn}, using $w_{x}(t_{m})=0$~\cite{fraser1969optimum}, we obtain
\bqa
w_{x}&=&\sqrt{\frac{2\gamma\eta}{1+2K}}\frac{e^{\gamma \tau/2}}{1-\eta+\eta e^{-\gamma(t_{m}-\tau)}}\times\nonumber\\
&&\int_{\tau}^{t_{m}}e^{-\gamma\tau'/2}y(\tau')d\tau',
\eqa
which can be verified by inspection.
Together with \erf{varVxx}, this determines the $x$-quadrature mean of the POVM via $x=v_{xx}w_{x}$.  Before providing the expression for $x$, we extend the integration back to $\tau=0$ and set the measurement turn-off time to $t_{m}=t$, as appropriate for considering the compiled measurement from $[0,t]$.  This leads to
\beq
x=\sqrt{\frac{\gamma (1+2K)}{2\eta}}\frac{1}{1-e^{-\gamma t}}\int_{0}^{t}e^{-\gamma\tau'/2}y(\tau')d\tau'.\label{xPOVMfinal}
\eeq

The reader is invited to observe how \erf{dwEqtn} dramatically simplifies for $\eta\rightarrow 1$.  It is therefore perhaps surprising that the only manifestation of non-perfect efficiency detection in the expression for $x$ is in the scalar coefficient. 

We can verify the accuracy of \erf{xPOVMfinal} by comparison with the POVM obtained via the thermo-entangled state representation.  That is, \erf{finalPOVM} is used together with the POVM parameters of \erf{POVMHomParams}.
To perform the comparison we note that $\hat{W}_{{\bf d}}$ of \erf{finalPOVM} is most easily converted to a Q-phase space function while the adjoint equation method here gives the first and second moments of the Wigner function (resulting from a symmetric ordering of operators).  The variances of these distributions are related by the Q-function variance being half a unit of vacuum noise larger than that of the Wigner function.  After this adjustment, the variances using the two methods agree.  We also find the mean values agree, with $2\sqrt{2}x=-{\rm d}/{\rm L}^{\prime\prime}$.

This approach, of using the adjoint equation, reproduces the mean and variance, but does not directly reproduce the non-operator stochastic dependence of the POVM. This is because a normalized version of the adjoint equation, with a specific final condition, is being treated.  However, the non-operator dependence could be obtained via the methods of \srf{simplePOVM}, if desired.


\blk

\subsection{Optomechanical position measurement with squeezing}
\label{optMech}

In recent work~\cite{warszawski2019tomography}, two of the current authors analyzed optomechanical position measurement, with a primary focus on quantum state tomography of the initial state of the mechanical oscillator.  The authors worked, for example, in the bad-cavity regime at both zero and blue-detuning~\cite{revCavOptMech}, in which it is possible to obtain a SME for the mechanics alone.  By incorporating squeezing (parametric amplification) alongside the measurement, it was found that effectively a homodyne limited measurement can be performed on the mechanical oscillator, despite operating in the weak measurement regime.  This is in contrast to the heterodyne limited measurement performed in the absence of squeezing, as the weak measurement does not allow localization of the mechanical position on the timescale of its period of motion.  The purpose of the current subsection is to use the theory developed in this paper to obtain expressions describing the quality of the optomechanical tomographic measurement in the zero-detuned limit.  To do so, we frame the problem in terms of the generic solutions we have provided in this paper and find the POVM.  The reader is referred to Ref.~\cite{aspelmeyer2014cavity} for an optomechanical review, and to Ref.~\cite{warszawski2019tomography} for more specific details relating to the system described here. 

We will examine the optomechanical system in the `zero-detuned' regime, which refers to the local oscillator being on resonance with the cavity that it illuminates.  The cavity is then coupled via radiation pressure with the mechanical oscillator (see figure 1(a) in \cite{warszawski2019tomography}).  We consider a measurement strength $\mu$, a thermal bath with coupling $\gamma$ and thermal phonon occupation $K_{{\rm th}}$, and parametric amplification of strength $\chi$ inducing squeezing in the quadrature defined by the angle $\theta$.  The system Hamiltonian is therefore
\beq
\hat{H}=\frac{i\chi}{4}\left(e^{-i \theta}\hat{a}^2-e^{i \theta}\hat{a}^{\dagger 2}\right),
\label{squeezeH}
\eeq
with $\theta=0$ defining squeezing in the $x$-quadrature.  The rationale behind the introduction of squeezing is that it makes the amplified (anti-squeezed) system quadrature more visible.  In terms of tomography, measurement in the presence of squeezing gives a better estimate of the anti-squeezed quadrature of the initial state.  Our analysis will indicate that by varying the squeezing angle over multiple trials, and using standard tomographic data analysis techniques~\cite{tomRevLvovsky}, the full system state can be determined in a more efficient manner than if squeezing was absent.

The linear SME for the mechanical oscillator, after the optical cavity has been adiabatically eliminated, is then~\cite{PhysRevLett.107.213603,sme}
\bqa
d\bar{\rho}(t) &=& \frac{\chi}{4}\left[e^{-i \theta}\hat{a}^2-e^{i \theta}\hat{a}^{\dagger 2},\bar{\rho}(t)\right]dt\nonumber\\
&&+[\gamma(K+1)+\mu^{\prime}]\mathcal{D}[\hat{a}]\bar{\rho}(t) dt\nonumber\\
&&+(\gamma K+\mu^{\prime})\mathcal{D}[\hat{a}^\dagger]\bar{\rho}(t) dt 
\nonumber \\
&&+\sqrt{\mu^{\prime}}y_{x}(t)dt\bar{{\cal H}}[\hat{x}] \bar{\rho}(t)\nonumber\\
&&+\sqrt{\mu^{\prime}}y_{p}(t)dt\bar{{\cal H}}[\hat{y}] \bar{\rho}(t).
\label{sme1}
\eqa
Here $\mu^{\prime}=\mu\eta$ represents an effective measurement strength, $K=K_{{\rm th}}+\mu(1-\eta)/\gamma$ is an effective bath temperature, 
$\hat{x}=(\hat{a}+\hat{a}^{\dag})/\sqrt{2},\hat{p}=i(\hat{a}^{\dag}-\hat{a})/\sqrt{2}$ are quadrature operators and $\{y_{x},y_{p}\}$ are real-valued stochastic quadrature measurement results.  The parameters $\{\mu^{\prime},K\}$ are introduced in order to simplify resultant expressions.

Before proceeding with solving the SME, we perform the canonical transformation of \erf{canon} with $\hat{a}^{\prime}=e^{-\theta/2}\hat{a}$.  This transformation will lead to a purely real representation of $\hat{Q}$ after the linear SME has been vectorized and factorized.  The transformed linear SME is (we suppress bosonic operator primes for simplicity of display) is
\bqa
d\bar{\rho}(t) &=& \frac{\chi}{4}\left[\hat{a}^2-\hat{a}^{\dagger 2},\bar{\rho}(t)\right]dt\nonumber\\
&&+[\gamma(K+1)+\mu^{\prime}]\mathcal{D}[\hat{a}]\bar{\rho}(t) dt\nonumber\\
&&+(\gamma K+\mu^{\prime})\mathcal{D}[\hat{a}^\dagger]\bar{\rho}(t) dt 
\nonumber \\
&&
+\sqrt{\mu^{\prime}}y_{x}(t)dt\bar{{\cal H}}[\hat{x}_{\theta}]\bar{\rho}(t)\nonumber\\
&&
+\sqrt{\mu^{\prime}}y_{p}(t)dt\bar{{\cal H}}[\hat{y}_{\theta}]\bar{\rho}(t),
\label{sme2}
\eqa
with $\hat{x}_{\theta}=\hat{a}e^{i\theta/2}+\hat{a}^{\dg}e^{-i\theta/2}$ and $\hat{y}_{\theta}=-i\hat{a}e^{i\theta/2}+i\hat{a}^{\dg}e^{-i\theta/2}$.

To place \erf{sme2} in the $M$-representation, we specify the following 3-component column vector of Lindblad operators 
\bqa
\hat{{\bm c}}&=&{\bm C}\hat{{\bm x}}\nonumber\\
&=&
\left[\begin{array}{cc}
    \sqrt{\gamma K+\mu^{\prime}}     & 0 \\
  0	 & \sqrt{\gamma K+\mu^{\prime}}  \\
  \sqrt{\frac{\gamma}{2}} & i \sqrt{\frac{\gamma}{2}}
\end{array}\right]
\hat{{\bm x}}\nonumber\\
&=& \sqrt{\gamma K+\mu^{\prime}}\begin{bmatrix}
\hat{x}\\
\hat{p}\\
 \sqrt{\frac{\gamma}{\gamma K+\mu^{\prime}}}\hat{a}
\label{LVecOpt}
\end{bmatrix}.
\eqa
We associate with each of the Lindblads a potentially complex valued measurement current.  In our case, the current associated with the $\hat{a}$ Lindblad is not measured, and those associated with $\hat{x}$ and $\hat{p}$ are both real.  The vector measurement current can, therefore, be written as ${\bm y}(t)^{{\rm T}}=(y_{x}(t),y_{p}(t),0,0,0,0)$.  The measurement setting, ${\bm M}$, associated with the currents is given by 
\bqa
\sqrt{\frac{\mu^{\prime}}{\gamma K+\mu^{\prime}}}
\left[\begin{array}{cccccc}
   \cos(\theta/2)     & \sin(\theta/2)&0&0&0&0  \\
 - \sin(\theta/2) 	 & \cos(\theta/2) &0&0&0&0   \\
 0&0&0&0&0&0 
\end{array}\right].
\eqa
The Hamiltonian of \erf{squeezeH} is in the form of \erf{hamiltonianF}, so we convert using \erf{convertG} to obtain
\bqa
{\bm G}=-\frac{\chi}{2}
\left[\begin{array}{cc}
 0     & 1 \\
1 & 0 
\end{array}\right].
\eqa

Given the matrices $\{{\bm C}, {\bm M},{\bm G}\}$ it is straightforward to determine the parameterization of $\hat{Q}$ and $d\hat{L}(t)$, either by using the formulae of \arf{RDLdldrDescription} or by direct vectorization and factorization.  We obtain,
\bqa
{\rm L}&=&\chi/4,\label{Q1}\\
\breve{{\rm L}}&=&\gamma(K+1)/2, \\
{\rm R}&=&-\chi/4, \\
\breve{{\rm R}}&=&\gamma K/2, \\
{\rm D}&=&-\gamma(2K+1)/2-2\mu^{\prime}, \\
\breve{D}&=&0,\label{Q2}\\
d{\rm l}&=&\sqrt{\mu^{\prime}}y^{*}dte^{i\theta/2}, \\
d{\rm r}&=&\sqrt{\mu^{\prime}}ydte^{-i\theta/2},
\eqa
where we have formed the complex measurement current, $ydt=\left(y_{x}+iy_{y} \right)dt/\sqrt{2}$, for convenience.  Note that the parameters describing $\hat{Q}$ are real.

In order to gauge the effectiveness of squeezing, as relates to tomography, we use an initial coherent state as a proxy that allows analytic results to be obtained.  That is, we wish to determine the probability distribution, $\wp( \alpha|{\rm d} )$ of the initial coherent amplitude, $\alpha$, given a measurement result ${\rm d}$.  From \erf{bayesNoh}, we know that this can be determined via the POVM.

To obtain the POVM, the expressions of \erfs{Q1}{Q2} are used in the matrix that is exponentiated in \erf{LouiMat}, with the result being the matrix elements on the RHS of that equation.  As the representation of $\hat{Q}$ is completely real, we can calculate the disentanglement parameters by substituting $\{q,s,u,v,w,x,y,z\}$ into \erfs{greekDis1}{greekDis}.  Similarly, the parameters $\{{\rm r}^{\prime},{\rm l}^{\prime}\}$ are obtained from integrating \erfs{linMove2}{linMove}.  The POVM then follows from \erfs{greekDis5}{greekDis} and \erf{R2}, together with the replacements of \erf{transform}. 

As the POVM is of direct utility in \cite{warszawski2019tomography}, we provide the POVM parameters
\bqa
&& {\rm L}^{\prime\prime}=
2\mu^{\prime}\left( 
\frac{1}{\gamma+4\mu^{\prime}-\chi+\Gamma_{-}\coth\left[\Gamma_{-}t/2 \right]}\right.\nonumber\\
&&\left.\quad\quad-
\frac{1}{\gamma+4\mu^{\prime}+\chi+\Gamma_{+}\coth\left[\Gamma_{+}t/2 \right]}
\right)\\
&&\breve{{\rm L}}^{\prime\prime}=
2\mu^{\prime}\left( 
\frac{1}{\gamma+4\mu^{\prime}-\chi+\Gamma_{-}\coth\left[\Gamma_{-}t/2 \right]}\right.
\nonumber\\
&&\left.\quad\quad +
\frac{1}{\gamma+4\mu^{\prime}+\chi+\Gamma_{+}\coth\left[\Gamma_{+}t/2 \right]}
\right)\\
&&{\rm d}=\nonumber\\
&&\frac{\sqrt{\mu^{\prime} }  (\gamma -\Gamma_{+}+4 \mu^{\prime} -\chi+4 \gamma  K)}{ 2\gamma  K- \chi } \int^{t}_{0}e^{-\Gamma_{+} \tau/2} y_{x}(\tau)d\tau\nonumber\\
&&-i\frac{\sqrt{\mu^{\prime} }  (\gamma -\Gamma_{-}+4\mu^{\prime} +\chi +4 \gamma  K)}{2 \gamma  K+\chi }\int^{t}_{0}e^{-\Gamma_{-} \tau/2} y_{p}(\tau)d\tau,\nonumber\\
\eqa
with rates \blk $\Gamma_{\pm} = \sqrt{(\gamma \pm\chi)^2+
8\mu^{\prime} \gamma(1+2K)+16\mu^{\prime 2}}$.  For simplicity, ${\rm d}$ is \blk provided in the large $t$ limit, which represents a measurement carried out until the final state is uncorrelated with the initial state.  


From the POVM, the variance in the retrodictive estimates of the $x$ and $p$-quadratures for an initial coherent state, 
which we denote as $\sigma_{x,p}^2$, can be found.  Bearing in mind that $\left\{{\rm L}^{\prime\prime},\breve{ {\rm L}}^{\prime\prime}\right\}\in {\mathbb R}$, it follows from \erf{genPDF} that 
\bqa
\sigma_{x}^2&=&\frac{1}{2\left(\breve{ {\rm L}}^{\prime\prime}-{\rm L}^{\prime\prime} \right)}=
\frac{1}{2}+\frac{\gamma+\chi +\Gamma_{+} \coth (\Gamma_{+} T/2)}{8 \mu ^{\prime}}\nonumber\\
\sigma_{p}^2&=&\frac{1}{2\left(\breve{ {\rm L}}^{\prime\prime}+{\rm L}^{\prime\prime} \right)}
=\frac{1}{2}+\frac{\gamma -\chi +\Gamma_{-} \coth (\Gamma_{-} T/2)}{8 \mu ^{\prime}},
\nonumber\\
\eqa
which are the results used extensively in \cite{warszawski2019tomography}.  The variance $\sigma_{p}^2$ is strictly smaller than $\sigma_{x}^2$, so the squeezing is having the anticipated effect of making the antisqueezed quadrature more visible.

\section{Discussion and Conclusion}
\label{conclusion}
In this paper we have shown how the evolution of continuously monitored quantum systems that possess linear Heisenberg-picture dynamics can be solved (provided they possess a time-independent Hamiltonian and measurement setting, ${\bm M}$). 
The treatment of mixed quantum states represents a non-trivial generalization of the previously existing literature~\cite{wisQTraj,jacLin,jacSte}, which was limited to solving the stochastic Schr\"odinger equation.
As a corollary to obtaining the stochastic master equation (SME) solution, the {\it deterministic} master equation (for linear quantum systems) is also solved, using a non-phase space method.  Our method of solution was to use the thermo-entangled state representation together with techniques from Lie algebra. The obtained SME solution is in the form of an evolution superoperator that is dependent upon $2N$ complex-valued measurement-dependent stochastic integrals ($N$ being the number of physical modes).

There are a number of uses for such an analytic solution, some of which we now detail. Firstly, one can calculate the possible states, and their probability distribution, resulting from combined Hamiltonian and measurement dynamics. Likewise, a simple method to numerically simulate Lindbladian evolution is obtained by sampling the distribution, in contrast to integrating the SME in the infinite dimensional Hilbert space (where some form of truncation is required).  
This should facilitate applications in quantum control, in particular state-based feedback control, as well as dissipative quantum state engineering. 
Finally, the statistics of the compiled measurement record are captured in a POVM, which is directly utilized in quantum state tomography, by which the initial state is determined via the measurement record.  Given that there has been much experimental progress towards manipulating quantum systems, verification of their states is of paramount importance.  One of the major techniques to achieve this is homodyne tomography, so our POVM would have direct application. 
It is worth noting that a SME POVM has been investigated previously in the literature, using the adjoint equation approach; we have shown that the two approaches agree for a simple case, as expected.

The reader may be curious as to why the multimode SME was chosen as the starting point given that its analytic tractability is limited in concert with the solubility of the relevant higher order polynomials (which are found to be of degree $2N$ for $N$ physical modes).  Apart from the symmetries of particular higher degree systems leaving the possibility of solution, a reason is that \eqref{inftyRho} may provide a very efficient launching pad for numeric investigation of multimode SME bosonic systems.  In~\cite{Gilmore} the finite dimensional representation of multimode bosonic operators is given and the benefits of numerical integration within the finite representation of the group are detailed.  The context in that work is unitary evolution but there is no barrier preventing the extension to stochastic non-unitary dynamics. The alternative of using a  Fock space representation is not feasible since the size of the space grows exponentially with the number of modes, and even then an approximation must be made limiting the evolution to a finite number of basis states (in say, the energy eigenbasis). By contrast, the proposed algebraic approach avoids any energy cut-off and utilizes matrices that grow in size only quadratically in $N$, specifically as $(4N+2)\times (4N+2)$. In this context it is worth noting the work by Galitski~\cite{galit2} which explores Lie methods in multi mode systems and emphasizes that a trajectory within a Lie algebra is in some ways more natural than a Hilbert space representation.  

Before considering other classes of systems that might be analogously solved, it is important to reiterate why the methods we used were successful in solving the SME for linear quantum systems.  The thermo-entangled state representation for fermions has been developed and is only a trivial extension to the bosonic case~\cite{Kosov2}, so this is not the critical ingredient for success.  Similarly, any nonlinear SME can be cast into a linear SME form, so the specific system does not impact upon this~\cite{WisMil10}.  In fact, the only step in our solution method that presents a potential stumbling block is that of `composition', by which exponential operators are reordered and reformed.  We now discuss this step in more detail.

For the case of linear quantum systems, the process of composition is facilitated by the fact that the operators contained in the deterministic and stochastic pieces of the infinitesimal evolution operator (see \erf{littlev}) coincide with the subalgebras $\mathfrak{q}$ (quadratic bosonic operators) and $\mathfrak{l}$ (linear bosonic operators), respectively.  Because $\mathfrak{l}$ is an ideal of the algebra $\mathfrak{q}\cup\mathfrak{l}$ (so that $[\mathfrak{l},\mathfrak{q}]\in\mathfrak{l}$), we were able to reorder the linear and quadratic exponential operators such that those that are strictly quadratic remained deterministic.  Additionally, the individual closure of both $\mathfrak{q}$ and $\mathfrak{l}$ allowed the separation of the linear and quadratic evolution pieces to be maintained when the operator disentanglement (or re-entanglement) was performed.  From these comments, one sees that it is important to consider the structure of the evolution operator imposed by the SME dynamics as well as the commutation relations of the full algebra.  For the systems of this paper, this could be restated as $\mathfrak{q}$ and $\mathfrak{l}$ being of individual relevance in addition to it being important to study the algebra $\mathfrak{q}\cup\mathfrak{l}\subset \mathfrak{sp}(4N+2)$.

As we have seen, the exponential evolution operator for linear quantum systems contains terms that are at most quadratic in the annihilation and creation bosonic mode operators; consequently, the algebra that they form closes.  In contrast, cubic operators (and beyond) give an algebra that expands under commutation, ultimately becoming infinite.  However, it was only for $N\leq 2$ that we were able to perform composition.  Consequently, operators closing under commutation is not sufficient for composition to be possible in the way we have described.

 An obvious consideration for future work is whether the same methods could be applied to other classes of single mode SMEs, such as for a spin $\frac{1}{2}$ particle.  For example, the Pauli operators provide an obvious representation to work in.  One complication is that the time-dependent evolutions (e.g. measurement terms) are now present in a more complex group, $SU(2)$, than the Heisenberg-Weyl algebra.  It is interesting to note that there exist some analytic results in the literature for qubits subjected to heterodyne detection~\cite{campagne2016observing,sarlette2017deterministic}.  Whether these bear any relation to our results could be investigated.

 Another generalization that could be considered is photo-detection instead of `dyne' detection.  With photo-detection, the evolution between detection events is deterministic so can be solved similarly to a standard master equation.  For example, in~\cite{Kosov} the waiting time distribution is calculated utilizing a thermo-entangled state approach for a particular system.  
In~\cite{3rdQuant} general deterministic evolution of a Lindblad form is treated, but the evolution between detections is not of this type. A final, presumably very restricted, generalization to consider would be the possibility of time dependent parameters for the Hamiltonian or measurement setting.  The latter would encompass adaptive measurement schemes, which are of great interest~\cite{berry2000optimal,mahler2013adaptive}.

It has been commented that the adjoint equation to the SME provides an alternative path for finding the POVM. It would be of interest to perform a detailed comparison, for higher dimensional systems, of the difficulty of solving the Kalman filter as compared with using the finite dimensional representation to find the POVM parameters.  This could be both from an analytic and numeric perspective.  Numerically, the Kalman filter has efficient solution algorithms~\cite{welch1995introduction}, which is perhaps not surprising given that it also reduces the system to a finite number of variables.  

When using the adjoint equation, evolution is attributed to the system operators rather than the system state, in analogy to the Heisenberg and Schr\"odinger pictures.  As a final consideration for future work, we query whether the class of soluble SMEs could be expanded (beyond quadratic Hamiltonian with linear measurements) by moving to an interaction picture, for which the evolution is split between operators and the state. The purpose would be to isolate more simple Lie groups that could be independently solved, before recombining the evolution to obtain the system solution.

\begin{acknowledgments}
We wish to thank Kiarn Laverick for very helpful discussions regarding the adjoint equation.  This work was supported by the Australian Research Council via discovery project number DP130103715, via the Centre of Excellence in Engineered Quantum Systems (EQuS), project number CE110001013 and CE170100009, and 
via the Centre for Quantum Computation and Communication Technology (CQC2T), project number CE110001027 and  CE170100012, and by the University of Sydney Faculty of Science via a Postgraduate Scholarship.  We acknowledge the traditional owners of the land on which this work was undertaken at Griffith University, the Yuggera people, and the traditional owners of the land on which this work was undertaken at the University of Sydney, the Gadigal people of the Eora Nation.
\end{acknowledgments}

\appendix

\section{Ostensible distribution}
\label{ostensible}
To explain the origin of the ostensible distribution, $\wp_{{\rm ost}}({\bm Y}(t))$, let us first introduce some concepts from quantum measurement. A quantum operation, ${\cal O}_{{\bm y}(t)}$, is defined as conditioning an a-priori system state, $\rho(t)$, on a measurement result ${\bm y}(t)$, to give an a-posteriori system state, $\rho_{{\rm c}}(t+dt)$.  The subscript `c' has been used to indicate conditioning upon a measurement result, considtent with the notation of \erf{smeGenNonLin}.  The superoperator, ${\cal O}_{{\bm y}(t)}$, is a completely positive convex linear map that is trace-preserving or trace-decreasing~\cite{WisMil10}.  Under these constraints, the operation can represent a physical process.  To make contact with with the context of our work, we have considered the measurement, ${\bm y}(t)$, resulting from a `dyne' measurement in an infinitesimal time slice, $dt$. However, a quantum operation is a general concept is applicable to an arbitrary measurement over a possibly finite time. The operation acts as follows
\beq
\rho_{{\rm c}}(t+dt)
={\cal O}_{{\bm y}(t)}\rho(t)=\sum_{j}\hat{\Omega}_{{\bm y}(t),j}\rho(t)\hat{\Omega}_{{\bm y}(t),j}^{\dag},
\label{KRAUS}
\eeq
for some set of Kraus operators $\{\hat{\Omega}_{y,j}:j\}$, producing an unnormalized density matrix, $\rho_{{\rm c}}(t+dt)$, even if $\rho(t)$ is normalized.  The probability of obtaining the measurement result ${\bm y}(t)$ is given by
\beq
\wp_{{\bm y}(t)} =\Tr[\rho_{{\rm c}}(t+d t)]= \Tr[{\cal O}_{{\bm y}(t)}\rho(t)],
\label{normProb}
\eeq
which could be used to give a normalized state, ${\cal O}_{{\bm y}(t)}\rho(t)/\Tr[{\cal O}_{{\bm y}(t)}\rho(t)]$, if desired. 

Importantly, it is possible to formulate quantum measurement theory slightly differently from that indicated by \erf{KRAUS}, by introducing an `ostensible' probability, $\wp_{{\rm ost}}({\bm y}(t))$, for the result ${\bm y}(t)$~\cite{wisQTraj}.  This involves the use of a rescaled operation $\bar{{\cal O}}_{{\bm y}(t)}\blk={\cal O}_{{\bm y}(t)}/{\wp_{{\rm ost}}({\bm y}(t))}$, so that the probability of obtaining the result ${\bm y}(t)$ becomes
\beq
\wp_{{\bm y}(t)} = \Tr[\bar{\rho}_{{\rm c}}(t+d t)]{\wp_{{\rm ost}}({\bm y}(t))},
\label{pZ}
\eeq
where 
\beq
\bar{\rho}_{{\rm c}}(t+d t)=\bar{{\cal O}}_{{\bm y}(t)}\rho (t).\label{lineOstEqn}
\eeq
The bar over $\rho_{{\rm c}}$ is meant to alert the reader to the fact that the norm of $\bar{\rho}_{{\rm c}}$ is not to be interpreted in the sense of \erf{normProb}, due to the introduction of ostensible statistics for ${\bm y}(t)$.
The utility of such a formulation is that there is freedom of choice regarding the ostensible distribution; this can be exploited so as to make the linear equation for $\bar{\rho}_{{\rm c}}(t+d t)$ as simple as possible.  

In \erf{lineOstEqn}, a linear equation for an unnormalized state has been obtained.  However, for $\bar{\rho}_{{\rm c}}(t+dt)$ to average to the correct unconditioned state, $\rho(t+dt)$, that is determined by unconditioned ME evolution, the result ${\bm y}(t)$ needs to be chosen according to the `ostensible' statistics, defined by ${\wp_{{\rm ost}}({\bm y}(t))}$, rather than its actual statistics.

Inspection of the nonlinear SME, in \erf{smeGenNonLin}, reveals that that the nonlinear piece
(contained in the ${\cal H}\left[ {\bm M}^{\dag}\hat{{\bm c}}\right]\rho(t)$ term, as seen from \erf{hSuper}) only affects the state normalization.  By dropping these linear terms, we can then immediately write a linear evolution equation for the density matrix that, when normalized, produces the correct system state. However, rather than continuing to work with the measurement noise, $d{\bm w}^{{\rm T}}(t) $, we use \erf{measCurrent} to write the obtained linear equation in terms of the current ${\bm y}(t)$.  Unfortunately, this spawns a nonlinear term, 
\beq
-\langle {\bm M}^{\dag} \hat{{\bm c}}+{\bm M}^{{\rm T}}
 \hat{{\bm c}}^{\ddagger}\rangle dt\left(
 {\bm M}^{\dag}\hat{{\bm c}}\bar{\rho}(t)+
  \bar{\rho}(t){\bm M}^{{\rm T}}
 \hat{{\bm c}}^{\ddagger}\right),
\eeq
that would appear to do more than affect the state normalization.  Despite its appearance, the term in question can be canceled by a pure rescaling of the density matrix
\bqa
\rho(t+dt)& \rightarrow &\nonumber \\
&&\hspace{-8em} \rho(t+dt)
\exp\left[
d{\bm w}^{{\rm T}}(t) \langle {\bm M}^{\dag} \hat{{\bm c}}+{\bm M}^{{\rm T}}
 \hat{{\bm c}}^{\ddagger}\rangle
\right].
\eqa
After this transformation, the nonlinear SME has been linearized to become the SME of \erf{smeGen}.

We have seen how to move to the particular linear SME that we work with in our paper, but this has not been a constructive procedure in terms of obtaining ${\wp_{{\rm ost}}({\bm y}(t))}$.  Rather than derive it, it is sufficient to show that the ostensible distribution
\beq
{\wp_{{\rm ost}}({\bm y}(t))}=\left(\frac{dt}{2\pi}\right)^{L}\exp\left[-\smallhalf dt  {\bm y}^{{\rm T}}{\bm y}\right]\label{yost}
\eeq
allows reproduction of the deterministic ME and provides the correct statistics for ${\bm y}(t))$.
The ostensible distribution for each of the $2L$ components of ${\bm y}(t))$ is equal to that of a zero-mean Gaussian process with variance equal to $1/dt$.

To show that the ME is reproduced, we need to average the linear SME density matrix increment, $d\bar{\rho}_{{\rm c}}(t)$, over the ostensible distribution.  This is particularly straightforward as the measurement term in $d\bar{\rho}_{{\rm c}}(t)$ is linearly dependent upon the current and averages to zero. Removing the measurement term from \erf{smeGen} leaves the ME, as expected.

The actual statistics for ${\bm y}(t))$ are determined by \erf{pZ}.  Assuming that the density matrix is normalized at time $t$, we obtain for the prefactor
\bqa
&&\Tr[\bar{\rho}_{{\rm c}}(t+dt)]\nonumber\\
&=& 1+\Tr[d\bar{\rho}_{{\rm c}}(t)],\\
&=&
1+ {\bm y}^{{\rm T}}(t)dt \langle {\bm M}^{\dag}\hat{{\bm c}}+{\bm M}^{{\rm T}}
 \hat{{\bm c}}^{\ddagger}\rangle
\\
&=&\exp\left[{\bm y}^{{\rm T}}(t)dt\langle {\bm M}^{\dag}\hat{{\bm c}}+{\bm M}^{{\rm T}} \hat{{\bm c}}^{\ddagger}
\rangle\right.\nonumber\\
&&\left.-\smallhalf \langle {\bm M}^{\dag}\hat{{\bm c}}+{\bm M}^{{\rm T}}
 \hat{{\bm c}}^{\ddagger}\rangle^{{\rm T}}\langle {\bm M}^{\dag}\hat{{\bm c}}+{\bm M}^{{\rm T}}
 \hat{{\bm c}}^{\ddagger}\rangle dt\right],
\label{pZ2}
\eqa
where the last expression holds to order $dt$, as can be seen when the \ito rule of \erf{actExpY2} is remembered.  The prefactor can be combined with ${\wp_{{\rm ost}}({\bm y}(t))}$, as per \erf{pZ}, to give
\bqa
\wp_{{\bm y}(t)} &=&\left(\frac{dt}{2\pi}\right)^{L}
\exp\left[-\smallhalf dt  
\left({\bm y}-\langle {\bm M}^{\dag}\hat{{\bm c}}+{\bm M}^{{\rm T}}
\hat{{\bm c}}^{\ddagger}\rangle\right)^{{\rm T}}\times\right.\nonumber\\
&&\left.
\quad\quad\quad\quad\left({\bm y}-\langle {\bm M}^{\dag}\hat{{\bm c}}+{\bm M}^{{\rm T}}
 \hat{{\bm c}}^{\ddagger}\rangle\right)\right].
\eqa
That is, we have shown that the linear SME of \erf{smeGen} together with the ostensible distribution of the form of \erf{yost}, predicts statistics for ${\bm y}(t))$ are those of a Gaussian process having mean $\langle {\bm M}^{\dag}\hat{{\bm c}}+{\bm M}^{{\rm T}}
\hat{{\bm c}}^{\ddagger}\rangle$ and variance $1/dt$.  This agrees with the correct actual statistics given in \erfs{actExpY}{actExpY2}.  Thus, we have the correct ostensible distribution for our chosen linear SME.  The ostensible distribution for the finite time is obtained simply as the product of the ostensible distributions for each of the $J$ time slices,
\beq
\wp_{{\rm ost}}({\bm Y}(t))={\wp_{{\rm ost}}({\bm y}(dt))}{\wp_{{\rm ost}}({\bm y}(2dt))}
\cdots{\wp_{{\rm ost}}({\bm y}(Jdt))},\label{yost222}
\eeq
with $Jdt=t$.

\section{Finite dimensional representations of Lie algebras}
\label{LieAlgebraSec}

In \srf{lieAlg}, the algebra $\mathfrak{g}$, being a subalgebra of the symplectic algebra $\mathfrak{sp}(4N+2)$~\cite{Gilmore,GilmoreCoh}, was identified as being the algebra relevant to the solution of the SME for linear quantum systems.  Specifically, the evolution operator, belongs to the Lie group,  $\hat{V}_{{\bm Y}(t)}\in G$, that is associated with the Lie algebra $\mathfrak{g}$, as per
\bqa
\hat{V}_{{\bm Y}(t)}&=&e^{\hat{Q}t}e^{\hat{L}^{\prime}(t)}\\
&=&e^{X},
\eqa
with $\{X,\hat{Q},\hat{L}^{\prime}(t)\}\in\mathfrak{g}$.  It was explained that a finite dimensional matrix representation of $\mathfrak{g}$, that respects the algebra's commutation relations, could be used to perform disentangling and reordering calculations. \blk Results obtained by explicit calculation in the representation can then be lifted to the abstract level.  Finite dimensional representations have long been understood as calculational tools in quantum optics~\cite{Mufti,Caves,Gilmore} and we refer the reader to Ref.~\cite{GilmoreCoh} for a particularly authoritative review.  In this section we describe the operational details relating to finite dimensional representations in the context of solving SMEs and obtaining the associated POVMs.

From a Lie algebra perspective, the tilde bosonic operators are on the same footing as non-tilde (physical mode) operators.  As in \srf{SMEsolPrimed}, we emphasize this by writing a combined, `mixed-mode' annihilation operator 
\beq
\hat{{\bm b}}=\left(\hat{{\bm a}};\tilde{{\bm a}}\right)=\left(\hat{a}_{1},...,\hat{a}_{N},\tilde{a}_{1},...,\tilde{a}_{N}\right)^T.
\label{bVecApp}
\eeq
We can then write the elements of the algebra $\mathfrak{g}$ as
\beq
\left\{\hat{1},\hat{{\bm b}}_{\mu},\hat{{\bm b}}^{\dagger}_{\mu},\hat{{\bm b}}_{\mu}\hat{{\bm b}}_{\nu},\hat{{\bm b}}_{\mu}^{\dagger}\hat{{\bm b}}_{\nu}^{\dagger},\hat{{\bm b}}_{\mu}^{\dagger}\hat{{\bm b}}_{\nu}+\smallhalf\delta_{\mu,\nu}
\right\},
\label{opAlgebra}
\eeq 
for $ \mu,\nu\in 1,...,2N$. The algebra of interest contains further structure: $\left\{\hat{1},\hat{{\bm b}}_{\mu},\hat{{\bm b}}^{\dagger}_{\mu}\right\}$ form an ideal and $\left\{\hat{{\bm b}}_{\mu}^{\dagger}\hat{{\bm b}}_{\nu}^{\dagger},\hat{{\bm b}}_{\mu}^{\dagger}\hat{{\bm b}}_{\nu}+\smallhalf\delta_{\mu,\nu},\hat{{\bm b}}_{\mu}\hat{{\bm b}}_{\nu}\right\}$ a subalgebra.  Also, note that $\left\{\hat{{\bm b}}_{\mu}^{\dagger}\hat{{\bm b}}_{\nu}+\smallhalf\delta_{\mu,\nu}\right\}$ alone forms a closed algebra, $\mathfrak{u}(2N)$~\cite{Gilmore}.  

\subsection{Specifying the representation}

\label{GilmoreRep}

The smallest faithful finite dimensional representation of $\mathfrak{sp}(4N+2)$ are square matrices of size $(4N+2)$.  A smaller representation would mean that we lose distinction between some algebra elements and would not be `faithful'.  In this paper, the real valued, size $(4N+2)$, representation from \cite{Gilmore} is used.  We reproduce it in \trf{GilmoreTable}, for the reader's convenience. This representation of the algebra has the property that each element is either nilpotent (with index 2) or idempotent (when columns and rows entirely of zeroes are stripped away), thus making the exponentiation, via a Taylor expansion, simple.  

\begin{table}[!t]
\caption{\label{GilmoreTable}Matrix representation of bosonic operators}
\begin{ruledtabular}
\begin{tabular}{cc}
Operator & Matrix representation
\footnote{The matrices, ${\bm M}$, (being completely unrelated to the matrix describing the measurement setting of the SME) are of dimension $(4N+2)\times (4N+2)$ and have their rows and columns labelled from $0$ to $2N$ and then from $-2N$ to -$0$.  ${\bm M}_{\mu\nu}$ denotes a matrix that has only a single non-zero component, being equal to one, at the $\mu$th row and $\nu$th column.  Note that a sign correction, to the table contained in~\cite{Gilmore}, has been made in the second row. 
}
\\ \hline
\rule{0pt}{4ex}
 $\hat{{\bm b}}^{\dagger}_{\mu}\hat{{\bm b}}_{\nu}+\smallhalf\delta_{\mu,\nu}$ & ${\bm M}_{\mu\nu}
 -{\bm M}_{-\nu -\mu}$\\
 $\hat{{\bm b}}^{\dagger}_{\mu}\hat{{\bm b}}^{\dagger}_{\nu}$ &  ${\bm M}_{\mu -\nu}
 +{\bm M}_{\nu -\mu}$\\
  $\hat{{\bm b}}^{\dagger}_{\mu}\hat{{\bm b}}^{\dagger}_{\mu}$ & $2{\bm M}_{\mu -\mu}$\\
   $\hat{{\bm b}}_{\mu}\hat{{\bm b}}_{\nu}$ & $-{\bm M}_{-\mu\nu}
 -{\bm M}_{-\nu \mu}$\\
      $\hat{{\bm b}}_{\mu}\hat{{\bm b}}_{\mu}$ & $-2{\bm M}_{-\mu\mu}$\\
           $\hat{{\bm b}}^{\dagger}_{\mu}$ &  ${\bm M}_{\mu 0}
 -{\bm M}_{-0 -\mu}$\\
           $\hat{{\bm b}}_{\mu}$ &  $-{\bm M}_{-\mu 0}
 -{\bm M}_{-0 \mu}$\\
                                 $\hat{1}$ & $-2{\bm M}_{-00}$
\end{tabular}
\end{ruledtabular}
\end{table}

To help reduce confusion, we given an example operator matrix for $N=1$.  In this case there are two creation operators, one each for the physical and unphysical modes.  Arbitrarily, we choose to display the representation of $\hat{ a}^{\dagger}\tilde{a}^{\dagger}$, which is given by
\begin{align}
\hat{ a}^{\dagger}\tilde{a}^{\dagger}=
 \begin{bmatrix}
  0    &0 &   0  & 0 & 0      & 0             \\
0 &    0        & 0 & 1                &0 &0 \\
0 &    0        & 0 & 0                &1 &0 \\
0 &    0        & 0 & 0                &0 &0 \\
0 &    0        & 0 & 0                &0 &0 \\
0 &    0        & 0 & 0                &0 &0 \\
\end{bmatrix},
\end{align}
as can be determined from \trf{GilmoreTable}.

\subsection{Disentanglement of a general element of $G$}

Disentanglement of an exponential operator, $g\in G$, refers to its splitting into a product of exponential operators, as per \erf{disEqn}.  There are, in general, many disentangled forms of $g$, with the appropriate form dependent upon the calculation being performed.  We choose to illustrate disentanglement of $g$, via the finite dimensional matrix representation, for the normally ordered form.  A general operator, $g\in G$, and its normally ordered form are given by
\bqa
g&=&\exp\left[\hat{{\bm b}}^{\dag}{\bm  r}+\hat{{\bm b}}^{\dag}{\bm R}\hat{{\bm b}}^{\ddagger}+\hat{{\bm b}}^{\dag}{\bm D}\hat{{\bm b}}
+\hat{{\bm b}}^{{\rm T}}{\bm L}\hat{{\bm b}}+{\bm  l}\hat{{\bm b}}
\right]
\label{gen}\\
&=&\exp\left[\hat{{\bm b}}^{\dag}\underline{\bm r}+\hat{{\bm b}}^{\dag}{\bm R}^{\prime}\hat{{\bm b}}^{\ddagger}\right]
\exp\left[\hat{{\bm b}}^{\dag}\underline{\bm D}\hat{{\bm b}}
+\delta^{\prime}\right]\nonumber\\
&&\times
\exp\left[\hat{{\bm b}}^{{\rm T}}{\bm L}^{\prime}\hat{{\bm b}}+\underline{\bm  l}\hat{{\bm b}} \right],
\label{disG2}
\eqa
where primes and underlines indicate that different functional forms are anticipated.

Given a particular choice of finite dimensional representation, $g$ can now be calculated.  We will make progress by doing this somewhat heuristically in the multimode case and then being more explicit in the case of a single mode (see \arf{singleModeSec}).  This is necessary as the multimode calculation involves solution of degree $2N$ polynomials and becomes intractable for large $N$.  The structure of $g$ when expressed as \erf{gen} (LHS) and \erf{disG2} (RHS), in the chosen representation of \trf{GilmoreTable}, is
\bqa
 &&\begin{bmatrix}
  1    &0 &   0  & 0         \\
 {\bm N}_{10} &     {\bm N}_{11}        &  {\bm N}_{1-1}  &0 \\
 {\bm N}_{-10} &    {\bm N}_{-11}       & {\bm N}_{-1-1} &0 \\
 {\bm N}_{-00} &    {\bm N}_{-01}        & {\bm N}_{-0-1}  &1  \\
\end{bmatrix}=\label{disMatrix}\\
&& \begin{bmatrix}
  1    &0 &   0  & 0     \\
\underline{\bm r}-2{\bm R}^{\prime}e^{-\underline{\bm D}^{{\rm T}}}\underline{\bm l}^{{\rm T}} &   e^{\underline{\bm D}}-4{\bm R}^{\prime} e^{-\underline{\bm D}^{{\rm T}}}{\bm L}^{\prime}       & 2{\bm R}^{\prime}e^{-\underline{\bm D}^{{\rm T}}}{\bm J} &0 \\
 -{\bm J} e^{-\underline{\bm D}^{{\rm T}}}\underline{\bm l}^{{\rm T}}& -2{\bm J} e^{-\underline{\bm D}^{{\rm T}}}{\bm L}^{\prime}&{\bm J}e^{-\underline{\bm D}^{{\rm T}}}{\bm J}              &0\\
-2\Delta^{\prime}+\underline{\bm r}^{{\rm T}} e^{-\underline{\bm D}^{{\rm T}}}\underline{\bm l}^{{\rm T}} &   -\underline{\bm l}+2\underline{\bm r}^{{\rm T}}e^{-\underline{\bm D}^{{\rm T}}}{\bm L}^{\prime}    &- \underline{\bm r}^{{\rm T}}e^{-\underline{\bm D}^{{\rm T}}}{\bm J}&1 \\
\end{bmatrix},\nonumber
\eqa
where on the LHS the row and column subscript symbols $\pm 0$ are unidimensional labels and the $\pm 1$ are $2N$ dimensional~\cite{Gilmore}.  That is, the ${\bm N}$'s represent matrices, vectors and a scalar depending on the subscript labels (and should not be confused with $N$, the number of physical modes).  On the RHS, the $2N\times 2N$ matrix ${\bm J}$ has been used, and is defined as
\begin{equation}
 {\bm J}_{\mu\nu}=\begin{cases}
    1, & \text{if $\mu+\nu=n+1$}.\\
    0, & \text{otherwise},
  \end{cases}
  \label{JMat}
\end{equation}
which is a matrix of ones on the anti-diagonal (and zeroes elsewhere).  
Note that our definitions of the matrices $\left\{{\bm R}^{\prime},{\bm L}^{\prime},\underline{\bm r},\underline{\bm l}\right\}$ differ slightly from those of \cite{Gilmore,GilmoreCoh}.  Specifically, our matrices have elements distributed in the standard fashion, for example ${\bm R}^{\prime}_{ij}$ is located in the $i$th row and $j$th column. The RHS of \erf{disMatrix} follows from the Taylor expansion of each of the 3 exponentials of \erf{disG2}, which either terminate at first order or are diagonal, in the chosen representation.

From \erf{disMatrix}, the disentanglement parameters $\left\{{\bm R}^{\prime},{\bm L}^{\prime},\underline{\bm D},\underline{\bm r},\underline{\bm l},\delta^{\prime}\right\}$ can easily be found in terms of the ${\bm N}$'s (by block matrix manipulation):
\bqa
\underline{\bm D}^{{\rm T}}&=&-{\bm J} \ln {\bm N}_{-1-1}{\bm J},\label{Dunder}\\
2{\bm R}^{\prime}&=&{\bm N}_{1-1}({\bm N}_{-1-1})^{-1}{\bm J},\\
2{\bm L}^{\prime}&=&-{\bm J}({\bm N}_{-1-1})^{-1}{\bm N}_{-11},\label{Lprimeapp}\\
\underline{\bm l}^{{\rm T}}&=&-{\bm J}({\bm N}_{-1-1})^{-1}{\bm N}_{-10},\\
\underline{\bm r}^{{\rm T}}&=&-{\bm N}_{-0-1}({\bm N}_{-1-1})^{-1}{\bm J}.\label{disParamsApp}
\eqa

As $e^{\hat{Q}t}$ is an element of ${\rm Sp}(4N+2)$, it has the same form as $g$ but with the linear pieces missing.  This can be directly seen from \erf{L}.  Consequently, a disentanglement of $e^{\hat{Q}t}$ has been achieved in terms of the ${\bm N}$ block matrices.  Note that a normal ordering can be performed
\beq
:\exp\left[
\hat{{\bm b}}^{\dag}\left(e^{\underline{{\bm D}}}-\eye_{2N}\right)\hat{{\bm b}} \right]:
=
\exp\left[\hat{{\bm b}}^{\dag}\underline{{\bm D}}\hat{{\bm b}}
\right],\label{opId}
\eeq
using an operator identity~\cite{FAMmulti}.


\subsection{Reordering of exponential operators}
 \label{reorderApp}
 
 \subsubsection{$e^{d\hat{L}(t)}e^{\hat{Q}t}$}
  \label{reorderAppLin1}
Having disentangled arbitrary elements of the group, we now move on to operator reordering.  Firstly, we consider the reordering of an exponential operator having a strictly linear exponent with that of an exponential operator having a strictly quadratic exponent.  Two operators that are sufficiently general are $e^{\hat{Q}t}$ and $e^{d\hat{L}(t)}$,
and we consider the reordering
\beq
 e^{d\hat{L}(t)}e^{\hat{Q}t}=e^{\hat{Q}t}e^{d\hat{L}^{\prime}(t)}.
\eeq
The ideal nature of $d\hat{L}(t)$ is evidenced by the quadratic term being unmodified when shifting through the linear term.  We can use the general form of \erf{disMatrix} for the finite dimensional representation of the reordering:
\bqa
&& \begin{bmatrix}
  1    &0 &   0  & 0     \\
d{\bm r}&  1    & 0 &0 \\
 - {\bm J}d{\bm l}^{{\rm T}}& 0&1             &0\\
d{\bm l}d{\bm r} &   -d{\bm l}    &-d{\bm r}^{{\rm T}}&1 \\
\end{bmatrix}
 \begin{bmatrix}
  1    &0 &   0  & 0         \\
 0 &     {\bm N}_{11}        &  {\bm N}_{1-1}  &0 \\
 0 &    {\bm N}_{-11}       & {\bm N}_{-1-1} &0 \\
 0 &    0        & 0  &1  \\
\end{bmatrix}=\nonumber\\
&& \begin{bmatrix}
  1    &0 &   0  & 0         \\
 0 &     {\bm N}_{11}        &  {\bm N}_{1-1}  &0 \\
 0 &    {\bm N}_{-11}       & {\bm N}_{-1-1} &0 \\
 0 &    0        & 0  &1  \\
\end{bmatrix}
 \begin{bmatrix}
  1    &0 &   0  & 0     \\
d{\bm r}^{\prime}&  1    & 0 &0 \\
 - {\bm J}d{\bm l}^{\prime {\rm T}}& 0&1             &0\\
d{\bm l}^{\prime }d{\bm r}^{\prime}&   -d{\bm l}^{\prime}    &-d{\bm r}^{\prime{\rm T}}&1 \\
\end{bmatrix}
\label{linMatrix}
\eqa
where we have chosen to represent $e^{\hat{Q}t}$ in terms of the blocks of its exponentiated form, rather than by its disentanglement parameters (which can be calculated in terms of the former anyway, as per \erf{disParamsApp}). Note that the ${\bm N}$'s are a function of $t$. The primed variables can now be solved, giving
\bal
d{\bm l}^{\prime}=&d{\bm l} {\bm N}_{11}+d{\bm r}^{{\rm T}}{\bm N}_{-11}
\\
d{\bm r}^{\prime}=& {\bm N}_{1-1}^{{\rm T}}d{\bm l}^{{\rm T}}+{\bm N}_{-1-1}^{{\rm T}}d{\bm r}.
\label{linPrime}
\eal
These form the increment $d\hat{L}^{\prime}(j dt)$ appearing in \erf{linComm} as per 
\beq
d\hat{L}^{\prime}(t)=\hat{{\bm b}}^{\dag}d{\bm r}^{\prime}+d{\bm l}^{\prime}\hat{{\bm b}}.
\label{dS2}
\eeq
Note that there are redundant equations in \erf{linMatrix} from which the primed variables can be solved.  This embodies the fact that the ${\bm N}$'s are interrelated.

 \subsubsection{Composition of $e^{\hat{L}^{\prime}(t)}$}
  \label{reorderAppLin2}

Given $d\hat{L}^{\prime}(j dt)$ from \erf{dS2}, we now wish to compose the product of linear exponentials, 
\beq
e^{\hat{L}^{\prime}(t)}=
\prod\limits_{j=1}^{J}\exp\left[d\hat{L}^{\prime}(j dt)\right],
\label{2Late2}
\eeq
that forms part of the evolution operator. 
The operators contained in $d\hat{L}^{\prime}(j dt)$ represent $2N$ copies of the Heisenberg-Weyl algebra, $\{\hat{1},\hat{a},\hat{a}^{\dag}\}$, and the composition can be treated simply with the Zassennhaus formula, \erf{zass}.  To explain further, the linear operators belong to $\mathfrak{l}$, meaning that commutators beyond first order are zero (see \erfs{linLin}{linLinLin}).  Note that the Zassennhaus formula is equivalent to the better known
Baker-Campbell-Hausdorff formula~\cite{Louisell} for the subalgebra $\mathfrak{l}$.  

Similarly to how the rightmost quadratic exponential was moved to the left (explained in \srf{Evolution}) we wish to move the rightmost term in the product of \erf{2Late2} to the left, combining the linear exponentials at each time slice.  Whenever two linear exponentials operators are combined, an exponential with an exponent equal to the sum of the linear operators and a correction proportional to the identity is obtained.  This can be seen from \erf{zass}, remembering that the exponents commutation relation obeys \erf{linLin}.  Consider now the product of the exponential operator obtained for $j$ combined time slices, with that of the single $j+1$ time slice.  This is given by
\bqa
&&\exp\left[\hat{{\bm b}}^{\dag}d{\bm r}^{\prime j+1} 
+d{\bm l}^{\prime j+1} \hat{{\bm b}} \right]\times\nonumber\\
&&\exp\left[\delta_{{\bm Y}(\tau)}+  \hat{{\bm b}}^{\dag}\sum\limits_{k=1}^{j}d{\bm r}^{\prime k} 
+\sum\limits_{k=1}^{j}d{\bm l}^{\prime k} \hat{{\bm b}} \right]\nonumber\\
&&=
\exp\left[\delta_{{\bm Y}(\tau+dt)}
+\hat{{\bm b}}^{\dag}\sum\limits_{k=1}^{j+1}d{\bm r}^{\prime k} 
+\sum\limits_{k=1}^{j+1}d{\bm l}^{\prime k} \hat{{\bm b}} \right],\label{linJth}
\eqa
with $\tau=jdt$ and 
\beq
\delta_{{\bm Y}(\tau+dt)}-
\delta_{{\bm Y}(\tau)}=\frac{1}{2}\sum_{k=1}^{j}\left(d{\bm l}^{\prime j+1}d{\bm r}^{\prime k}-
d{\bm l}^{\prime k}d{\bm r}^{\prime j+1} \right).
\eeq 
Each time slice, when it is absorbed, spawns a term proportional to the identity, as this expression for $\delta_{{\bm Y}(\tau+dt)}$ indicates.  Note that $\delta_{{\bm Y}(0)}=0$. 

To find $e^{\hat{L}^{\prime}(t)}$ we have to sum the the contributions to $\delta_{{\bm Y}(t)}$ over the index $j$, as well as extending the summations in \erf{linJth} to $k=J$.  As a final step, we choose to give a normally ordered expression, which involves a final elementary disentanglement.  This gives
\bqa
e^{\hat{L}^{\prime}(t)}&=&\exp\left[\frac{1}{2}\sum\limits_{j=1}^{J}d{\bm l}^{\prime j}d{\bm r}^{\prime j}+\sum\limits_{j=1}^{J}\sum\limits_{k=1}^{j}d{\bm l}^{\prime j}d{\bm r}^{\prime k}\right]\nonumber\\
&&\times \exp\left[\hat{{\bm b}}^{\dag}\sum\limits_{j=1}^{J}d{\bm r}^{\prime j}   \right]
\exp\left[ \sum\limits_{j=1}^{J}d{\bm l}^{\prime j} \hat{{\bm b}} \right].\label{e2theS}
\eqa
The single index summation that is quadratic in $\left\{d{\bm l}^{\prime j},d{\bm r}^{\prime j}\right\}$ (thus quadratic in noise) will have contributing summands of $\mathcal{O}(dt)$ and be of a deterministic nature due to \ito's rule.  The double index summation will contain both stochastic ($k\neq j$) and deterministic  summands ($k= j$), as will the sums attached to the operators.  

The summations of \erf{e2theS} should be identified with integrals, but in converting them we need to take care in the case where the summand will lead to an integrand that is itself a function of the noise.  This is not the case for the single summations.  For the double summation, there are terms containing the product of correlated stochastic increments.  To convert to {\ito } integrals we must {\it adapt} the summand to separate these terms~\cite{GarBook}.  That is,
\begin{equation}
\begin{aligned}
\sum\limits_{j=1}^{J} d{\bm l}^{\prime j}\sum\limits_{k=1}^{j}d{\bm r}^{\prime k}=&\sum\limits_{j=1}^{J} d{\bm l}^{\prime j}\left(d{\bm r}^{\prime j}+\sum\limits_{k=1}^{j-1}d{\bm r}^{\prime k}\right)\\
=&\int\limits_{0}^{t}d{\bm l}^{\prime}(\tau)d{\bm r}^{\prime}(\tau)+\int\limits_{0}^{t} d{\bm l}^{\prime}(\tau){\bm r}^{\prime}(\tau) 
\end{aligned}
\end{equation}
with 
\beq
{\bm r}^{\prime}(\tau)=\int\limits_{0}^{\tau}d{\bm r}^{\prime}(s)
\label{intNote}
\eeq
for notational convenience. By \ito's rule $d{\bm l}^{\prime}(\tau)d{\bm r}^{\prime}(\tau)$ is $\mathcal{O}(d\tau)$ so the first integral is deterministic, whilst the second term retains its stochasticity.  

For clarity, the complete linear component of \erf{solSumm} is
\bqa
e^{\hat{L}^{\prime}(t)}&=&\exp\left[\frac{3}{2}\int\limits_{0}^{t}d{\bm l}^{\prime}(\tau)d{\bm r}^{\prime}(\tau)+\int\limits_{0}^{t}d{\bm l}^{\prime}(\tau)r'(\tau) \right]
\nonumber\\
&&\times \exp\left[  \hat{{\bm b}}^{\dag}{\bm  r}^{\prime}(t)  \right]
\exp\left[{\bm l}^{\prime}(t)\hat{{\bm b}}\right]\label{defineh}\\
&=&e^{h(t)}e^{\hat{{\bm b}}^{\dag} {\bm r}^{\prime}(t)}e^{{\bm l}^{\prime} (t)\hat{{\bm b}}},
\label{linPiece2}
\eqa
with
$h$ being a scalar non-Gaussian complex-valued stochastic integral. 
The explicit time dependence of $\{h,{\bm l}^{\prime},{\bm r}^{\prime}\}$ in the main text has been suppressed for display purposes.

 \subsubsection{$e^{\hat{Q}t}e^{\hat{L}^{\prime}(t)}$}
\label{acute}

To normally order the linear pieces of the evolution operator, and move from \erf{preoper} to \erf{fullevoopeqn}, the following reordering is required
\beq
e^{\hat{{\bm b}}^{\dag} \underline{\bm r}}e^{\hat{Q}t}
e^{ \underline{\bm l}\hat{{\bm b}}}
=e^{\hat{Q}t}e^{\hat{{\bm b}}^{\dag} {\bm r}^{\prime}}e^{{\bm l}^{\prime}\hat{{\bm b}}},
\eeq
with $\{\underline{\bm r},\underline{\bm l} \}$ to be found. The matrix representation of this is
\bqa
&& \begin{bmatrix}
  1    &0 &   0  & 0     \\
\underline{\bm r}&  1    & 0 &0 \\
0& 0&1             &0\\
0 &   0    &-\underline{\bm r}^{{\rm T}}&1 \\
\end{bmatrix}
 \begin{bmatrix}
  1    &0 &   0  & 0         \\
 0 &     {\bm N}_{11}        &  {\bm N}_{1-1}  &0 \\
 0 &    {\bm N}_{-11}       & {\bm N}_{-1-1} &0 \\
 0 &    0        & 0  &1  \\
\end{bmatrix}
\begin{bmatrix}
  1    &0 &   0  & 0     \\
0&  1    & 0 &0 \\
 - {\bm J}\underline{\bm l}^{{\rm T}}& 0&1             &0\\
0 &   -\underline{\bm l}    &0&1 \\
\end{bmatrix}\nonumber\\
&& =\begin{bmatrix}
  1    &0 &   0  & 0         \\
 0 &     {\bm N}_{11}        &  {\bm N}_{1-1}  &0 \\
 0 &    {\bm N}_{-11}       & {\bm N}_{-1-1} &0 \\
 0 &    0        & 0  &1  \\
\end{bmatrix}
 \begin{bmatrix}
  1    &0 &   0  & 0     \\
{\bm r}^{\prime}&  1    & 0 &0 \\
 - {\bm J}{\bm l}^{\prime {\rm T}}& 0&1             &0\\
{\bm l}^{\prime }{\bm r}^{\prime}&   -{\bm l}^{\prime}    &-{\bm r}^{\prime{\rm T}}&1 \\
\end{bmatrix}.
\label{linMatrixNormOrder}
\eqa
Solving for $\{\underline{\bm r},\underline{\bm l} \}$ gives
\bqa
\underline{\bm l}&=&{\bm l}^{\prime}-{\bm r}^{\prime{\rm T}}\left({\bm N}_{-1-1}^{{\rm T}}\right)^{-1}{\bm N}_{-11}\label{normLittlel}\\
\underline{\bm r}&=&\left({\bm N}_{-1-1}^{{\rm T}}\right)^{-1}{\bm r}^{\prime}.\label{normLittlel2}
\eqa

 \subsubsection{POVM reorderings}
  \label{reorderAppLin4}

We now move on to another example. In \srf{POVMsec}, the finite dimensional representation was also used to calculate a reordering of $e^{\tilde{{\bm a}}^{{\rm T}}\hat{{\bm a}}}e^{\hat{Q}t}$, as per \erf{povmDis}), when obtaining the POVM. 
This reordering can be found from the disentanglement of of $e^{\hat{Q}t}$ with minimal work.  If we write $\tilde{{\bm a}}^{{\rm T}}\hat{{\bm a}}=\smallhalf\hat{{\bm b}}^{{\rm T}}\bar{\pmb{I}}\hat{{\bm b}}$ (see \erf{iBar} for $\bar{\pmb{I}}$) then the finite dimensional representation of $e^{\tilde{{\bm a}}^{{\rm T}}\hat{{\bm a}}}e^{\hat{Q}t}$ is
\bqa
e^{\tilde{{\bm a}}^{{\rm T}}\hat{{\bm a}}}e^{\hat{Q}t}&=&
 \begin{bmatrix}
  1    &0 &   0  & 0     \\
0& \eye_{2N}    & 0 &0 \\
0& -\bar{\pmb{I}}&\eye_{2N}              &0\\
0&   0   &0&1 \\
\end{bmatrix}
 \begin{bmatrix}
  1    &0 &   0  & 0         \\
 0 &     {\bm N}_{11}        &  {\bm N}_{1-1}  &0 \\
 0 &    {\bm N}_{-11}       & {\bm N}_{-1-1} &0 \\
 0 &    0        & 0  &1  \\
\end{bmatrix},\nonumber\\
&=&
 \begin{bmatrix}
  1    &0 &   0  & 0         \\
 0 &     {\bm N}_{11}        &  {\bm N}_{1-1}  &0 \\
 0 &    {\bm N}_{-11}-\bar{\pmb{I}} {\bm N}_{11}      & {\bm N}_{-1-1}-\bar{\pmb{I}} {\bm N}_{1-1}  &0 \\
 0 &    0        & 0  &1  \\
\end{bmatrix}.\nonumber\\
\label{linMatrix2}
\eqa
The RHS of \eqref{povmDis} actually has the same functional form as that of \eqref{disG} as the $e^{\hat{{\bm b}}^{{\rm T}}\bar{\pmb{I}}\hat{{\bm b}}/2}$ term can be absorbed to give $e^{\hat{{\bm b}}^{{\rm T}}\left({\bm L}^{\prime\prime}+\bar{\pmb{I}}/2\right)\hat{{\bm b}}}$.  Thus the reordering is the same as the disentanglement of \erf{disParamsApp}, but with the following replacements:
\bqa
{\bm N}_{-11}&\rightarrow &{\bm N}_{-11}-\bar{\pmb{I}}{\bm N}_{11},\nonumber\\
\quad {\bm N}_{-1-1}&\rightarrow &{\bm N}_{-1-1}-\bar{\pmb{I}} {\bm N}_{1-1},\nonumber\\
{\bm L}^{\prime}&\rightarrow &{\bm L}^{\prime\prime}+\bar{\pmb{I}}/2. 
\label{transform} \\ \nonumber
\eqa

The calculation of ${\bf d}$, which appears in \erfs{pom}{R}, is similarly done by solving a matrix equation.  However, there exists the simplification that the re-ordering is amongst physical mode operators only, so that ${\rm Sp}(2N+2)$ elements need be considered, rather than ${\rm Sp}(4N+2)$.

\subsection{Further details for a single mode}
\label{singleModeSec}
Reduction to a single {\it physical} mode, $N=1$, will allow further clarification of our methods and facilitate the provision of some single mode examples.  


For $N=1$, the relevant algebra is the two-mode (one of which is unphysical) double (and single)-photon algebra, which is a semisimple subalgebra of $\mathfrak{sp}(6)$ \cite{Gilmore,Twa93}.  It is 15 dimensional, consisting of the elements: 
\beq
\left\{\hat{1},\hat{a},\hat{a}^\dagger,\hat{a}^2,\hat{a}^{\dagger 2},\hat{a}^\dagger \hat{a},\tilde{a},\tilde{a}^\dagger,\tilde{a}^2,\tilde{a}^{\dagger 2},\tilde{a}^\dagger \tilde{a},\hat{a} \tilde{a}, \hat{a}^\dagger \tilde{a}^\dagger,\hat{a} \tilde{a}^\dagger,\hat{a}^\dagger \tilde{a}\right\}.
\eeq  
The single mode versions of equations \erf{L} and \erf{dS} (tilde operators will be kept explicit rather than combined into a vector with non-tildes) are
\bqa
\hat{Q}t&=&{\rm L}\hat{a}^{2}+{\rm L}^{*} \tilde{a}^{2}+
2\breve{{\rm L}} \hat{a}\tilde{a}+
{\rm R} \hat{a}^{\dagger 2}+{\rm R}^{*} \tilde{a}^{\dagger 2}+
2\breve{{\rm R}}  \hat{a}\dg\tilde{a}^{\dagger}\nonumber\\
&&+
{\rm D} \hat{a}\dg \hat{a}+{\rm D}^{*}\tilde{a}\dg \tilde{a} +\breve{{\rm D}}\hat{a}\dg\tilde{a}+
\breve{{\rm D}}^{*}\hat{a}\tilde{a}^{\dagger}\label{LQT}
\\
d\hat{L}(t)&=&d {\rm l} \hat{a}+d {\rm l}^{*} \tilde{a}+d {\rm r} \hat{a}\dg+d {\rm r}^{*}\tilde{a}^{\dagger},
\label{sNl}
\eqa
with the relationship of $\left\{{\rm R},\breve{{\rm R}},{\rm L},\breve{{\rm L}}, {\rm D},\breve{{\rm D}},d{\rm r},d{\rm l}\right\}$ to the system parameters of the SME given in \arf{RDLdldrDescription}.  Note that the Hermiticity preservation conditions of the SME have been utilized; there is only one independent parameter in each of $\{d{\bm l},d{\bm r}\}$, which we have written as $\left\{d{\rm l},d{\rm r}\right\}$.  That is, $d{\rm l}=d{\bm l}_{1}$ and $d{\rm l}^{*}=d{\bm l}_{2}$.  Similarly, we have set ${\rm L}={\bm L}_{11}$, $\breve{{\rm L}}={\bm L}_{12}$, together with analogous assignments for $\{{\rm R},{\rm D}\}$.  

When $e^{\hat{Q}t}$ is calculated in the real, finite $(4N+2)\times(4N+2)$ dimensional representation of~\cite{Gilmore} (with $N=1$) it has the following symmetry (due to Hermiticity preservation of the evolution):
\bqa
e^{\hat{Q}\tau}&=&
\exp\left[
\begin{array}{cccccc}
0  & 0 & 0 & 0 & 0 & 0 \\
 0 &  {\rm D} &  \breve{{\rm D}}& 2 \breve{{\rm R}}   & 2 {\rm R} & 0 \\
 0 &  \breve{{\rm D}}^{*}  & {\rm D}^{*} & 2   {\rm R}^{*} & 2 \breve{{\rm R}}   & 0 \\
 0 & -2 \breve{{\rm L}}  & -2 {\rm L}^{*} & - {\rm D}^{*} & -  \breve{{\rm D}} & 0 \\
 0 & -2 {\rm L} & -2\breve{{\rm L}} & - \breve{{\rm D}}   & - {\rm D}  & 0 \\
 -2 C  & 0 & 0 & 0 & 0 & 0 \\
\end{array}
\right]\nonumber\\
&=&
 \begin{bmatrix}
  1    &0 &   0  & 0 & 0      & 0             \\
0 &   q        & s & u                &v &0 \\
 0 &  s^*&q^*               &v^* & u^* &0\\
 0 &   w    &x               &  y& z&0 \\
0 &    x^*&w^*              & z^*& y^* &0 \\
   c   &0 &     0      & 0& 0 &1 \\
\end{bmatrix}.
\label{LouiMat}
\eqa
The $t$ dependent matrix elements $\{q,s,u,v,w,x,y,z\}$ (which form the ${\bm N}$ matrices of \erf{disMatrix}) are inter-related and, in general, have a complicated dependence upon $\{{\bm L},{\bm R},{\bm D}\}$.  There is no difficulty in their accurate calculation with a symbolic manipulator, although their length prohibits their display here.  From \erf{LouiMat}, it can be seen that the calculation reduces to the exponentiation of the inner $4\times 4$ block.  Further simplifying matters is that the characteristic polynomial is a depressed quartic (no cubic term).  This is important as the Cayley-Hamilton theorem states that a matrix satisfies its own eigenvalue equation.  The power series expansion for the exponential has no higher than cubic powers of the matrix.  The eigenvalues of the matrix define the time scales of the deterministic evolution.  As it will often prove useful to consider the simplification that the representation of $e^{\hat{Q}t}$ is real valued (the system quadratic parameters being real), we provide the eigenvalues for this case.  The $4$ eigenvalues ($\pm\lambda_{\pm}$) of the inner block of $\hat{Q}t$ are then defined by
\beq
\lambda_{\pm}=  \sqrt{({\rm D}+\breve{{\rm D}})^2-4({\rm L}\pm \breve{{\rm L}}) ( {\rm R}\pm \breve{{\rm R}})}.
\eeq
The division of the $4$ eigenvalues into $2$ pairs illustrates the Hermiticity symmetry which leads to the collapsing of the $4\times 4$ matrix analysis into multiple $2\times 2$ operations.  This can be made explicit, when $\hat{Q}t$ is real valued, by a unitary transform of the representation of $\hat{Q}t$ to a $2\times 2$ block diagonal form.

The single mode, towards normal order, form of $e^{\hat{Q}t}$ is
\bqa
e^{\hat{Q}t} &= &
e^{\delta^{\prime}} \exp\left[2\breve{{\rm R}^{\prime}} \hat{a}^\dagger \tilde{a}^\dagger+{\rm R}^{\prime} \hat{a}^{\dagger2}+{\rm R}^{\prime *} \tilde{a}^{\dagger2}\right]\\
&&\times\exp\left[\underline{{\rm D}} \hat{a} \tilde{a}^\dagger+\underline{{\rm D}}^{ *} \hat{a}^\dagger \tilde{a}+\underline{{\rm D}} \hat{a}^\dagger \hat{a}+\underline{{\rm D}}^{ *}\tilde{a}^\dagger \tilde{a}\right]\nonumber\\
&&\times\exp\left[2\breve{{\rm L}^{\prime}} \hat{a}\tilde{a}+{\rm L}^{\prime} \hat{a}^2+{\rm L}^{\prime *} \tilde{a}^2\right].
\label{quadDis}
\eqa
In analogy to \erf{disParamsApp}, the disentanglement parameters can be found
\bqa
\underline{{\rm D}}&=&-\frac{1}{2}\log \left(y^2-z^2\right),\label{greekDis1}\\
\underline{\breve{{\rm D}}}&=& \frac{1}{2} \log \left(\frac{y-z}{y+z}\right),\label{greekDis2}\\
{\rm R}^{\prime} &=&\frac{v y-u z}{2( y^2- z^2)},\label{greekDis3}\\
\breve{{\rm R}}^{\prime} &=& \frac{u y-v z}{2(y^2-z^2)},\label{greekDis4}\\
{\rm L}^{\prime} &=&\frac{w z-x y}{2( y^2-z^2)},\label{greekDis5}\\
\breve{{\rm L}}^{\prime}&=&\frac{x z-w y}{2(y^2-z^2)},
\label{greekDis}
\eqa
where we have assumed that any complex parameters in $\hat{Q}$ can be made real by transformation (see the end of \srf{homdyneg}).  This assumption is performed for simplicity of display, not necessity.

Similarly, the single mode version of \erf{linPrime} can be found after the linear exponentials, $e^{d\hat{L}(t)}$ and $e^{d\hat{L}^{\prime}(t)}$, are calculated in the finite dimensional representation.  This leads to 
\bqa
d {\rm l}^{\prime}&=&q d{\rm l} + s^*d {\rm l} ^{*}+w d {\rm r}+x^*d {\rm r}^{*} \label{linMove2} \\
d {\rm r} ^{\prime}&=&u d {\rm l} +v^*d {\rm l} ^{*}+y d {\rm r}+z^*d{\rm r} ^{*},
\label{linMove}
\eqa
together with $d {\rm l}^{\prime *}=d {\bm l}_{2}^{\prime}$ and $d{\rm r}^{\prime *}=d {\bm r}_{2}^{\prime}$.

Parallel to the multi-mode case, the calculation of the single mode POVM requires the reordering of $e^{\tilde{a}\hat{a}}e^{\hat{Q}t}$, which has the finite dimensional representation
\beq
\left[
\begin{array}{cccccc}
 1 & 0 & 0 & 0 & 0 & 0 \\
 0& q & s & u & v& 0  \\
 0 & s^* & q^* & v^* & u^* & 0 \\
 0 & w-q & x-s & y-u & z-v& 0 \\
 0 & x^*-s^* & w^*-q^* & z^*-v^* & y^*-u^*  & 0\\
 c & 0 & 0 & 0 & 0& 1 \\
\end{array}
\right].
\eeq
The disentanglement order chosen in \erf{povmDis} allows the $e^{\hat{a}\tilde{a}}$ term to be absorbed into $e^{2\breve{{\rm L}}^{\prime\prime} \hat{a}\tilde{a}}$ to give $e^{(2\breve{{\rm L}}^{\prime\prime}+1)\hat{a}\tilde{a}}$, so that the disentanglement is the same as in \erf{greekDis} but with the following replacements:
\begin{align}
& w\rightarrow w-q, \quad x\rightarrow x-s,\quad y\rightarrow y-u, \nonumber \\ 
& z\rightarrow z-v,\quad \breve{{\rm L}} ^{\prime}\rightarrow \breve{{\rm L}}^{\prime\prime} +\smallhalf.
\label{transform}
\end{align}
This process provides the POVM parameters ${\rm L}^{\prime\prime},\breve{{\rm L}}^{\prime\prime}$.  The final POVM parameter, 
\beq
{\rm d}= {\rm l}^{\prime *}+
2 {\rm L}^{\prime\prime}{}^{*}{\rm r}^{\prime *}+
\left(1+2\breve{ {\rm L}}^{\prime\prime}\right){\rm r}^{\prime},
\label{R2}
\eeq 
is also found using the finite dimensional representation

We can then show the single mode POVM element,
\bqa
\hat{W}_{{\rm d}}&=&{\cal N}
\exp\left[ 
-2|\langle\alpha\rangle_{ {\rm d}}|^{2}\breve{ {\rm L}}^{\prime\prime}
+\langle \alpha\rangle_{ {\rm d}}^{2} {\rm L}^{\prime\prime}+
\langle \alpha\rangle_{ {\rm d}}^{* 2} {\rm L}^{\prime\prime *}
\right]\nonumber\\
&&\times\exp\left[
{\rm d}\hat{ a}^{\dag} +
 {\rm L}^{\prime\prime}\hat{a}^{\dag 2}
\right]
:\exp\left[
2\breve{ {\rm L}}^{\prime\prime}\hat{ a}^{\dag}\hat{ a}
\right]:\nonumber\\
&&\times
\exp\left[
{\rm L}^{\prime\prime *}\hat{a}^{2} 
+ {\rm d}^{*}\hat{a}
\right],\label{pomSingle}
\eqa 
with the real and imaginary parts of $\langle\alpha\rangle_{ {\rm d}}$ related to those of ${\rm d}$ by
\beq
\langle\alpha\rangle_{ {\rm d},{\mathfrak R}}=
\frac{{\rm d}_{\mathfrak R}}{2\left(\breve{ {\rm L}}^{\prime\prime}-  {\rm L}^{\prime\prime}_{\mathfrak R}\right)}
\eeq
and
\beq
\langle\alpha\rangle_{ {\rm d},{\mathfrak I}}=
\frac{{\rm d}_{\mathfrak I}}{2\left(\breve{ {\rm L}}^{\prime\prime}+  {\rm L}^{\prime\prime}_{\mathfrak R}\right)}.
\eeq

\section{Switching between \texorpdfstring{$\{\hat{H},\hat{{\bm c}},{\bm M}\}$ and $\{{\bm R},{\bm D},{\bm L},d{\bm l},d{\bm r}\}$}{TEXT} system descriptions }
\label{RDLdldrDescription}

In this appendix we give the relationships between $\{{\bm G},{\bm C},{\bm M}\}$ and $\{{\bm R},{\bm D},{\bm L},d{\bm l},d{\bm r}\}$, with the former being the initial parameterization of the linear SME given in \erf{smeGen} and the latter set being a convenient parameterization used for its solution in \erfs{L}{dS}.

To begin, we move to thermo-entangled state representation of \erf{smeGen}, given in \erf{dMap}.  Equating the RHS of \erf{vExp} with \erfs{L}{dS} will allow the relationships between $\{{\bm G},{\bm C},{\bm M}\}$ and $\{{\bm R},{\bm D},{\bm L},d{\bm l},d{\bm r}\}$ to be inferred.  It is clear that we firstly need to convert from the operator $\hat{{\bm c}}$ to $\{\hat{{\bm b}},\hat{{\bm b}}^{\dag}\}$.  Assuming that we are given $\hat{{\bm c}}$ in the form $\hat{{\bm c}}={\bm C}\hat{{\bm x}}$, we need to express $\hat{{\bm x}}$ in terms of $\{\hat{{\bm b}},\hat{{\bm b}}^{\dag}\}$.  To this end, we write
\beq
\hat{{\bm x}}=\underline{{\bm X}}\hat{{\bm b}}+\underline{{\bm X}}^{*}\hat{{\bm b}}^{\ddagger},
\eeq
with $\underline{{\bm X}}$ a $2N\times 2N$ matrix whose elements are defined by
\beq
   \sqrt{2} \underline{{\bm X}}_{mn}=
    \begin{cases}
      1, & \text{if}\ m=2n-1 \\
      -i, & \text{if}\ m=2n \\
      0  & \text{otherwise}
    \end{cases},
\eeq
which, of course, gives
\beq
\hat{{\bm c}}=
{\bm C}\left(\underline{{\bm X}}\hat{{\bm b}}+\underline{{\bm X}}^{*}\hat{{\bm b}}^{\ddagger}\right).
\eeq
We have used the underline notation to differentiate $\underline{{\bm X}}$ from the similarly purposed ${\bm X}$, that is defined differently in \erf{XMat}.
It is convenient to define the matrix
\beq
\bar{\pmb{I}}=
\left[\begin{array}{cc}
    {\bm 0}     & \eye_{N} \\
  \eye_{N}	 & {\bm 0} 
\end{array}\right],
\label{iBar}
\eeq
such that 
\beq
\tilde{{\bm b}}=\bar{\pmb{I}}\hat{{\bm b}}.
\eeq
For compactness, we also define 
\bqa
{\bm B}&=&{\bm C}^{\dag}{\bm C}\\
{\bm F}&=&{\bm C}^{{\rm T}}{\bm M}^{*}{\bm M}^{\dag}{\bm C}\\
{\bm K}&=&{\bm C}^{{\rm T}}{\bm M}^{*}{\bm M}^{{\rm T}}{\bm C}^{*}
\eqa
and, for any matrix ${\bm A}$ (excluding ${\bm A}=\pmb{I}$, for which the barring expression is defined in \erf{iBar}),
\bqa
{\bm A}_{{\rm T}}&=&\underline{{\bm X}}^{{\rm T}}{\bm A}\underline{{\bm X}}\\
{\bm A}_{{\rm t}}&=&\underline{{\bm X}}^{{\rm T}}{\bm A}\underline{{\bm X}}^{*}\\
{\bm A}_{{\rm D}}&=&\underline{{\bm X}}^{\dagger}{\bm A}\underline{{\bm X}}\\
{\bm A}_{{\rm d}}&=&\underline{{\bm X}}^{\dagger}{\bm A}\underline{{\bm X}}^{*}\\
\bar{{\bm A}}&=&\bar{\pmb{I}}{\bm A}\bar{\pmb{I}}.
\eqa
Note that the order of operation is defined as superscript first, then subscript followed lastly by `baring', so that taking the transpose, Hermitian conjugate or complex conjugate will only act on the ${\bm A}$ matrix and leave the $\underline{{\bm X}}$ matrices unaffected.  Conjugation by $\bar{\pmb{I}}$ is performed as a final step.  For example, $\bar{{\bm A}}_{{\rm T}}^{\dag}=\bar{\pmb{I}}\underline{{\bm X}}^{{\rm T}}{\bm A}^{\dag}\underline{{\bm X}}\bar{\pmb{I}}$.  It is then a straightforward task to express \erf{vExp} in terms of $\{\hat{{\bm b}},\hat{{\bm b}}^{\dag}\}$ and then collect terms in order to compare with \erfs{L}{dS}. The results are
\bqa
{\bm L}&=&
-\frac{i}{2}\left({\bm G}_{{\rm T}}-\bar{{\bm G}}_{{\rm d}}\right)
+\bar{\pmb{I}}{\bm B}_{{\rm D}}-\smallhalf {\bm B}_{{\rm T}}\nonumber\\
&&-\smallhalf\left( \bar{{\bm B}}_{{\rm d}}^{*}
+{\bm F}_{{\rm T}}+{\bm K}_{{\rm t}}\bar{\pmb{I}}
+\bar{\pmb{I}}{\bm K}_{{\rm D}}^{*}+\bar{{\bm F}}_{{\rm d}}^{\dag}\right)\\
{\bm R}&=&
-\frac{i}{2}\left({\bm G}_{{\rm d}}-\bar{{\bm G}}_{{\rm T}}\right)
+\bar{\pmb{I}}{\bm B}_{{\rm t}}-\smallhalf {\bm B}_{{\rm d}}\nonumber\\
&&
-\smallhalf \bar{{\bm B}}_{{\rm T}}^{*}
+{\bm F}_{{\rm d}}+{\bm K}_{{\rm D}}\bar{\pmb{I}}
+\bar{\pmb{I}}{\bm K}_{{\rm t}}^{*}+\bar{{\bm F}}_{{\rm T}}^{\dag}\\
{\bm D}&=&
-i\left({\bm G}_{{\rm D}}-\bar{{\bm G}}_{{\rm t}}\right)
+\bar{\pmb{I}}{\bm B}_{{\rm T}}+{\bm B}_{{\rm d}}^{*}\bar{\pmb{I}}\nonumber\\
&&-\smallhalf\left({\bm B}_{{\rm D}}+{\bm B}_{{\rm D}}^{*}+
\bar{{\bm B}}_{{\rm t}}+\bar{{\bm B}}_{{\rm t}}^{*} \right)\nonumber\\
&&-{\bm F}_{{\rm D}}-{\bm K}_{{\rm d}}\bar{\pmb{I}}
-\bar{\pmb{I}}{\bm K}_{{\rm T}}^{*}-\bar{{\bm F}}_{{\rm t}}^{\dag}\\
d{\bm l}&=&{\bm y}^{{\rm T}}dt\left({\bm M}^{\dag}{\bm C} \underline{{\bm X}} 
+{\bm M}^{{\rm T}}{\bm C}^{*} \underline{{\bm X}} ^{*}\bar{\pmb{I}}
\right)\\
d{\bm r}^{{\rm T}}&=&{\bm y}^{{\rm T}}dt\left({\bm M}^{\dag}{\bm C} \underline{{\bm X}} ^{*}
+{\bm M}^{{\rm T}}{\bm C}^{*} \underline{{\bm X}}\bar{\pmb{I}}\right).
\eqa

\section{Integrating out \texorpdfstring{$h$}{TEXT} leaves Gaussian statistics for \texorpdfstring{${\bf d}$}{TEXT}}
\label{intAOut}

We wish to show that for an initial multi-mode coherent state, ${\bf d}$ has Gaussian statistics.  The reason for this choice of initial state is that it evaluates the $Q$-function of the POVM.  If the $Q$-function is Gaussian, then the POVM is Gaussian, by definition. \blk
From \erf{pNoA}, assuming $\rho_{0}=\ket{{\bm \alpha}_{0}}\bra{{\bm \alpha}_{0}}$,
\bqa
\wp( {\bf d}|{\bm \alpha}_{0})&=&
e^{\Delta+{\bm \alpha}_{0}^{\dag}{\bf d}+{\bm \alpha}_{0}^{\dag}{\mathbf L}^{*}{\bm \alpha}_{0}^{\ddagger}
+2{\bm \alpha}_{0}^{\dag}\breve{{\bf L}}{\bm \alpha}_{0} +
{\bm \alpha}_{0}^{{\rm T}}{\mathbf L}{\bm \alpha}_{0}+{\bf d}^{\dag}{\bm \alpha}_{0}}
\nonumber\\
&&\times\int
e^{h}\wp_{{\rm ost}}(h, {\bf d})d^{2}h.
\eqa
The first exponential contains terms that are linear in ${\bf d}$, which will shift the Gaussian mean of the distribution (provided that the remaining factors are Gaussian, of course), and also terms independent of ${\bf d}$, which contribute to the normalization only.  Thus, it is sufficient to show that 
\beq
\wp^{\prime}( {\bf d}|{\bm \alpha}_{0})=
\int
e^{h}\wp_{{\rm ost}}(h, {\bf d})d^{2}h
\label{thisGauss}
\eeq
is Gaussian in ${\bf d}$, in order to prove that $\wp( {\bf d}|{\bm \alpha}_{0})$ is itself Gaussian in ${\bf d}$.

Let us begin by writing the ostensible distribution for the random variables $\{h, {\bf d}\}$ as an integral over all possible measurement records:
\beq
\wp_{{\rm ost}}(h, {\bf d})=
\int \wp_{{\rm ost}}({\bm Y})
{\bm \delta}^{2}\left(h-h_{s} \right)
{\bm \delta}^{2}\left({\bf d}-{\bf d}_{s} \right)
{\bm Y}\bm{dt},
\label{pOstHD}
\eeq
where $\{h_{s}, {\bf d}_{s}\}$ are complex valued stochastic integrals, detailed in \erf{defineh} (which defines $h_{s}$) and \erf{R}.  The measurement record over all time is represented by ${\bm Y}$, although there has been a minor abuse of notation.  In \erf{pOstHD}, we use ${\bm Y}\bm{dt}$ to represents the product of the infinitesimal quantities that compose the set ${\bm Y}\bm{dt}$.
The bold font Dirac delta function represents a product of delta functions, two for each of the complex-valued vector argument's components; for an arbitrary vector ${\bm v}$, of length $N$, it is given by
\beq
{\bm \delta}({\bm v})=\delta({\bm v}_{1})\delta({\bm v}_{2}),...,\delta({\bm v}_{N}).
\eeq
As can be seen from \erf{yost} and \erf{yost222}, the expression for $\wp_{{\rm ost}}({\bm Y})$ is given by the product of Gaussians
\beq
\wp_{{\rm ost}}({\bm Y})=
\left(\frac{dt}{2\pi} \right)^{JL}\exp\left[-\smallhalf 
 \sum\limits_{j=1}^{J}{\bm y}_{j}^{{\rm T}}{\bm y}_{j} dt
\right],
\label{fullRecDiscrete}
\eeq
where ${\bm y}_{j}\equiv\left( {\bm y}_{j,1}{\bm y}_{j,2}...{\bm y}_{j,2L}\right)^{{\rm T}}$ is a column vector of measurement results at the time $jdt$ corresponding to the monitoring of the $L$ Lindblad channels. 

Substituting \erf{pOstHD} into \erf{thisGauss}, we see that the integrals over $h$ are collapsed by the Dirac delta functions. This leaves
\beq
\wp^{\prime}( {\bf d}|{\bm \alpha}_{0})=
\int 
e^{h_{s}}\wp_{{\rm ost}}({\bm Y})
{\bm \delta}^{2}\left({\bf d}-{\bf d}_{s} \right)
{\bm Y}\bm{dt}.
\label{pPrime}
\eeq
To progress, the integral is discretized into a very large number, $J$, of time slices, such that $t=Jd t$.   
The stochastic integrals $\{h_{s}, {\bf d}_{s}\}$ can also be discretized.  We drop any deterministic contributions to $\{h_{s}, {\bf d}_{s}\}$ that affect the normalization of $\wp( {\bf d}|{\bm \alpha}_{0})$ only (as opposed to its moments) and express them as  
\bqa
h_{s}&=&
\sum_{j,k=1}^{J}{\bm y}_{j}^{{\rm T}}dt{\bm H}_{j,k}{\bm y}_{k}dt\label{hQuad}\\
{\bf d}_{s}&=&\sum_{j=1}^{J}{\bm D}_{j}{\bm y}_{j}dt,
\eqa
where the newly introduced ${\bm H}_{j,k}$ and ${\bm D}_{j}$ are matrices of dimension $2L\times 2L$ and $N\times 2L$ respectively (for each $j,k$).

From \erf{hQuad}, we see that $h_{s}$ is the sum of chi-squared random variables.
As we are only trying to prove that ${\bf d}$ has Gaussian statistics, and not find what the mean and variance actually are, we do not try to specify $\{{\bm H}_{j,k},{\bm D}_{j} \}$, apart from noting that they are deterministic.  The dimensions of ${\bm H}_{j,k}$, for given $j,k$, are $2L\times 2L$, while ${\bm D}_{j}$ is an $N\times 2L$ matrix.  

The remaining $2N$ Dirac delta functions in \erf{pPrime} can be used to collapse a further $2N$ of the measurement record integrals.
For simplicity, we choose to collapse the first time slice, corresponding to the integrals over ${\bm y}_{1}$, and also only consider $N=L$, so that all the ${\bm y}_{1}$ integrals are collapsed (and no others).  The case where $N\neq L$ presents only further notational difficulties.  For simplicity, we assume that ${\bm D}_{1}$ has a left inverse, so that the Dirac delta functions collapse the integrals onto the following value 
\beq
{\bm y}_{1}dt={\bm D}_{1}^{-1}\left({\bf d}-\sum_{j=2}^{J}{\bm D}_{j}{\bm y}_{j}dt\right).
\eeq
The important point is that now ${\bm y}_{1}$ is set as a {\it linear} function of both ${\bf d}$ and the remaining measurement records.  In \erf{fullRecDiscrete} and \erf{hQuad}, there are terms that are respectively linear and quadratic (and independent) in ${\bm y}_{1}$, which will lead to terms that are linear and quadratic in ${\bf d}$.  Substituting for ${\bm y}_{1}$ in \erf{fullRecDiscrete} and \erf{hQuad}, and then placing these expressions back into \erf{pPrime}, we obtain the following form
\bqa
\wp^{\prime}( {\bf d}|{\bm \alpha}_{0})&=&
\int 
\exp\left[
-\smallhalf\sum_{j,k=2}^{J}{\bm y}_{j}^{{\rm T}}dt{\bm U}_{j,k}{\bm y}_{k}dt\right.
\nonumber\\
&&\left.+\sum_{j=2}^{J}{\bm v}_{j}^{{\rm T}}({\bf d}){\bm y}_{j}dt+w({\bf d})
\right]
{\bm Y}^{\prime}{\bf d}{\bm t},
\label{pPrimeNoDirac}
\eqa
with ${\bm Y}^{\prime}\equiv {\bm y}_{2},...,{\bm y}_{J}$.  The (here unspecified) $\{{\bm U}_{j,k},{\bm v}_{j}({\bf d}),w({\bf d})\}$ are independent of the measurement record, but $\{{\bm v}_{j}({\bf d}),w({\bf d})\}$ do have dependence upon ${\bf d}$.  ${\bm v}_{j}({\bf d})$ will be at most linear in ${\bf d}$, while $w({\bf d})$ will be at most quadratic.  To make contact with standard multidimensional Gaussian integrals, we re-express \erf{pPrimeNoDirac} as
\bqa
\wp^{\prime}( {\bf d}|{\bm \alpha}_{0})&=&
\int 
\exp\left[
-\smallhalf {\bm y}^{\prime{\rm T}}dt{\bm U}^{\prime}{\bm y}^{\prime}dt\right.\nonumber\\
&&\left.+{\bm v}^{\prime {\rm T}}({\bf d}){\bm y}^{\prime}dt+w({\bf d})
\right]
{\bm Y}^{\prime}\bm{dt},
\label{pPrimeNoDirac2}
\eqa
where the dimensions of ${\bm y}^{\prime},{\bm U}^{\prime},{\bm v}^{\prime}$ are respectively $2L(J-1)\times 1,2L(J-1)\times 2L(J-1), 1\times2L(J-1)$.  The multidimensional Gaussian integral in \erf{pPrimeNoDirac2} can be evaluated, giving
\bqa
\wp^{\prime}( {\bf d}|{\bm \alpha}_{0})&=&
\sqrt{\frac{(2\pi)^{2L(J-1)}}{{\rm det}\,\,{\bm U}^{\prime}}}\nonumber\\
&&\times\exp\left[ 
w({\bf d})+\smallhalf {\bm v}^{\prime {\rm T}}({\bf d}) {\bm U}^{\prime -1}{\bm v}^{\prime}({\bf d})
\right].\nonumber\\
\eqa
Due to ${\bm v}^{\prime {\rm T}}({\bf d})$ being linear in ${\bf d}$ and $w({\bf d})$ being quadratic in ${\bf d}$, $\wp^{\prime}( {\bf d}|{\bm \alpha}_{0})$ is a Gaussian distribution, which in turn, implies that $\wp( {\bf d}|{\bm \alpha}_{0})$ is also Gaussian in ${\bf d}$.

\vspace{1cm}

\bibliographystyle{apsrev4-1}
\bibliography{bibliographySubmit}

\begin{thebibliography}{116}%
\makeatletter
\providecommand \@ifxundefined [1]{%
 \@ifx{#1\undefined}
}%
\providecommand \@ifnum [1]{%
 \ifnum #1\expandafter \@firstoftwo
 \else \expandafter \@secondoftwo
 \fi
}%
\providecommand \@ifx [1]{%
 \ifx #1\expandafter \@firstoftwo
 \else \expandafter \@secondoftwo
 \fi
}%
\providecommand \natexlab [1]{#1}%
\providecommand \enquote  [1]{``#1''}%
\providecommand \bibnamefont  [1]{#1}%
\providecommand \bibfnamefont [1]{#1}%
\providecommand \citenamefont [1]{#1}%
\providecommand \href@noop [0]{\@secondoftwo}%
\providecommand \href [0]{\begingroup \@sanitize@url \@href}%
\providecommand \@href[1]{\@@startlink{#1}\@@href}%
\providecommand \@@href[1]{\endgroup#1\@@endlink}%
\providecommand \@sanitize@url [0]{\catcode `\\12\catcode `\$12\catcode
  `\&12\catcode `\#12\catcode `\^12\catcode `\_12\catcode `\%12\relax}%
\providecommand \@@startlink[1]{}%
\providecommand \@@endlink[0]{}%
\providecommand \url  [0]{\begingroup\@sanitize@url \@url }%
\providecommand \@url [1]{\endgroup\@href {#1}{\urlprefix }}%
\providecommand \urlprefix  [0]{URL }%
\providecommand \Eprint [0]{\href }%
\providecommand \doibase [0]{http://dx.doi.org/}%
\providecommand \selectlanguage [0]{\@gobble}%
\providecommand \bibinfo  [0]{\@secondoftwo}%
\providecommand \bibfield  [0]{\@secondoftwo}%
\providecommand \translation [1]{[#1]}%
\providecommand \BibitemOpen [0]{}%
\providecommand \bibitemStop [0]{}%
\providecommand \bibitemNoStop [0]{.\EOS\space}%
\providecommand \EOS [0]{\spacefactor3000\relax}%
\providecommand \BibitemShut  [1]{\csname bibitem#1\endcsname}%
\let\auto@bib@innerbib\@empty
\bibitem [{\citenamefont {Belavkin}(1999)}]{belavkin1999measurement}%
  \BibitemOpen
  \bibfield  {author} {\bibinfo {author} {\bibfnamefont {V.}~\bibnamefont
  {Belavkin}},\ }\href@noop {} {\bibfield  {journal} {\bibinfo  {journal} {Rep.
  Math. Phys.}\ }\textbf {\bibinfo {volume} {43}},\ \bibinfo {pages} {A405}
  (\bibinfo {year} {1999})}\BibitemShut {NoStop}%
\bibitem [{\citenamefont {Belavkin}\ and\ \citenamefont
  {Staszewski}(1989)}]{belavkin1989quantum}%
  \BibitemOpen
  \bibfield  {author} {\bibinfo {author} {\bibfnamefont {V.}~\bibnamefont
  {Belavkin}}\ and\ \bibinfo {author} {\bibfnamefont {P.}~\bibnamefont
  {Staszewski}},\ }\href@noop {} {\bibfield  {journal} {\bibinfo  {journal}
  {Phys. Lett. A}\ }\textbf {\bibinfo {volume} {140}},\ \bibinfo {pages} {359}
  (\bibinfo {year} {1989})}\BibitemShut {NoStop}%
\bibitem [{\citenamefont {Carmichael}(1600)}]{CarQTraj}%
  \BibitemOpen
  \bibfield  {author} {\bibinfo {author} {\bibfnamefont {H.}~\bibnamefont
  {Carmichael}},\ }\href@noop {} {\emph {\bibinfo {title} {An Open Systems
  Approach to Quantum Optics: Lectures Presented at the Université Libre de
  Bruxelles, October 28 to November 4, 1991 (Lecture Notes in Physics
  Monographs) 1993 edition by Carmichael, Howard (2014) Paperback}}},\ \bibinfo
  {edition} {1993rd}\ ed.\ (\bibinfo  {publisher} {Springer},\ \bibinfo {year}
  {1600})\BibitemShut {NoStop}%
\bibitem [{\citenamefont {Wiseman}\ and\ \citenamefont
  {Milburn}(2014)}]{WisMil10}%
  \BibitemOpen
  \bibfield  {author} {\bibinfo {author} {\bibfnamefont {H.~M.}\ \bibnamefont
  {Wiseman}}\ and\ \bibinfo {author} {\bibfnamefont {G.~J.}\ \bibnamefont
  {Milburn}},\ }\href {http://amazon.com/o/ASIN/1107424151/} {\emph {\bibinfo
  {title} {Quantum Measurement and Control}}},\ \bibinfo {edition} {1st}\ ed.\
  (\bibinfo  {publisher} {Cambridge University Press},\ \bibinfo {year}
  {2014})\BibitemShut {NoStop}%
\bibitem [{\citenamefont {Wiseman}(1996)}]{wisQTraj}%
  \BibitemOpen
  \bibfield  {author} {\bibinfo {author} {\bibfnamefont {H.~M.}\ \bibnamefont
  {Wiseman}},\ }\href {http://stacks.iop.org/1355-5111/8/i=1/a=015} {\bibfield
  {journal} {\bibinfo  {journal} {Quantum and Semiclass. Opt.}\ }\textbf
  {\bibinfo {volume} {8}},\ \bibinfo {pages} {205} (\bibinfo {year}
  {1996})}\BibitemShut {NoStop}%
\bibitem [{\citenamefont {Herkommer}\ \emph {et~al.}(1996)\citenamefont
  {Herkommer}, \citenamefont {Carmichael},\ and\ \citenamefont
  {Schleich}}]{herkommer1996localization}%
  \BibitemOpen
  \bibfield  {author} {\bibinfo {author} {\bibfnamefont {A.}~\bibnamefont
  {Herkommer}}, \bibinfo {author} {\bibfnamefont {H.}~\bibnamefont
  {Carmichael}}, \ and\ \bibinfo {author} {\bibfnamefont {W.}~\bibnamefont
  {Schleich}},\ }\href@noop {} {\bibfield  {journal} {\bibinfo  {journal}
  {Quantum and Semiclass. Opt.}\ }\textbf {\bibinfo {volume} {8}},\ \bibinfo
  {pages} {189} (\bibinfo {year} {1996})}\BibitemShut {NoStop}%
\bibitem [{\citenamefont {Jacobs}\ and\ \citenamefont
  {Steck}(2006{\natexlab{a}})}]{jacSteck}%
  \BibitemOpen
  \bibfield  {author} {\bibinfo {author} {\bibfnamefont {K.}~\bibnamefont
  {Jacobs}}\ and\ \bibinfo {author} {\bibfnamefont {D.~A.}\ \bibnamefont
  {Steck}},\ }\href@noop {} {\bibfield  {journal} {\bibinfo  {journal}
  {Contemp. Phys.}\ }\textbf {\bibinfo {volume} {47}},\ \bibinfo {pages} {279}
  (\bibinfo {year} {2006}{\natexlab{a}})}\BibitemShut {NoStop}%
\bibitem [{\citenamefont {Brun}(2002)}]{brun2002simple}%
  \BibitemOpen
  \bibfield  {author} {\bibinfo {author} {\bibfnamefont {T.~A.}\ \bibnamefont
  {Brun}},\ }\href@noop {} {\bibfield  {journal} {\bibinfo  {journal} {Am. J.
  Phys.}\ }\textbf {\bibinfo {volume} {70}},\ \bibinfo {pages} {719} (\bibinfo
  {year} {2002})}\BibitemShut {NoStop}%
\bibitem [{\citenamefont {Caves}\ and\ \citenamefont
  {Milburn}(1987)}]{caves1987quantum}%
  \BibitemOpen
  \bibfield  {author} {\bibinfo {author} {\bibfnamefont {C.~M.}\ \bibnamefont
  {Caves}}\ and\ \bibinfo {author} {\bibfnamefont {G.~J.}\ \bibnamefont
  {Milburn}},\ }\href@noop {} {\bibfield  {journal} {\bibinfo  {journal} {Phys.
  Rev. A}\ }\textbf {\bibinfo {volume} {36}},\ \bibinfo {pages} {5543}
  (\bibinfo {year} {1987})}\BibitemShut {NoStop}%
\bibitem [{\citenamefont {Di{\'o}si}(1988)}]{diosi1988continuous}%
  \BibitemOpen
  \bibfield  {author} {\bibinfo {author} {\bibfnamefont {L.}~\bibnamefont
  {Di{\'o}si}},\ }\href@noop {} {\bibfield  {journal} {\bibinfo  {journal}
  {Phys. Lett. A}\ }\textbf {\bibinfo {volume} {129}},\ \bibinfo {pages} {419}
  (\bibinfo {year} {1988})}\BibitemShut {NoStop}%
\bibitem [{\citenamefont {Dalibard}\ \emph {et~al.}(1992)\citenamefont
  {Dalibard}, \citenamefont {Castin},\ and\ \citenamefont
  {M{\o}lmer}}]{dalibard1992wave}%
  \BibitemOpen
  \bibfield  {author} {\bibinfo {author} {\bibfnamefont {J.}~\bibnamefont
  {Dalibard}}, \bibinfo {author} {\bibfnamefont {Y.}~\bibnamefont {Castin}}, \
  and\ \bibinfo {author} {\bibfnamefont {K.}~\bibnamefont {M{\o}lmer}},\
  }\href@noop {} {\bibfield  {journal} {\bibinfo  {journal} {Phys. Rev. Lett.}\
  }\textbf {\bibinfo {volume} {68}},\ \bibinfo {pages} {580} (\bibinfo {year}
  {1992})}\BibitemShut {NoStop}%
\bibitem [{\citenamefont {Daley}(2014)}]{daley2014quantum}%
  \BibitemOpen
  \bibfield  {author} {\bibinfo {author} {\bibfnamefont {A.~J.}\ \bibnamefont
  {Daley}},\ }\href@noop {} {\bibfield  {journal} {\bibinfo  {journal} {Adv.
  Phys.}\ }\textbf {\bibinfo {volume} {63}},\ \bibinfo {pages} {77} (\bibinfo
  {year} {2014})}\BibitemShut {NoStop}%
\bibitem [{\citenamefont {Sarlette}\ and\ \citenamefont
  {Rouchon}(2017)}]{sarlette2017deterministic}%
  \BibitemOpen
  \bibfield  {author} {\bibinfo {author} {\bibfnamefont {A.}~\bibnamefont
  {Sarlette}}\ and\ \bibinfo {author} {\bibfnamefont {P.}~\bibnamefont
  {Rouchon}},\ }\href@noop {} {\bibfield  {journal} {\bibinfo  {journal} {J.
  Math. Phys.}\ }\textbf {\bibinfo {volume} {58}},\ \bibinfo {pages} {062106}
  (\bibinfo {year} {2017})}\BibitemShut {NoStop}%
\bibitem [{\citenamefont {Aspect}\ \emph {et~al.}(1988)\citenamefont {Aspect},
  \citenamefont {Arimondo}, \citenamefont {Kaiser}, \citenamefont
  {Vansteenkiste},\ and\ \citenamefont {Cohen-Tannoudji}}]{PhysRevLett.61.826}%
  \BibitemOpen
  \bibfield  {author} {\bibinfo {author} {\bibfnamefont {A.}~\bibnamefont
  {Aspect}}, \bibinfo {author} {\bibfnamefont {E.}~\bibnamefont {Arimondo}},
  \bibinfo {author} {\bibfnamefont {R.}~\bibnamefont {Kaiser}}, \bibinfo
  {author} {\bibfnamefont {N.}~\bibnamefont {Vansteenkiste}}, \ and\ \bibinfo
  {author} {\bibfnamefont {C.}~\bibnamefont {Cohen-Tannoudji}},\ }\href
  {\doibase 10.1103/PhysRevLett.61.826} {\bibfield  {journal} {\bibinfo
  {journal} {Phys. Rev. Lett.}\ }\textbf {\bibinfo {volume} {61}},\ \bibinfo
  {pages} {826} (\bibinfo {year} {1988})}\BibitemShut {NoStop}%
\bibitem [{\citenamefont {Brune}\ \emph {et~al.}(1996)\citenamefont {Brune},
  \citenamefont {Hagley}, \citenamefont {Dreyer}, \citenamefont {Ma\^{\i}tre},
  \citenamefont {Maali}, \citenamefont {Wunderlich}, \citenamefont {Raimond},\
  and\ \citenamefont {Haroche}}]{PhysRevLett.77.4887}%
  \BibitemOpen
  \bibfield  {author} {\bibinfo {author} {\bibfnamefont {M.}~\bibnamefont
  {Brune}}, \bibinfo {author} {\bibfnamefont {E.}~\bibnamefont {Hagley}},
  \bibinfo {author} {\bibfnamefont {J.}~\bibnamefont {Dreyer}}, \bibinfo
  {author} {\bibfnamefont {X.}~\bibnamefont {Ma\^{\i}tre}}, \bibinfo {author}
  {\bibfnamefont {A.}~\bibnamefont {Maali}}, \bibinfo {author} {\bibfnamefont
  {C.}~\bibnamefont {Wunderlich}}, \bibinfo {author} {\bibfnamefont {J.~M.}\
  \bibnamefont {Raimond}}, \ and\ \bibinfo {author} {\bibfnamefont
  {S.}~\bibnamefont {Haroche}},\ }\href {\doibase 10.1103/PhysRevLett.77.4887}
  {\bibfield  {journal} {\bibinfo  {journal} {Phys. Rev. Lett.}\ }\textbf
  {\bibinfo {volume} {77}},\ \bibinfo {pages} {4887} (\bibinfo {year}
  {1996})}\BibitemShut {NoStop}%
\bibitem [{\citenamefont {Peil}\ and\ \citenamefont
  {Gabrielse}(1999)}]{peil1999observing}%
  \BibitemOpen
  \bibfield  {author} {\bibinfo {author} {\bibfnamefont {S.}~\bibnamefont
  {Peil}}\ and\ \bibinfo {author} {\bibfnamefont {G.}~\bibnamefont
  {Gabrielse}},\ }\href@noop {} {\bibfield  {journal} {\bibinfo  {journal}
  {Phys. Rev. Lett.}\ }\textbf {\bibinfo {volume} {83}},\ \bibinfo {pages}
  {1287} (\bibinfo {year} {1999})}\BibitemShut {NoStop}%
\bibitem [{\citenamefont {Lu}\ \emph {et~al.}(2003)\citenamefont {Lu},
  \citenamefont {Ji}, \citenamefont {Pfeiffer}, \citenamefont {West},\ and\
  \citenamefont {Rimberg}}]{lu2003real}%
  \BibitemOpen
  \bibfield  {author} {\bibinfo {author} {\bibfnamefont {W.}~\bibnamefont
  {Lu}}, \bibinfo {author} {\bibfnamefont {Z.}~\bibnamefont {Ji}}, \bibinfo
  {author} {\bibfnamefont {L.}~\bibnamefont {Pfeiffer}}, \bibinfo {author}
  {\bibfnamefont {K.}~\bibnamefont {West}}, \ and\ \bibinfo {author}
  {\bibfnamefont {A.}~\bibnamefont {Rimberg}},\ }\href@noop {} {\bibfield
  {journal} {\bibinfo  {journal} {Nature}\ }\textbf {\bibinfo {volume} {423}},\
  \bibinfo {pages} {422} (\bibinfo {year} {2003})}\BibitemShut {NoStop}%
\bibitem [{\citenamefont {Elzerman}\ \emph {et~al.}(2004)\citenamefont
  {Elzerman}, \citenamefont {Hanson}, \citenamefont {Van~Beveren},
  \citenamefont {Witkamp}, \citenamefont {Vandersypen},\ and\ \citenamefont
  {Kouwenhoven}}]{elzerman2004single}%
  \BibitemOpen
  \bibfield  {author} {\bibinfo {author} {\bibfnamefont {J.}~\bibnamefont
  {Elzerman}}, \bibinfo {author} {\bibfnamefont {R.}~\bibnamefont {Hanson}},
  \bibinfo {author} {\bibfnamefont {L.~W.}\ \bibnamefont {Van~Beveren}},
  \bibinfo {author} {\bibfnamefont {B.}~\bibnamefont {Witkamp}}, \bibinfo
  {author} {\bibfnamefont {L.}~\bibnamefont {Vandersypen}}, \ and\ \bibinfo
  {author} {\bibfnamefont {L.~P.}\ \bibnamefont {Kouwenhoven}},\ }\href@noop {}
  {\bibfield  {journal} {\bibinfo  {journal} {Nature}\ }\textbf {\bibinfo
  {volume} {430}},\ \bibinfo {pages} {431} (\bibinfo {year}
  {2004})}\BibitemShut {NoStop}%
\bibitem [{\citenamefont {Nagourney}\ \emph {et~al.}(1986)\citenamefont
  {Nagourney}, \citenamefont {Sandberg},\ and\ \citenamefont
  {Dehmelt}}]{nagourney1986shelved}%
  \BibitemOpen
  \bibfield  {author} {\bibinfo {author} {\bibfnamefont {W.}~\bibnamefont
  {Nagourney}}, \bibinfo {author} {\bibfnamefont {J.}~\bibnamefont {Sandberg}},
  \ and\ \bibinfo {author} {\bibfnamefont {H.}~\bibnamefont {Dehmelt}},\
  }\href@noop {} {\bibfield  {journal} {\bibinfo  {journal} {Phys. Rev. Lett.}\
  }\textbf {\bibinfo {volume} {56}},\ \bibinfo {pages} {2797} (\bibinfo {year}
  {1986})}\BibitemShut {NoStop}%
\bibitem [{\citenamefont {Sauter}\ \emph {et~al.}(1986)\citenamefont {Sauter},
  \citenamefont {Neuhauser}, \citenamefont {Blatt},\ and\ \citenamefont
  {Toschek}}]{PhysRevLett.57.1696}%
  \BibitemOpen
  \bibfield  {author} {\bibinfo {author} {\bibfnamefont {T.}~\bibnamefont
  {Sauter}}, \bibinfo {author} {\bibfnamefont {W.}~\bibnamefont {Neuhauser}},
  \bibinfo {author} {\bibfnamefont {R.}~\bibnamefont {Blatt}}, \ and\ \bibinfo
  {author} {\bibfnamefont {P.~E.}\ \bibnamefont {Toschek}},\ }\href {\doibase
  10.1103/PhysRevLett.57.1696} {\bibfield  {journal} {\bibinfo  {journal}
  {Phys. Rev. Lett.}\ }\textbf {\bibinfo {volume} {57}},\ \bibinfo {pages}
  {1696} (\bibinfo {year} {1986})}\BibitemShut {NoStop}%
\bibitem [{\citenamefont {Bergquist}\ \emph {et~al.}(1986)\citenamefont
  {Bergquist}, \citenamefont {Hulet}, \citenamefont {Itano},\ and\
  \citenamefont {Wineland}}]{PhysRevLett.57.1699}%
  \BibitemOpen
  \bibfield  {author} {\bibinfo {author} {\bibfnamefont {J.~C.}\ \bibnamefont
  {Bergquist}}, \bibinfo {author} {\bibfnamefont {R.~G.}\ \bibnamefont
  {Hulet}}, \bibinfo {author} {\bibfnamefont {W.~M.}\ \bibnamefont {Itano}}, \
  and\ \bibinfo {author} {\bibfnamefont {D.~J.}\ \bibnamefont {Wineland}},\
  }\href {\doibase 10.1103/PhysRevLett.57.1699} {\bibfield  {journal} {\bibinfo
   {journal} {Phys. Rev. Lett.}\ }\textbf {\bibinfo {volume} {57}},\ \bibinfo
  {pages} {1699} (\bibinfo {year} {1986})}\BibitemShut {NoStop}%
\bibitem [{\citenamefont {Gleyzes}\ \emph {et~al.}(2007)\citenamefont
  {Gleyzes}, \citenamefont {Kuhr}, \citenamefont {Guerlin}, \citenamefont
  {Bernu}, \citenamefont {Deleglise}, \citenamefont {Hoff}, \citenamefont
  {Brune}, \citenamefont {Raimond},\ and\ \citenamefont
  {Haroche}}]{gleyzes2007quantum}%
  \BibitemOpen
  \bibfield  {author} {\bibinfo {author} {\bibfnamefont {S.}~\bibnamefont
  {Gleyzes}}, \bibinfo {author} {\bibfnamefont {S.}~\bibnamefont {Kuhr}},
  \bibinfo {author} {\bibfnamefont {C.}~\bibnamefont {Guerlin}}, \bibinfo
  {author} {\bibfnamefont {J.}~\bibnamefont {Bernu}}, \bibinfo {author}
  {\bibfnamefont {S.}~\bibnamefont {Deleglise}}, \bibinfo {author}
  {\bibfnamefont {U.~B.}\ \bibnamefont {Hoff}}, \bibinfo {author}
  {\bibfnamefont {M.}~\bibnamefont {Brune}}, \bibinfo {author} {\bibfnamefont
  {J.-M.}\ \bibnamefont {Raimond}}, \ and\ \bibinfo {author} {\bibfnamefont
  {S.}~\bibnamefont {Haroche}},\ }\href@noop {} {\bibfield  {journal} {\bibinfo
   {journal} {Nature}\ }\textbf {\bibinfo {volume} {446}},\ \bibinfo {pages}
  {297} (\bibinfo {year} {2007})}\BibitemShut {NoStop}%
\bibitem [{\citenamefont {Neumann}\ \emph {et~al.}(2010)\citenamefont
  {Neumann}, \citenamefont {Beck}, \citenamefont {Steiner}, \citenamefont
  {Rempp}, \citenamefont {Fedder}, \citenamefont {Hemmer}, \citenamefont
  {Wrachtrup},\ and\ \citenamefont {Jelezko}}]{neumann2010single}%
  \BibitemOpen
  \bibfield  {author} {\bibinfo {author} {\bibfnamefont {P.}~\bibnamefont
  {Neumann}}, \bibinfo {author} {\bibfnamefont {J.}~\bibnamefont {Beck}},
  \bibinfo {author} {\bibfnamefont {M.}~\bibnamefont {Steiner}}, \bibinfo
  {author} {\bibfnamefont {F.}~\bibnamefont {Rempp}}, \bibinfo {author}
  {\bibfnamefont {H.}~\bibnamefont {Fedder}}, \bibinfo {author} {\bibfnamefont
  {P.~R.}\ \bibnamefont {Hemmer}}, \bibinfo {author} {\bibfnamefont
  {J.}~\bibnamefont {Wrachtrup}}, \ and\ \bibinfo {author} {\bibfnamefont
  {F.}~\bibnamefont {Jelezko}},\ }\href@noop {} {\bibfield  {journal} {\bibinfo
   {journal} {Science}\ }\textbf {\bibinfo {volume} {329}},\ \bibinfo {pages}
  {542} (\bibinfo {year} {2010})}\BibitemShut {NoStop}%
\bibitem [{\citenamefont {Sayrin}\ \emph {et~al.}(2011)\citenamefont {Sayrin},
  \citenamefont {Dotsenko}, \citenamefont {Zhou}, \citenamefont {Peaudecerf},
  \citenamefont {Rybarczyk}, \citenamefont {Gleyzes}, \citenamefont {Rouchon},
  \citenamefont {Mirrahimi}, \citenamefont {Amini}, \citenamefont {Brune} \emph
  {et~al.}}]{sayrin2011real}%
  \BibitemOpen
  \bibfield  {author} {\bibinfo {author} {\bibfnamefont {C.}~\bibnamefont
  {Sayrin}}, \bibinfo {author} {\bibfnamefont {I.}~\bibnamefont {Dotsenko}},
  \bibinfo {author} {\bibfnamefont {X.}~\bibnamefont {Zhou}}, \bibinfo {author}
  {\bibfnamefont {B.}~\bibnamefont {Peaudecerf}}, \bibinfo {author}
  {\bibfnamefont {T.}~\bibnamefont {Rybarczyk}}, \bibinfo {author}
  {\bibfnamefont {S.}~\bibnamefont {Gleyzes}}, \bibinfo {author} {\bibfnamefont
  {P.}~\bibnamefont {Rouchon}}, \bibinfo {author} {\bibfnamefont
  {M.}~\bibnamefont {Mirrahimi}}, \bibinfo {author} {\bibfnamefont
  {H.}~\bibnamefont {Amini}}, \bibinfo {author} {\bibfnamefont
  {M.}~\bibnamefont {Brune}},  \emph {et~al.},\ }\href@noop {} {\bibfield
  {journal} {\bibinfo  {journal} {Nature}\ }\textbf {\bibinfo {volume} {477}},\
  \bibinfo {pages} {73} (\bibinfo {year} {2011})}\BibitemShut {NoStop}%
\bibitem [{\citenamefont {Sun}\ \emph {et~al.}(2014)\citenamefont {Sun},
  \citenamefont {Petrenko}, \citenamefont {Leghtas}, \citenamefont {Vlastakis},
  \citenamefont {Kirchmair}, \citenamefont {Sliwa}, \citenamefont {Narla},
  \citenamefont {Hatridge}, \citenamefont {Shankar}, \citenamefont {Blumoff}
  \emph {et~al.}}]{sun2014tracking}%
  \BibitemOpen
  \bibfield  {author} {\bibinfo {author} {\bibfnamefont {L.}~\bibnamefont
  {Sun}}, \bibinfo {author} {\bibfnamefont {A.}~\bibnamefont {Petrenko}},
  \bibinfo {author} {\bibfnamefont {Z.}~\bibnamefont {Leghtas}}, \bibinfo
  {author} {\bibfnamefont {B.}~\bibnamefont {Vlastakis}}, \bibinfo {author}
  {\bibfnamefont {G.}~\bibnamefont {Kirchmair}}, \bibinfo {author}
  {\bibfnamefont {K.}~\bibnamefont {Sliwa}}, \bibinfo {author} {\bibfnamefont
  {A.}~\bibnamefont {Narla}}, \bibinfo {author} {\bibfnamefont
  {M.}~\bibnamefont {Hatridge}}, \bibinfo {author} {\bibfnamefont
  {S.}~\bibnamefont {Shankar}}, \bibinfo {author} {\bibfnamefont
  {J.}~\bibnamefont {Blumoff}},  \emph {et~al.},\ }\href@noop {} {\bibfield
  {journal} {\bibinfo  {journal} {Nature}\ }\textbf {\bibinfo {volume} {511}},\
  \bibinfo {pages} {444} (\bibinfo {year} {2014})}\BibitemShut {NoStop}%
\bibitem [{\citenamefont {Minev}\ \emph {et~al.}(2019)\citenamefont {Minev},
  \citenamefont {Mundhada}, \citenamefont {Shankar}, \citenamefont {Reinhold},
  \citenamefont {Guti{\'e}rrez-J{\'a}uregui}, \citenamefont {Schoelkopf},
  \citenamefont {Mirrahimi}, \citenamefont {Carmichael},\ and\ \citenamefont
  {Devoret}}]{minev2019catch}%
  \BibitemOpen
  \bibfield  {author} {\bibinfo {author} {\bibfnamefont {Z.}~\bibnamefont
  {Minev}}, \bibinfo {author} {\bibfnamefont {S.}~\bibnamefont {Mundhada}},
  \bibinfo {author} {\bibfnamefont {S.}~\bibnamefont {Shankar}}, \bibinfo
  {author} {\bibfnamefont {P.}~\bibnamefont {Reinhold}}, \bibinfo {author}
  {\bibfnamefont {R.}~\bibnamefont {Guti{\'e}rrez-J{\'a}uregui}}, \bibinfo
  {author} {\bibfnamefont {R.}~\bibnamefont {Schoelkopf}}, \bibinfo {author}
  {\bibfnamefont {M.}~\bibnamefont {Mirrahimi}}, \bibinfo {author}
  {\bibfnamefont {H.}~\bibnamefont {Carmichael}}, \ and\ \bibinfo {author}
  {\bibfnamefont {M.}~\bibnamefont {Devoret}},\ }\href@noop {} {\bibfield
  {journal} {\bibinfo  {journal} {Nature}\ ,\ \bibinfo {pages} {1}} (\bibinfo
  {year} {2019})}\BibitemShut {NoStop}%
\bibitem [{\citenamefont {Campagne-Ibarcq}\ \emph {et~al.}(2016)\citenamefont
  {Campagne-Ibarcq}, \citenamefont {Six}, \citenamefont {Bretheau},
  \citenamefont {Sarlette}, \citenamefont {Mirrahimi}, \citenamefont
  {Rouchon},\ and\ \citenamefont {Huard}}]{campagne2016observing}%
  \BibitemOpen
  \bibfield  {author} {\bibinfo {author} {\bibfnamefont {P.}~\bibnamefont
  {Campagne-Ibarcq}}, \bibinfo {author} {\bibfnamefont {P.}~\bibnamefont
  {Six}}, \bibinfo {author} {\bibfnamefont {L.}~\bibnamefont {Bretheau}},
  \bibinfo {author} {\bibfnamefont {A.}~\bibnamefont {Sarlette}}, \bibinfo
  {author} {\bibfnamefont {M.}~\bibnamefont {Mirrahimi}}, \bibinfo {author}
  {\bibfnamefont {P.}~\bibnamefont {Rouchon}}, \ and\ \bibinfo {author}
  {\bibfnamefont {B.}~\bibnamefont {Huard}},\ }\href@noop {} {\bibfield
  {journal} {\bibinfo  {journal} {Phys. Rev. X}\ }\textbf {\bibinfo {volume}
  {6}},\ \bibinfo {pages} {011002} (\bibinfo {year} {2016})}\BibitemShut
  {NoStop}%
\bibitem [{\citenamefont {Vijay}\ \emph {et~al.}(2011)\citenamefont {Vijay},
  \citenamefont {Slichter},\ and\ \citenamefont
  {Siddiqi}}]{vijay2011observation}%
  \BibitemOpen
  \bibfield  {author} {\bibinfo {author} {\bibfnamefont {R.}~\bibnamefont
  {Vijay}}, \bibinfo {author} {\bibfnamefont {D.}~\bibnamefont {Slichter}}, \
  and\ \bibinfo {author} {\bibfnamefont {I.}~\bibnamefont {Siddiqi}},\
  }\href@noop {} {\bibfield  {journal} {\bibinfo  {journal} {Phys. Rev. Lett.}\
  }\textbf {\bibinfo {volume} {106}},\ \bibinfo {pages} {110502} (\bibinfo
  {year} {2011})}\BibitemShut {NoStop}%
\bibitem [{\citenamefont {Murch}\ \emph {et~al.}(2013)\citenamefont {Murch},
  \citenamefont {Weber}, \citenamefont {Macklin},\ and\ \citenamefont
  {Siddiqi}}]{murch2013observing}%
  \BibitemOpen
  \bibfield  {author} {\bibinfo {author} {\bibfnamefont {K.}~\bibnamefont
  {Murch}}, \bibinfo {author} {\bibfnamefont {S.}~\bibnamefont {Weber}},
  \bibinfo {author} {\bibfnamefont {C.}~\bibnamefont {Macklin}}, \ and\
  \bibinfo {author} {\bibfnamefont {I.}~\bibnamefont {Siddiqi}},\ }\href@noop
  {} {\bibfield  {journal} {\bibinfo  {journal} {Nature}\ }\textbf {\bibinfo
  {volume} {502}},\ \bibinfo {pages} {211} (\bibinfo {year}
  {2013})}\BibitemShut {NoStop}%
\bibitem [{\citenamefont {De~Lange}\ \emph {et~al.}(2014)\citenamefont
  {De~Lange}, \citenamefont {Rist{\`e}}, \citenamefont {Tiggelman},
  \citenamefont {Eichler}, \citenamefont {Tornberg}, \citenamefont {Johansson},
  \citenamefont {Wallraff}, \citenamefont {Schouten},\ and\ \citenamefont
  {DiCarlo}}]{de2014reversing}%
  \BibitemOpen
  \bibfield  {author} {\bibinfo {author} {\bibfnamefont {G.}~\bibnamefont
  {De~Lange}}, \bibinfo {author} {\bibfnamefont {D.}~\bibnamefont {Rist{\`e}}},
  \bibinfo {author} {\bibfnamefont {M.}~\bibnamefont {Tiggelman}}, \bibinfo
  {author} {\bibfnamefont {C.}~\bibnamefont {Eichler}}, \bibinfo {author}
  {\bibfnamefont {L.}~\bibnamefont {Tornberg}}, \bibinfo {author}
  {\bibfnamefont {G.}~\bibnamefont {Johansson}}, \bibinfo {author}
  {\bibfnamefont {A.}~\bibnamefont {Wallraff}}, \bibinfo {author}
  {\bibfnamefont {R.}~\bibnamefont {Schouten}}, \ and\ \bibinfo {author}
  {\bibfnamefont {L.}~\bibnamefont {DiCarlo}},\ }\href@noop {} {\bibfield
  {journal} {\bibinfo  {journal} {Phys. Rev. Lett.}\ }\textbf {\bibinfo
  {volume} {112}},\ \bibinfo {pages} {080501} (\bibinfo {year}
  {2014})}\BibitemShut {NoStop}%
\bibitem [{\citenamefont {Huang}\ and\ \citenamefont
  {Sarovar}(2018)}]{huang2018smoothing}%
  \BibitemOpen
  \bibfield  {author} {\bibinfo {author} {\bibfnamefont {Z.}~\bibnamefont
  {Huang}}\ and\ \bibinfo {author} {\bibfnamefont {M.}~\bibnamefont
  {Sarovar}},\ }\href@noop {} {\bibfield  {journal} {\bibinfo  {journal} {Phys.
  Rev. A}\ }\textbf {\bibinfo {volume} {97}},\ \bibinfo {pages} {042106}
  (\bibinfo {year} {2018})}\BibitemShut {NoStop}%
\bibitem [{\citenamefont {Zhang}\ and\ \citenamefont
  {M{\o}lmer}(2017)}]{zhang2017prediction}%
  \BibitemOpen
  \bibfield  {author} {\bibinfo {author} {\bibfnamefont {J.}~\bibnamefont
  {Zhang}}\ and\ \bibinfo {author} {\bibfnamefont {K.}~\bibnamefont
  {M{\o}lmer}},\ }\href@noop {} {\bibfield  {journal} {\bibinfo  {journal}
  {Phys. Rev. A}\ }\textbf {\bibinfo {volume} {96}},\ \bibinfo {pages} {062131}
  (\bibinfo {year} {2017})}\BibitemShut {NoStop}%
\bibitem [{\citenamefont {Braunstein}\ and\ \citenamefont
  {Van~Loock}(2005)}]{braunstein2005quantum}%
  \BibitemOpen
  \bibfield  {author} {\bibinfo {author} {\bibfnamefont {S.~L.}\ \bibnamefont
  {Braunstein}}\ and\ \bibinfo {author} {\bibfnamefont {P.}~\bibnamefont
  {Van~Loock}},\ }\href@noop {} {\bibfield  {journal} {\bibinfo  {journal}
  {Rev. Mod. Phys.}\ }\textbf {\bibinfo {volume} {77}},\ \bibinfo {pages} {513}
  (\bibinfo {year} {2005})}\BibitemShut {NoStop}%
\bibitem [{\citenamefont {Weedbrook}\ \emph {et~al.}(2012)\citenamefont
  {Weedbrook}, \citenamefont {Pirandola}, \citenamefont
  {Garc{\'\i}a-Patr{\'o}n}, \citenamefont {Cerf}, \citenamefont {Ralph},
  \citenamefont {Shapiro},\ and\ \citenamefont
  {Lloyd}}]{weedbrook2012gaussian}%
  \BibitemOpen
  \bibfield  {author} {\bibinfo {author} {\bibfnamefont {C.}~\bibnamefont
  {Weedbrook}}, \bibinfo {author} {\bibfnamefont {S.}~\bibnamefont
  {Pirandola}}, \bibinfo {author} {\bibfnamefont {R.}~\bibnamefont
  {Garc{\'\i}a-Patr{\'o}n}}, \bibinfo {author} {\bibfnamefont {N.~J.}\
  \bibnamefont {Cerf}}, \bibinfo {author} {\bibfnamefont {T.~C.}\ \bibnamefont
  {Ralph}}, \bibinfo {author} {\bibfnamefont {J.~H.}\ \bibnamefont {Shapiro}},
  \ and\ \bibinfo {author} {\bibfnamefont {S.}~\bibnamefont {Lloyd}},\
  }\href@noop {} {\bibfield  {journal} {\bibinfo  {journal} {Rev. Mod. Phys.}\
  }\textbf {\bibinfo {volume} {84}},\ \bibinfo {pages} {621} (\bibinfo {year}
  {2012})}\BibitemShut {NoStop}%
\bibitem [{\citenamefont {Wiseman}\ and\ \citenamefont
  {Doherty}(2005)}]{PhysRevLett.94.070405}%
  \BibitemOpen
  \bibfield  {author} {\bibinfo {author} {\bibfnamefont {H.~M.}\ \bibnamefont
  {Wiseman}}\ and\ \bibinfo {author} {\bibfnamefont {A.~C.}\ \bibnamefont
  {Doherty}},\ }\href {\doibase 10.1103/PhysRevLett.94.070405} {\bibfield
  {journal} {\bibinfo  {journal} {Phys. Rev. Lett.}\ }\textbf {\bibinfo
  {volume} {94}},\ \bibinfo {pages} {070405} (\bibinfo {year}
  {2005})}\BibitemShut {NoStop}%
\bibitem [{\citenamefont {Zhang}\ \emph {et~al.}(2012)\citenamefont {Zhang},
  \citenamefont {Wiederhecker}, \citenamefont {Manipatruni}, \citenamefont
  {Barnard}, \citenamefont {McEuen},\ and\ \citenamefont
  {Lipson}}]{PhysRevLett.109.233906}%
  \BibitemOpen
  \bibfield  {author} {\bibinfo {author} {\bibfnamefont {M.}~\bibnamefont
  {Zhang}}, \bibinfo {author} {\bibfnamefont {G.~S.}\ \bibnamefont
  {Wiederhecker}}, \bibinfo {author} {\bibfnamefont {S.}~\bibnamefont
  {Manipatruni}}, \bibinfo {author} {\bibfnamefont {A.}~\bibnamefont
  {Barnard}}, \bibinfo {author} {\bibfnamefont {P.}~\bibnamefont {McEuen}}, \
  and\ \bibinfo {author} {\bibfnamefont {M.}~\bibnamefont {Lipson}},\ }\href
  {\doibase 10.1103/PhysRevLett.109.233906} {\bibfield  {journal} {\bibinfo
  {journal} {Phys. Rev. Lett.}\ }\textbf {\bibinfo {volume} {109}},\ \bibinfo
  {pages} {233906} (\bibinfo {year} {2012})}\BibitemShut {NoStop}%
\bibitem [{\citenamefont {Wieczorek}\ \emph {et~al.}(2015)\citenamefont
  {Wieczorek}, \citenamefont {Hofer}, \citenamefont {Hoelscher-Obermaier},
  \citenamefont {Riedinger}, \citenamefont {Hammerer},\ and\ \citenamefont
  {Aspelmeyer}}]{wieczorek2015optimal}%
  \BibitemOpen
  \bibfield  {author} {\bibinfo {author} {\bibfnamefont {W.}~\bibnamefont
  {Wieczorek}}, \bibinfo {author} {\bibfnamefont {S.~G.}\ \bibnamefont
  {Hofer}}, \bibinfo {author} {\bibfnamefont {J.}~\bibnamefont
  {Hoelscher-Obermaier}}, \bibinfo {author} {\bibfnamefont {R.}~\bibnamefont
  {Riedinger}}, \bibinfo {author} {\bibfnamefont {K.}~\bibnamefont {Hammerer}},
  \ and\ \bibinfo {author} {\bibfnamefont {M.}~\bibnamefont {Aspelmeyer}},\
  }\href@noop {} {\bibfield  {journal} {\bibinfo  {journal} {Phys. Rev. Lett.}\
  }\textbf {\bibinfo {volume} {114}},\ \bibinfo {pages} {223601} (\bibinfo
  {year} {2015})}\BibitemShut {NoStop}%
\bibitem [{\citenamefont {Ockeloen-Korppi}\ \emph {et~al.}(2018)\citenamefont
  {Ockeloen-Korppi}, \citenamefont {Damsk{\"a}gg}, \citenamefont
  {Pirkkalainen}, \citenamefont {Asjad}, \citenamefont {Clerk}, \citenamefont
  {Massel}, \citenamefont {Woolley},\ and\ \citenamefont
  {Sillanp{\"a}{\"a}}}]{ockeloen2018stabilized}%
  \BibitemOpen
  \bibfield  {author} {\bibinfo {author} {\bibfnamefont {C.}~\bibnamefont
  {Ockeloen-Korppi}}, \bibinfo {author} {\bibfnamefont {E.}~\bibnamefont
  {Damsk{\"a}gg}}, \bibinfo {author} {\bibfnamefont {J.-M.}\ \bibnamefont
  {Pirkkalainen}}, \bibinfo {author} {\bibfnamefont {M.}~\bibnamefont {Asjad}},
  \bibinfo {author} {\bibfnamefont {A.}~\bibnamefont {Clerk}}, \bibinfo
  {author} {\bibfnamefont {F.}~\bibnamefont {Massel}}, \bibinfo {author}
  {\bibfnamefont {M.}~\bibnamefont {Woolley}}, \ and\ \bibinfo {author}
  {\bibfnamefont {M.}~\bibnamefont {Sillanp{\"a}{\"a}}},\ }\href@noop {}
  {\bibfield  {journal} {\bibinfo  {journal} {Nature}\ }\textbf {\bibinfo
  {volume} {556}},\ \bibinfo {pages} {478} (\bibinfo {year}
  {2018})}\BibitemShut {NoStop}%
\bibitem [{\citenamefont {Kohler}\ \emph {et~al.}(2018)\citenamefont {Kohler},
  \citenamefont {Gerber}, \citenamefont {Dowd},\ and\ \citenamefont
  {Stamper-Kurn}}]{kohler2018negative}%
  \BibitemOpen
  \bibfield  {author} {\bibinfo {author} {\bibfnamefont {J.}~\bibnamefont
  {Kohler}}, \bibinfo {author} {\bibfnamefont {J.~A.}\ \bibnamefont {Gerber}},
  \bibinfo {author} {\bibfnamefont {E.}~\bibnamefont {Dowd}}, \ and\ \bibinfo
  {author} {\bibfnamefont {D.~M.}\ \bibnamefont {Stamper-Kurn}},\ }\href@noop
  {} {\bibfield  {journal} {\bibinfo  {journal} {Phys. Rev. Lett.}\ }\textbf
  {\bibinfo {volume} {120}},\ \bibinfo {pages} {013601} (\bibinfo {year}
  {2018})}\BibitemShut {NoStop}%
\bibitem [{\citenamefont {Wade}\ \emph {et~al.}(2015)\citenamefont {Wade},
  \citenamefont {Sherson},\ and\ \citenamefont
  {M{\o}lmer}}]{wade2015squeezing}%
  \BibitemOpen
  \bibfield  {author} {\bibinfo {author} {\bibfnamefont {A.~C.}\ \bibnamefont
  {Wade}}, \bibinfo {author} {\bibfnamefont {J.~F.}\ \bibnamefont {Sherson}}, \
  and\ \bibinfo {author} {\bibfnamefont {K.}~\bibnamefont {M{\o}lmer}},\
  }\href@noop {} {\bibfield  {journal} {\bibinfo  {journal} {Phys. Rev. Lett.}\
  }\textbf {\bibinfo {volume} {115}},\ \bibinfo {pages} {060401} (\bibinfo
  {year} {2015})}\BibitemShut {NoStop}%
\bibitem [{\citenamefont {Laverick}\ \emph {et~al.}(2019)\citenamefont
  {Laverick}, \citenamefont {Chantasri},\ and\ \citenamefont
  {Wiseman}}]{laverick2019quantum}%
  \BibitemOpen
  \bibfield  {author} {\bibinfo {author} {\bibfnamefont {K.~T.}\ \bibnamefont
  {Laverick}}, \bibinfo {author} {\bibfnamefont {A.}~\bibnamefont {Chantasri}},
  \ and\ \bibinfo {author} {\bibfnamefont {H.~M.}\ \bibnamefont {Wiseman}},\
  }\href@noop {} {\bibfield  {journal} {\bibinfo  {journal} {Phys. Rev. Lett.}\
  }\textbf {\bibinfo {volume} {122}},\ \bibinfo {pages} {190402} (\bibinfo
  {year} {2019})}\BibitemShut {NoStop}%
\bibitem [{\citenamefont {Petersen}(2010)}]{petersen2010quantum}%
  \BibitemOpen
  \bibfield  {author} {\bibinfo {author} {\bibfnamefont {I.~R.}\ \bibnamefont
  {Petersen}},\ }in\ \href@noop {} {\emph {\bibinfo {booktitle} {Proceedings of
  the 19th international symposium on mathematical theory of networks and
  systems}}}\ (\bibinfo {year} {2010})\BibitemShut {NoStop}%
\bibitem [{\citenamefont {James}\ \emph {et~al.}(2008)\citenamefont {James},
  \citenamefont {Nurdin},\ and\ \citenamefont {Petersen}}]{james2008h}%
  \BibitemOpen
  \bibfield  {author} {\bibinfo {author} {\bibfnamefont {M.~R.}\ \bibnamefont
  {James}}, \bibinfo {author} {\bibfnamefont {H.~I.}\ \bibnamefont {Nurdin}}, \
  and\ \bibinfo {author} {\bibfnamefont {I.~R.}\ \bibnamefont {Petersen}},\
  }\href@noop {} {\bibfield  {journal} {\bibinfo  {journal} {IEEE Trans. Autom.
  Control}\ }\textbf {\bibinfo {volume} {53}},\ \bibinfo {pages} {1787}
  (\bibinfo {year} {2008})}\BibitemShut {NoStop}%
\bibitem [{\citenamefont {Mirrahimi}\ \emph {et~al.}(2014)\citenamefont
  {Mirrahimi}, \citenamefont {Leghtas}, \citenamefont {Albert}, \citenamefont
  {Touzard}, \citenamefont {Schoelkopf}, \citenamefont {Jiang},\ and\
  \citenamefont {Devoret}}]{mirrahimi2014dynamically}%
  \BibitemOpen
  \bibfield  {author} {\bibinfo {author} {\bibfnamefont {M.}~\bibnamefont
  {Mirrahimi}}, \bibinfo {author} {\bibfnamefont {Z.}~\bibnamefont {Leghtas}},
  \bibinfo {author} {\bibfnamefont {V.~V.}\ \bibnamefont {Albert}}, \bibinfo
  {author} {\bibfnamefont {S.}~\bibnamefont {Touzard}}, \bibinfo {author}
  {\bibfnamefont {R.~J.}\ \bibnamefont {Schoelkopf}}, \bibinfo {author}
  {\bibfnamefont {L.}~\bibnamefont {Jiang}}, \ and\ \bibinfo {author}
  {\bibfnamefont {M.~H.}\ \bibnamefont {Devoret}},\ }\href@noop {} {\bibfield
  {journal} {\bibinfo  {journal} {New J. Phys}\ }\textbf {\bibinfo {volume}
  {16}},\ \bibinfo {pages} {045014} (\bibinfo {year} {2014})}\BibitemShut
  {NoStop}%
\bibitem [{\citenamefont {Gough}\ \emph {et~al.}(2010)\citenamefont {Gough},
  \citenamefont {James},\ and\ \citenamefont {Nurdin}}]{gough2010squeezing}%
  \BibitemOpen
  \bibfield  {author} {\bibinfo {author} {\bibfnamefont {J.~E.}\ \bibnamefont
  {Gough}}, \bibinfo {author} {\bibfnamefont {M.~R.}\ \bibnamefont {James}}, \
  and\ \bibinfo {author} {\bibfnamefont {H.~I.}\ \bibnamefont {Nurdin}},\
  }\href@noop {} {\bibfield  {journal} {\bibinfo  {journal} {Phys. Rev. A}\
  }\textbf {\bibinfo {volume} {81}},\ \bibinfo {pages} {023804} (\bibinfo
  {year} {2010})}\BibitemShut {NoStop}%
\bibitem [{\citenamefont {Combes}\ \emph {et~al.}(2017)\citenamefont {Combes},
  \citenamefont {Kerckhoff},\ and\ \citenamefont {Sarovar}}]{combes2017slh}%
  \BibitemOpen
  \bibfield  {author} {\bibinfo {author} {\bibfnamefont {J.}~\bibnamefont
  {Combes}}, \bibinfo {author} {\bibfnamefont {J.}~\bibnamefont {Kerckhoff}}, \
  and\ \bibinfo {author} {\bibfnamefont {M.}~\bibnamefont {Sarovar}},\
  }\href@noop {} {\bibfield  {journal} {\bibinfo  {journal} {Adv. Phys.: X}\
  }\textbf {\bibinfo {volume} {2}},\ \bibinfo {pages} {784} (\bibinfo {year}
  {2017})}\BibitemShut {NoStop}%
\bibitem [{\citenamefont {Wiseman}\ and\ \citenamefont
  {Milburn}(1993)}]{wiseman1993quantum}%
  \BibitemOpen
  \bibfield  {author} {\bibinfo {author} {\bibfnamefont {H.~M.}\ \bibnamefont
  {Wiseman}}\ and\ \bibinfo {author} {\bibfnamefont {G.~J.}\ \bibnamefont
  {Milburn}},\ }\href@noop {} {\bibfield  {journal} {\bibinfo  {journal}
  {Physical review A}\ }\textbf {\bibinfo {volume} {47}},\ \bibinfo {pages}
  {642} (\bibinfo {year} {1993})}\BibitemShut {NoStop}%
\bibitem [{\citenamefont {Wiseman}\ and\ \citenamefont
  {Di{\'o}si}(2001)}]{WisDiosi}%
  \BibitemOpen
  \bibfield  {author} {\bibinfo {author} {\bibfnamefont {H.~M.}\ \bibnamefont
  {Wiseman}}\ and\ \bibinfo {author} {\bibfnamefont {L.}~\bibnamefont
  {Di{\'o}si}},\ }\href@noop {} {\bibfield  {journal} {\bibinfo  {journal}
  {Chem. Phys.}\ }\textbf {\bibinfo {volume} {268}},\ \bibinfo {pages} {91}
  (\bibinfo {year} {2001})}\BibitemShut {NoStop}%
\bibitem [{\citenamefont {Doherty}\ and\ \citenamefont
  {Jacobs}(1999)}]{DohJac}%
  \BibitemOpen
  \bibfield  {author} {\bibinfo {author} {\bibfnamefont {A.~C.}\ \bibnamefont
  {Doherty}}\ and\ \bibinfo {author} {\bibfnamefont {K.}~\bibnamefont
  {Jacobs}},\ }\href@noop {} {\bibfield  {journal} {\bibinfo  {journal} {Phys.
  Rev. A}\ }\textbf {\bibinfo {volume} {60}},\ \bibinfo {pages} {2700}
  (\bibinfo {year} {1999})}\BibitemShut {NoStop}%
\bibitem [{\citenamefont {Verstraete}\ \emph {et~al.}(2009)\citenamefont
  {Verstraete}, \citenamefont {Wolf},\ and\ \citenamefont
  {Cirac}}]{verstraete2009quantum}%
  \BibitemOpen
  \bibfield  {author} {\bibinfo {author} {\bibfnamefont {F.}~\bibnamefont
  {Verstraete}}, \bibinfo {author} {\bibfnamefont {M.~M.}\ \bibnamefont
  {Wolf}}, \ and\ \bibinfo {author} {\bibfnamefont {J.~I.}\ \bibnamefont
  {Cirac}},\ }\href@noop {} {\bibfield  {journal} {\bibinfo  {journal} {Nature
  physics}\ }\textbf {\bibinfo {volume} {5}},\ \bibinfo {pages} {633} (\bibinfo
  {year} {2009})}\BibitemShut {NoStop}%
\bibitem [{\citenamefont {Toth}\ \emph {et~al.}(2017)\citenamefont {Toth},
  \citenamefont {Bernier}, \citenamefont {Nunnenkamp}, \citenamefont
  {Feofanov},\ and\ \citenamefont {Kippenberg}}]{toth2017dissipative}%
  \BibitemOpen
  \bibfield  {author} {\bibinfo {author} {\bibfnamefont {L.~D.}\ \bibnamefont
  {Toth}}, \bibinfo {author} {\bibfnamefont {N.~R.}\ \bibnamefont {Bernier}},
  \bibinfo {author} {\bibfnamefont {A.}~\bibnamefont {Nunnenkamp}}, \bibinfo
  {author} {\bibfnamefont {A.}~\bibnamefont {Feofanov}}, \ and\ \bibinfo
  {author} {\bibfnamefont {T.}~\bibnamefont {Kippenberg}},\ }\href@noop {}
  {\bibfield  {journal} {\bibinfo  {journal} {Nature Physics}\ }\textbf
  {\bibinfo {volume} {13}},\ \bibinfo {pages} {787} (\bibinfo {year}
  {2017})}\BibitemShut {NoStop}%
\bibitem [{\citenamefont {Moore}\ \emph {et~al.}(2017)\citenamefont {Moore},
  \citenamefont {Houhou},\ and\ \citenamefont {Ferraro}}]{moore2017arbitrary}%
  \BibitemOpen
  \bibfield  {author} {\bibinfo {author} {\bibfnamefont {D.~W.}\ \bibnamefont
  {Moore}}, \bibinfo {author} {\bibfnamefont {O.}~\bibnamefont {Houhou}}, \
  and\ \bibinfo {author} {\bibfnamefont {A.}~\bibnamefont {Ferraro}},\
  }\href@noop {} {\bibfield  {journal} {\bibinfo  {journal} {Physical Review
  A}\ }\textbf {\bibinfo {volume} {96}},\ \bibinfo {pages} {022305} (\bibinfo
  {year} {2017})}\BibitemShut {NoStop}%
\bibitem [{\citenamefont {Six}\ \emph {et~al.}(2016)\citenamefont {Six},
  \citenamefont {Campagne-Ibarcq}, \citenamefont {Dotsenko}, \citenamefont
  {Sarlette}, \citenamefont {Huard},\ and\ \citenamefont
  {Rouchon}}]{six2016quantum}%
  \BibitemOpen
  \bibfield  {author} {\bibinfo {author} {\bibfnamefont {P.}~\bibnamefont
  {Six}}, \bibinfo {author} {\bibfnamefont {P.}~\bibnamefont
  {Campagne-Ibarcq}}, \bibinfo {author} {\bibfnamefont {I.}~\bibnamefont
  {Dotsenko}}, \bibinfo {author} {\bibfnamefont {A.}~\bibnamefont {Sarlette}},
  \bibinfo {author} {\bibfnamefont {B.}~\bibnamefont {Huard}}, \ and\ \bibinfo
  {author} {\bibfnamefont {P.}~\bibnamefont {Rouchon}},\ }\href@noop {}
  {\bibfield  {journal} {\bibinfo  {journal} {Phys. Rev. A}\ }\textbf {\bibinfo
  {volume} {93}},\ \bibinfo {pages} {012109} (\bibinfo {year}
  {2016})}\BibitemShut {NoStop}%
\bibitem [{\citenamefont {Chantasri}\ \emph {et~al.}(2019)\citenamefont
  {Chantasri}, \citenamefont {Pang}, \citenamefont {Chalermpusitarak},\ and\
  \citenamefont {Jordan}}]{Chantasri2019}%
  \BibitemOpen
  \bibfield  {author} {\bibinfo {author} {\bibfnamefont {A.}~\bibnamefont
  {Chantasri}}, \bibinfo {author} {\bibfnamefont {S.}~\bibnamefont {Pang}},
  \bibinfo {author} {\bibfnamefont {T.}~\bibnamefont {Chalermpusitarak}}, \
  and\ \bibinfo {author} {\bibfnamefont {A.~N.}\ \bibnamefont {Jordan}},\
  }\href {https://doi.org/10.1007/s40509-019-00198-2} {\bibfield  {journal}
  {\bibinfo  {journal} {Quantum Stud.: Math. Found.}\ } (\bibinfo {year}
  {2019})}\BibitemShut {NoStop}%
\bibitem [{\citenamefont {Warszawski}\ \emph {et~al.}(2019)\citenamefont
  {Warszawski}, \citenamefont {Szorkovszky}, \citenamefont {Bowen},\ and\
  \citenamefont {Doherty}}]{warszawski2019tomography}%
  \BibitemOpen
  \bibfield  {author} {\bibinfo {author} {\bibfnamefont {P.}~\bibnamefont
  {Warszawski}}, \bibinfo {author} {\bibfnamefont {A.}~\bibnamefont
  {Szorkovszky}}, \bibinfo {author} {\bibfnamefont {W.}~\bibnamefont {Bowen}},
  \ and\ \bibinfo {author} {\bibfnamefont {A.}~\bibnamefont {Doherty}},\
  }\href@noop {} {\bibfield  {journal} {\bibinfo  {journal} {New J. Phys}\
  }\textbf {\bibinfo {volume} {21}},\ \bibinfo {pages} {023020} (\bibinfo
  {year} {2019})}\BibitemShut {NoStop}%
\bibitem [{\citenamefont {Myatt}\ \emph {et~al.}(2000)\citenamefont {Myatt},
  \citenamefont {King}, \citenamefont {Turchette}, \citenamefont {Sackett},
  \citenamefont {Kielpinski}, \citenamefont {Itano}, \citenamefont {Monroe},\
  and\ \citenamefont {Wineland}}]{myatt2000decoherence}%
  \BibitemOpen
  \bibfield  {author} {\bibinfo {author} {\bibfnamefont {C.~J.}\ \bibnamefont
  {Myatt}}, \bibinfo {author} {\bibfnamefont {B.~E.}\ \bibnamefont {King}},
  \bibinfo {author} {\bibfnamefont {Q.~A.}\ \bibnamefont {Turchette}}, \bibinfo
  {author} {\bibfnamefont {C.~A.}\ \bibnamefont {Sackett}}, \bibinfo {author}
  {\bibfnamefont {D.}~\bibnamefont {Kielpinski}}, \bibinfo {author}
  {\bibfnamefont {W.~M.}\ \bibnamefont {Itano}}, \bibinfo {author}
  {\bibfnamefont {C.}~\bibnamefont {Monroe}}, \ and\ \bibinfo {author}
  {\bibfnamefont {D.~J.}\ \bibnamefont {Wineland}},\ }\href@noop {} {\bibfield
  {journal} {\bibinfo  {journal} {Nature}\ }\textbf {\bibinfo {volume} {403}},\
  \bibinfo {pages} {269} (\bibinfo {year} {2000})}\BibitemShut {NoStop}%
\bibitem [{\citenamefont {Paz}\ and\ \citenamefont
  {Roncaglia}(2008)}]{paz2008dynamics}%
  \BibitemOpen
  \bibfield  {author} {\bibinfo {author} {\bibfnamefont {J.~P.}\ \bibnamefont
  {Paz}}\ and\ \bibinfo {author} {\bibfnamefont {A.~J.}\ \bibnamefont
  {Roncaglia}},\ }\href@noop {} {\bibfield  {journal} {\bibinfo  {journal}
  {Phys. Rev. Lett.}\ }\textbf {\bibinfo {volume} {100}},\ \bibinfo {pages}
  {220401} (\bibinfo {year} {2008})}\BibitemShut {NoStop}%
\bibitem [{\citenamefont {Goetsch}\ and\ \citenamefont
  {Graham}(1994)}]{GoeGra}%
  \BibitemOpen
  \bibfield  {author} {\bibinfo {author} {\bibfnamefont {P.}~\bibnamefont
  {Goetsch}}\ and\ \bibinfo {author} {\bibfnamefont {R.}~\bibnamefont
  {Graham}},\ }\href@noop {} {\bibfield  {journal} {\bibinfo  {journal} {Phys.
  Rev. A}\ }\textbf {\bibinfo {volume} {50}},\ \bibinfo {pages} {5242}
  (\bibinfo {year} {1994})}\BibitemShut {NoStop}%
\bibitem [{\citenamefont {Jacobs}\ and\ \citenamefont {Knight}(1998)}]{jacLin}%
  \BibitemOpen
  \bibfield  {author} {\bibinfo {author} {\bibfnamefont {K.}~\bibnamefont
  {Jacobs}}\ and\ \bibinfo {author} {\bibfnamefont {P.}~\bibnamefont
  {Knight}},\ }\href@noop {} {\bibfield  {journal} {\bibinfo  {journal} {Phys.
  Rev. A}\ }\textbf {\bibinfo {volume} {57}},\ \bibinfo {pages} {2301}
  (\bibinfo {year} {1998})}\BibitemShut {NoStop}%
\bibitem [{\citenamefont {Jacobs}\ and\ \citenamefont
  {Steck}(2006{\natexlab{b}})}]{jacSte}%
  \BibitemOpen
  \bibfield  {author} {\bibinfo {author} {\bibfnamefont {K.}~\bibnamefont
  {Jacobs}}\ and\ \bibinfo {author} {\bibfnamefont {D.~A.}\ \bibnamefont
  {Steck}},\ }\href@noop {} {\bibfield  {journal} {\bibinfo  {journal}
  {Contemp. Phys.}\ }\textbf {\bibinfo {volume} {47}},\ \bibinfo {pages} {279}
  (\bibinfo {year} {2006}{\natexlab{b}})}\BibitemShut {NoStop}%
\bibitem [{\citenamefont {Gilmore}\ and\ \citenamefont {Yuan}(1989)}]{Gilmore}%
  \BibitemOpen
  \bibfield  {author} {\bibinfo {author} {\bibfnamefont {R.}~\bibnamefont
  {Gilmore}}\ and\ \bibinfo {author} {\bibfnamefont {J.-M.}\ \bibnamefont
  {Yuan}},\ }\href@noop {} {\bibfield  {journal} {\bibinfo  {journal} {J. Chem.
  Phys.}\ }\textbf {\bibinfo {volume} {91}},\ \bibinfo {pages} {917} (\bibinfo
  {year} {1989})}\BibitemShut {NoStop}%
\bibitem [{\citenamefont {Gilmore}\ and\ \citenamefont
  {Yuan}(1987)}]{GilmoreSingle}%
  \BibitemOpen
  \bibfield  {author} {\bibinfo {author} {\bibfnamefont {R.}~\bibnamefont
  {Gilmore}}\ and\ \bibinfo {author} {\bibfnamefont {J.-M.}\ \bibnamefont
  {Yuan}},\ }\href@noop {} {\bibfield  {journal} {\bibinfo  {journal} {J. Chem.
  Phys.}\ }\textbf {\bibinfo {volume} {86}},\ \bibinfo {pages} {130} (\bibinfo
  {year} {1987})}\BibitemShut {NoStop}%
\bibitem [{\citenamefont {Mufti}\ \emph {et~al.}(1993)\citenamefont {Mufti},
  \citenamefont {Schmitt},\ and\ \citenamefont {Sargent~III}}]{Mufti}%
  \BibitemOpen
  \bibfield  {author} {\bibinfo {author} {\bibfnamefont {A.}~\bibnamefont
  {Mufti}}, \bibinfo {author} {\bibfnamefont {H.}~\bibnamefont {Schmitt}}, \
  and\ \bibinfo {author} {\bibfnamefont {M.}~\bibnamefont {Sargent~III}},\
  }\href@noop {} {\bibfield  {journal} {\bibinfo  {journal} {Am. J. Phys.}\
  }\textbf {\bibinfo {volume} {61}},\ \bibinfo {pages} {729} (\bibinfo {year}
  {1993})}\BibitemShut {NoStop}%
\bibitem [{\citenamefont {Ban}(1993)}]{BanLie}%
  \BibitemOpen
  \bibfield  {author} {\bibinfo {author} {\bibfnamefont {M.}~\bibnamefont
  {Ban}},\ }\href@noop {} {\bibfield  {journal} {\bibinfo  {journal} {Phys.
  Rev. A}\ }\textbf {\bibinfo {volume} {47}},\ \bibinfo {pages} {5093}
  (\bibinfo {year} {1993})}\BibitemShut {NoStop}%
\bibitem [{\citenamefont {Galitski}(2011)}]{galit2}%
  \BibitemOpen
  \bibfield  {author} {\bibinfo {author} {\bibfnamefont {V.}~\bibnamefont
  {Galitski}},\ }\href@noop {} {\bibfield  {journal} {\bibinfo  {journal}
  {Phys. Rev. A}\ }\textbf {\bibinfo {volume} {84}},\ \bibinfo {pages} {012118}
  (\bibinfo {year} {2011})}\BibitemShut {NoStop}%
\bibitem [{\citenamefont {Wilson}\ \emph {et~al.}(2012)\citenamefont {Wilson},
  \citenamefont {Fregoso},\ and\ \citenamefont {Galitski}}]{galit1}%
  \BibitemOpen
  \bibfield  {author} {\bibinfo {author} {\bibfnamefont {J.~H.}\ \bibnamefont
  {Wilson}}, \bibinfo {author} {\bibfnamefont {B.~M.}\ \bibnamefont {Fregoso}},
  \ and\ \bibinfo {author} {\bibfnamefont {V.~M.}\ \bibnamefont {Galitski}},\
  }\href@noop {} {\bibfield  {journal} {\bibinfo  {journal} {Phys. Rev. B}\
  }\textbf {\bibinfo {volume} {85}},\ \bibinfo {pages} {174304} (\bibinfo
  {year} {2012})}\BibitemShut {NoStop}%
\bibitem [{dis()}]{disentanglingClarity}%
  \BibitemOpen
  \href@noop {} {}\bibinfo {note} {Operator disentangling is a mathematical
  process unrelated to `entanglement' in the quantum information
  sense.}\BibitemShut {Stop}%
\bibitem [{\citenamefont {Scholz}\ \emph {et~al.}(2010)\citenamefont {Scholz},
  \citenamefont {Voronov},\ and\ \citenamefont {Weyrauch}}]{SchoWey}%
  \BibitemOpen
  \bibfield  {author} {\bibinfo {author} {\bibfnamefont {D.}~\bibnamefont
  {Scholz}}, \bibinfo {author} {\bibfnamefont {V.~G.}\ \bibnamefont {Voronov}},
  \ and\ \bibinfo {author} {\bibfnamefont {M.}~\bibnamefont {Weyrauch}},\
  }\href@noop {} {\bibfield  {journal} {\bibinfo  {journal} {J. Math. Phys.}\
  }\textbf {\bibinfo {volume} {51}},\ \bibinfo {pages} {063513} (\bibinfo
  {year} {2010})}\BibitemShut {NoStop}%
\bibitem [{\citenamefont {Fan}\ and\ \citenamefont
  {Hu}(2009{\natexlab{a}})}]{FANdis}%
  \BibitemOpen
  \bibfield  {author} {\bibinfo {author} {\bibfnamefont {H.-Y.}\ \bibnamefont
  {Fan}}\ and\ \bibinfo {author} {\bibfnamefont {L.-Y.}\ \bibnamefont {Hu}},\
  }\href@noop {} {\bibfield  {journal} {\bibinfo  {journal} {Commun. Theor.
  Phys.}\ }\textbf {\bibinfo {volume} {51}},\ \bibinfo {pages} {321} (\bibinfo
  {year} {2009}{\natexlab{a}})}\BibitemShut {NoStop}%
\bibitem [{\citenamefont {Fan}\ and\ \citenamefont
  {Hu}(2009{\natexlab{b}})}]{FANdis2}%
  \BibitemOpen
  \bibfield  {author} {\bibinfo {author} {\bibfnamefont {H.-Y.}\ \bibnamefont
  {Fan}}\ and\ \bibinfo {author} {\bibfnamefont {L.-Y.}\ \bibnamefont {Hu}},\
  }\href@noop {} {\bibfield  {journal} {\bibinfo  {journal} {Commun. Theor.
  Phys.(Beijing, China)}\ }\textbf {\bibinfo {volume} {51}},\ \bibinfo {pages}
  {506} (\bibinfo {year} {2009}{\natexlab{b}})}\BibitemShut {NoStop}%
\bibitem [{\citenamefont {Vargas-Mart{\'\i}nez}\ \emph
  {et~al.}(2006)\citenamefont {Vargas-Mart{\'\i}nez}, \citenamefont
  {Moya-Cessa},\ and\ \citenamefont {Fern{\'a}ndez~Guasti}}]{VarMoy}%
  \BibitemOpen
  \bibfield  {author} {\bibinfo {author} {\bibfnamefont {J.}~\bibnamefont
  {Vargas-Mart{\'\i}nez}}, \bibinfo {author} {\bibfnamefont {H.}~\bibnamefont
  {Moya-Cessa}}, \ and\ \bibinfo {author} {\bibfnamefont {M.}~\bibnamefont
  {Fern{\'a}ndez~Guasti}},\ }\href@noop {} {\bibfield  {journal} {\bibinfo
  {journal} {Revista mexicana de f{\'\i}sica E}\ }\textbf {\bibinfo {volume}
  {52}},\ \bibinfo {pages} {13} (\bibinfo {year} {2006})}\BibitemShut {NoStop}%
\bibitem [{\citenamefont {Fern{\'a}ndez}(1989)}]{Fer89}%
  \BibitemOpen
  \bibfield  {author} {\bibinfo {author} {\bibfnamefont {F.~M.}\ \bibnamefont
  {Fern{\'a}ndez}},\ }\href@noop {} {\bibfield  {journal} {\bibinfo  {journal}
  {Phys. Rev. A}\ }\textbf {\bibinfo {volume} {40}},\ \bibinfo {pages} {41}
  (\bibinfo {year} {1989})}\BibitemShut {NoStop}%
\bibitem [{\citenamefont {W{\"u}nsche}(2001)}]{Wun01}%
  \BibitemOpen
  \bibfield  {author} {\bibinfo {author} {\bibfnamefont {A.}~\bibnamefont
  {W{\"u}nsche}},\ }\href@noop {} {\bibfield  {journal} {\bibinfo  {journal}
  {J. Opt. B: Quantum and Semiclass. Opt.}\ }\textbf {\bibinfo {volume} {4}},\
  \bibinfo {pages} {1} (\bibinfo {year} {2001})}\BibitemShut {NoStop}%
\bibitem [{\citenamefont {Wilcox}(1967)}]{Wil67}%
  \BibitemOpen
  \bibfield  {author} {\bibinfo {author} {\bibfnamefont {R.}~\bibnamefont
  {Wilcox}},\ }\href@noop {} {\bibfield  {journal} {\bibinfo  {journal} {J.
  Math. Phys.}\ }\textbf {\bibinfo {volume} {8}},\ \bibinfo {pages} {962}
  (\bibinfo {year} {1967})}\BibitemShut {NoStop}%
\bibitem [{\citenamefont {DasGupta}(1996)}]{Das96}%
  \BibitemOpen
  \bibfield  {author} {\bibinfo {author} {\bibfnamefont {A.}~\bibnamefont
  {DasGupta}},\ }\href@noop {} {\bibfield  {journal} {\bibinfo  {journal} {Am.
  J. Phys.}\ }\textbf {\bibinfo {volume} {64}},\ \bibinfo {pages} {1422}
  (\bibinfo {year} {1996})}\BibitemShut {NoStop}%
\bibitem [{\citenamefont {Twamley}(1993)}]{Twa93}%
  \BibitemOpen
  \bibfield  {author} {\bibinfo {author} {\bibfnamefont {J.}~\bibnamefont
  {Twamley}},\ }\href@noop {} {\bibfield  {journal} {\bibinfo  {journal} {Phys.
  Rev. A}\ }\textbf {\bibinfo {volume} {48}},\ \bibinfo {pages} {2627}
  (\bibinfo {year} {1993})}\BibitemShut {NoStop}%
\bibitem [{\citenamefont {Zhou}\ \emph {et~al.}(2011)\citenamefont {Zhou},
  \citenamefont {Fan},\ and\ \citenamefont {Song}}]{ZhoFan}%
  \BibitemOpen
  \bibfield  {author} {\bibinfo {author} {\bibfnamefont {J.}~\bibnamefont
  {Zhou}}, \bibinfo {author} {\bibfnamefont {H.-Y.}\ \bibnamefont {Fan}}, \
  and\ \bibinfo {author} {\bibfnamefont {J.}~\bibnamefont {Song}},\ }\href@noop
  {} {\bibfield  {journal} {\bibinfo  {journal} {Int. J. Theor. Phys.}\
  }\textbf {\bibinfo {volume} {50}},\ \bibinfo {pages} {3149} (\bibinfo {year}
  {2011})}\BibitemShut {NoStop}%
\bibitem [{\citenamefont {Hu}\ and\ \citenamefont {Fan}(2009)}]{FANthermo}%
  \BibitemOpen
  \bibfield  {author} {\bibinfo {author} {\bibfnamefont {L.-Y.}\ \bibnamefont
  {Hu}}\ and\ \bibinfo {author} {\bibfnamefont {H.-Y.}\ \bibnamefont {Fan}},\
  }\href@noop {} {\bibfield  {journal} {\bibinfo  {journal} {Int. J. Theor.
  Phys.}\ }\textbf {\bibinfo {volume} {48}},\ \bibinfo {pages} {3396} (\bibinfo
  {year} {2009})}\BibitemShut {NoStop}%
\bibitem [{\citenamefont {Fan}\ and\ \citenamefont {Lu}(2007)}]{FANthermo2}%
  \BibitemOpen
  \bibfield  {author} {\bibinfo {author} {\bibfnamefont {H.-Y.}\ \bibnamefont
  {Fan}}\ and\ \bibinfo {author} {\bibfnamefont {H.-L.}\ \bibnamefont {Lu}},\
  }\href@noop {} {\bibfield  {journal} {\bibinfo  {journal} {Mod. Phys. Lett.
  B}\ }\textbf {\bibinfo {volume} {21}},\ \bibinfo {pages} {183} (\bibinfo
  {year} {2007})}\BibitemShut {NoStop}%
\bibitem [{\citenamefont {Kosov}(2016)}]{Kosov}%
  \BibitemOpen
  \bibfield  {author} {\bibinfo {author} {\bibfnamefont {D.~S.}\ \bibnamefont
  {Kosov}},\ }\href@noop {} {\bibfield  {journal} {\bibinfo  {journal}
  {arXiv:1605.02170}\ } (\bibinfo {year} {2016})}\BibitemShut {NoStop}%
\bibitem [{\citenamefont {Fan}\ and\ \citenamefont {Fan}(1998)}]{FANthermEta}%
  \BibitemOpen
  \bibfield  {author} {\bibinfo {author} {\bibfnamefont {H.-Y.}\ \bibnamefont
  {Fan}}\ and\ \bibinfo {author} {\bibfnamefont {Y.}~\bibnamefont {Fan}},\
  }\href@noop {} {\bibfield  {journal} {\bibinfo  {journal} {Phys. Lett. A}\
  }\textbf {\bibinfo {volume} {246}},\ \bibinfo {pages} {242} (\bibinfo {year}
  {1998})}\BibitemShut {NoStop}%
\bibitem [{\citenamefont {Arimitsu}\ and\ \citenamefont
  {Umezawa}(1985)}]{UmeAri}%
  \BibitemOpen
  \bibfield  {author} {\bibinfo {author} {\bibfnamefont {T.}~\bibnamefont
  {Arimitsu}}\ and\ \bibinfo {author} {\bibfnamefont {H.}~\bibnamefont
  {Umezawa}},\ }\href@noop {} {\bibfield  {journal} {\bibinfo  {journal} {Prog.
  Theo. Phys.}\ }\textbf {\bibinfo {volume} {74}},\ \bibinfo {pages} {429}
  (\bibinfo {year} {1985})}\BibitemShut {NoStop}%
\bibitem [{\citenamefont {Umezawa}\ \emph {et~al.}(1982)\citenamefont
  {Umezawa}, \citenamefont {Matsumoto},\ and\ \citenamefont
  {Tachiki}}]{Umebook}%
  \BibitemOpen
  \bibfield  {author} {\bibinfo {author} {\bibfnamefont {H.}~\bibnamefont
  {Umezawa}}, \bibinfo {author} {\bibfnamefont {H.}~\bibnamefont {Matsumoto}},
  \ and\ \bibinfo {author} {\bibfnamefont {M.}~\bibnamefont {Tachiki}},\
  }\href@noop {} {\  (\bibinfo {year} {1982})}\BibitemShut {NoStop}%
\bibitem [{\citenamefont {Fan}\ and\ \citenamefont {Hu}(2008)}]{FanHu}%
  \BibitemOpen
  \bibfield  {author} {\bibinfo {author} {\bibfnamefont {H.-Y.}\ \bibnamefont
  {Fan}}\ and\ \bibinfo {author} {\bibfnamefont {L.-Y.}\ \bibnamefont {Hu}},\
  }\href@noop {} {\bibfield  {journal} {\bibinfo  {journal} {Opt. Comm.}\
  }\textbf {\bibinfo {volume} {281}},\ \bibinfo {pages} {5571} (\bibinfo {year}
  {2008})}\BibitemShut {NoStop}%
\bibitem [{\citenamefont {Corney}\ and\ \citenamefont
  {Drummond}(2003)}]{DruCor}%
  \BibitemOpen
  \bibfield  {author} {\bibinfo {author} {\bibfnamefont {J.~F.}\ \bibnamefont
  {Corney}}\ and\ \bibinfo {author} {\bibfnamefont {P.~D.}\ \bibnamefont
  {Drummond}},\ }\href@noop {} {\bibfield  {journal} {\bibinfo  {journal}
  {Phys. Rev. A}\ }\textbf {\bibinfo {volume} {68}},\ \bibinfo {pages} {063822}
  (\bibinfo {year} {2003})}\BibitemShut {NoStop}%
\bibitem [{\citenamefont {Prosen}\ and\ \citenamefont
  {Seligman}(2010)}]{3rdQuant}%
  \BibitemOpen
  \bibfield  {author} {\bibinfo {author} {\bibfnamefont {T.}~\bibnamefont
  {Prosen}}\ and\ \bibinfo {author} {\bibfnamefont {T.~H.}\ \bibnamefont
  {Seligman}},\ }\href@noop {} {\bibfield  {journal} {\bibinfo  {journal} {J.
  Phys. A: Math. Theo.}\ }\textbf {\bibinfo {volume} {43}},\ \bibinfo {pages}
  {392004} (\bibinfo {year} {2010})}\BibitemShut {NoStop}%
\bibitem [{\citenamefont {Bazrafkan}\ \emph {et~al.}(2014)\citenamefont
  {Bazrafkan}, \citenamefont {Ashrafi},\ and\ \citenamefont {Naghdi}}]{BazNag}%
  \BibitemOpen
  \bibfield  {author} {\bibinfo {author} {\bibfnamefont {M.~R.}\ \bibnamefont
  {Bazrafkan}}, \bibinfo {author} {\bibfnamefont {S.~M.}\ \bibnamefont
  {Ashrafi}}, \ and\ \bibinfo {author} {\bibfnamefont {F.}~\bibnamefont
  {Naghdi}},\ }\href@noop {} {\bibfield  {journal} {\bibinfo  {journal} {Chin.
  Phys. Lett.}\ }\textbf {\bibinfo {volume} {31}},\ \bibinfo {pages} {070303}
  (\bibinfo {year} {2014})}\BibitemShut {NoStop}%
\bibitem [{\citenamefont {Ban}(2009)}]{Ban}%
  \BibitemOpen
  \bibfield  {author} {\bibinfo {author} {\bibfnamefont {M.}~\bibnamefont
  {Ban}},\ }\href@noop {} {\bibfield  {journal} {\bibinfo  {journal} {J. Mod.
  Opt.}\ }\textbf {\bibinfo {volume} {56}},\ \bibinfo {pages} {577} (\bibinfo
  {year} {2009})}\BibitemShut {NoStop}%
\bibitem [{\citenamefont {Chia}\ and\ \citenamefont
  {Wiseman}(2011)}]{chia2011complete}%
  \BibitemOpen
  \bibfield  {author} {\bibinfo {author} {\bibfnamefont {A.}~\bibnamefont
  {Chia}}\ and\ \bibinfo {author} {\bibfnamefont {H.~M.}\ \bibnamefont
  {Wiseman}},\ }\href@noop {} {\bibfield  {journal} {\bibinfo  {journal} {Phys.
  Rev. A}\ }\textbf {\bibinfo {volume} {84}},\ \bibinfo {pages} {012119}
  (\bibinfo {year} {2011})}\BibitemShut {NoStop}%
\bibitem [{\citenamefont {Gardiner}(1985)}]{GarBook}%
  \BibitemOpen
  \bibfield  {author} {\bibinfo {author} {\bibfnamefont {C.~W.}\ \bibnamefont
  {Gardiner}},\ }\href@noop {} {\emph {\bibinfo {title} {Stochastic methods}}}\
  (\bibinfo  {publisher} {Springer-Verlag, Berlin--Heidelberg--New
  York--Tokyo},\ \bibinfo {year} {1985})\BibitemShut {NoStop}%
\bibitem [{\citenamefont {Hong-yi}\ and\ \citenamefont
  {Klauder}(1994)}]{PhysRevA.49.704}%
  \BibitemOpen
  \bibfield  {author} {\bibinfo {author} {\bibfnamefont {F.}~\bibnamefont
  {Hong-yi}}\ and\ \bibinfo {author} {\bibfnamefont {J.~R.}\ \bibnamefont
  {Klauder}},\ }\href {\doibase 10.1103/PhysRevA.49.704} {\bibfield  {journal}
  {\bibinfo  {journal} {Phys. Rev. A}\ }\textbf {\bibinfo {volume} {49}},\
  \bibinfo {pages} {704} (\bibinfo {year} {1994})}\BibitemShut {NoStop}%
\bibitem [{\citenamefont {Einstein}\ \emph {et~al.}(1935)\citenamefont
  {Einstein}, \citenamefont {Podolsky},\ and\ \citenamefont
  {Rosen}}]{PhysRev.47.777}%
  \BibitemOpen
  \bibfield  {author} {\bibinfo {author} {\bibfnamefont {A.}~\bibnamefont
  {Einstein}}, \bibinfo {author} {\bibfnamefont {B.}~\bibnamefont {Podolsky}},
  \ and\ \bibinfo {author} {\bibfnamefont {N.}~\bibnamefont {Rosen}},\ }\href
  {\doibase 10.1103/PhysRev.47.777} {\bibfield  {journal} {\bibinfo  {journal}
  {Phys. Rev.}\ }\textbf {\bibinfo {volume} {47}},\ \bibinfo {pages} {777}
  (\bibinfo {year} {1935})}\BibitemShut {NoStop}%
\bibitem [{\citenamefont {Zhang}\ \emph {et~al.}(1990)\citenamefont {Zhang},
  \citenamefont {Gilmore} \emph {et~al.}}]{GilmoreCoh}%
  \BibitemOpen
  \bibfield  {author} {\bibinfo {author} {\bibfnamefont {W.-M.}\ \bibnamefont
  {Zhang}}, \bibinfo {author} {\bibfnamefont {R.}~\bibnamefont {Gilmore}},
  \emph {et~al.},\ }\href@noop {} {\bibfield  {journal} {\bibinfo  {journal}
  {Rev. Mod. Phys.}\ }\textbf {\bibinfo {volume} {62}},\ \bibinfo {pages} {867}
  (\bibinfo {year} {1990})}\BibitemShut {NoStop}%
\bibitem [{\citenamefont {Hall}(2015)}]{hall2015lie}%
  \BibitemOpen
  \bibfield  {author} {\bibinfo {author} {\bibfnamefont {B.}~\bibnamefont
  {Hall}},\ }\href@noop {} {\emph {\bibinfo {title} {Lie groups, Lie algebras,
  and representations: an elementary introduction}}},\ Vol.\ \bibinfo {volume}
  {222}\ (\bibinfo  {publisher} {Springer},\ \bibinfo {year}
  {2015})\BibitemShut {NoStop}%
\bibitem [{\citenamefont {Gilmore}(2012)}]{gilmore2012lie}%
  \BibitemOpen
  \bibfield  {author} {\bibinfo {author} {\bibfnamefont {R.}~\bibnamefont
  {Gilmore}},\ }\href@noop {} {\emph {\bibinfo {title} {Lie groups, Lie
  algebras, and some of their applications}}}\ (\bibinfo  {publisher} {Courier
  Corporation},\ \bibinfo {year} {2012})\BibitemShut {NoStop}%
\bibitem [{not()}]{notationClean}%
  \BibitemOpen
  \href@noop {} {}\bibinfo {note} {The reader may question why \erf{splitExp}
  wasn't chosen with the quadratic exponential to the left of the linear
  exponential, given that the subsequent task is just to move it to the left
  again. The reason is purely to allow a simpler notation when illustrating the
  required reorderings. Specifically, after $j-1$ reorderings the combined
  quadratic exponent will be $j\hat{Q}dt$ and the linear exponent to the left
  will also have time argument $jdt$. This is to be understood in relation to
  \erf{linComm}.}\BibitemShut {Stop}%
\bibitem [{\citenamefont {H-Y}(1989)}]{FAMmulti}%
  \BibitemOpen
  \bibfield  {author} {\bibinfo {author} {\bibfnamefont {F.}~\bibnamefont
  {H-Y}},\ }\href@noop {} {\bibfield  {journal} {\bibinfo  {journal} {J. Phys.
  A: Math. Gen.}\ }\textbf {\bibinfo {volume} {22}},\ \bibinfo {pages} {1193}
  (\bibinfo {year} {1989})}\BibitemShut {NoStop}%
\bibitem [{\citenamefont {Gammelmark}\ \emph {et~al.}(2013)\citenamefont
  {Gammelmark}, \citenamefont {Julsgaard},\ and\ \citenamefont
  {M\o{}lmer}}]{PhysRevLett.111.160401}%
  \BibitemOpen
  \bibfield  {author} {\bibinfo {author} {\bibfnamefont {S.}~\bibnamefont
  {Gammelmark}}, \bibinfo {author} {\bibfnamefont {B.}~\bibnamefont
  {Julsgaard}}, \ and\ \bibinfo {author} {\bibfnamefont {K.}~\bibnamefont
  {M\o{}lmer}},\ }\href {\doibase 10.1103/PhysRevLett.111.160401} {\bibfield
  {journal} {\bibinfo  {journal} {Phys. Rev. Lett.}\ }\textbf {\bibinfo
  {volume} {111}},\ \bibinfo {pages} {160401} (\bibinfo {year}
  {2013})}\BibitemShut {NoStop}%
\bibitem [{\citenamefont {Ban}(2007)}]{ban2007quantum}%
  \BibitemOpen
  \bibfield  {author} {\bibinfo {author} {\bibfnamefont {M.}~\bibnamefont
  {Ban}},\ }\href@noop {} {\bibfield  {journal} {\bibinfo  {journal} {Int. J.
  Theor. Phys.}\ }\textbf {\bibinfo {volume} {46}},\ \bibinfo {pages} {184}
  (\bibinfo {year} {2007})}\BibitemShut {NoStop}%
\bibitem [{\citenamefont {Picinbono}(1996)}]{539051}%
  \BibitemOpen
  \bibfield  {author} {\bibinfo {author} {\bibfnamefont {B.}~\bibnamefont
  {Picinbono}},\ }\href {\doibase 10.1109/78.539051} {\bibfield  {journal}
  {\bibinfo  {journal} {IEEE Trans. Sig. Proc.}\ }\textbf {\bibinfo {volume}
  {44}},\ \bibinfo {pages} {2637} (\bibinfo {year} {1996})}\BibitemShut
  {NoStop}%
\bibitem [{inf()}]{infDim}%
  \BibitemOpen
  \href@noop {} {}\bibinfo {note} {Although complications exist in infinite
  dimensional Hilbert space~\cite{attal2014quantum} regarding the definition of
  the adjoint equation, we note that in all physical systems of interest there
  will some finite dimensional restriction, that accurately models the system,
  for which these difficulties may be resolved.}\BibitemShut {Stop}%
\bibitem [{\citenamefont {Lammers}(2018)}]{lammers2018state}%
  \BibitemOpen
  \bibfield  {author} {\bibinfo {author} {\bibfnamefont {J.}~\bibnamefont
  {Lammers}},\ }\emph {\bibinfo {title} {State preparation and verification in
  continuously measured quantum systems}},\ \href@noop {} {Ph.D. thesis},\
  \bibinfo  {school} {Hannover: Institutionelles Repositorium der Leibniz
  Universit{\"a}t Hannover} (\bibinfo {year} {2018})\BibitemShut {NoStop}%
\bibitem [{\citenamefont {Fraser}\ and\ \citenamefont
  {Potter}(1969)}]{fraser1969optimum}%
  \BibitemOpen
  \bibfield  {author} {\bibinfo {author} {\bibfnamefont {D.}~\bibnamefont
  {Fraser}}\ and\ \bibinfo {author} {\bibfnamefont {J.}~\bibnamefont
  {Potter}},\ }\href@noop {} {\bibfield  {journal} {\bibinfo  {journal} {IEEE
  Transactions on automatic control}\ }\textbf {\bibinfo {volume} {14}},\
  \bibinfo {pages} {387} (\bibinfo {year} {1969})}\BibitemShut {NoStop}%
\bibitem [{\citenamefont {Genoni}\ \emph {et~al.}(2014)\citenamefont {Genoni},
  \citenamefont {Mancini},\ and\ \citenamefont {Serafini}}]{Genoni}%
  \BibitemOpen
  \bibfield  {author} {\bibinfo {author} {\bibfnamefont {M.~G.}\ \bibnamefont
  {Genoni}}, \bibinfo {author} {\bibfnamefont {S.}~\bibnamefont {Mancini}}, \
  and\ \bibinfo {author} {\bibfnamefont {A.}~\bibnamefont {Serafini}},\
  }\href@noop {} {\bibfield  {journal} {\bibinfo  {journal} {Russian J. Math.
  Phys.}\ }\textbf {\bibinfo {volume} {21}},\ \bibinfo {pages} {329} (\bibinfo
  {year} {2014})}\BibitemShut {NoStop}%
\bibitem [{\citenamefont {Aspelmeyer}\ \emph
  {et~al.}(2014{\natexlab{a}})\citenamefont {Aspelmeyer}, \citenamefont
  {Kippenberg},\ and\ \citenamefont {Marquardt}}]{revCavOptMech}%
  \BibitemOpen
  \bibfield  {author} {\bibinfo {author} {\bibfnamefont {M.}~\bibnamefont
  {Aspelmeyer}}, \bibinfo {author} {\bibfnamefont {T.~J.}\ \bibnamefont
  {Kippenberg}}, \ and\ \bibinfo {author} {\bibfnamefont {F.}~\bibnamefont
  {Marquardt}},\ }\href {\doibase 10.1103/RevModPhys.86.1391} {\bibfield
  {journal} {\bibinfo  {journal} {Rev. Mod. Phys.}\ }\textbf {\bibinfo {volume}
  {86}},\ \bibinfo {pages} {1391} (\bibinfo {year}
  {2014}{\natexlab{a}})}\BibitemShut {NoStop}%
\bibitem [{\citenamefont {Aspelmeyer}\ \emph
  {et~al.}(2014{\natexlab{b}})\citenamefont {Aspelmeyer}, \citenamefont
  {Kippenberg},\ and\ \citenamefont {Marquardt}}]{aspelmeyer2014cavity}%
  \BibitemOpen
  \bibfield  {author} {\bibinfo {author} {\bibfnamefont {M.}~\bibnamefont
  {Aspelmeyer}}, \bibinfo {author} {\bibfnamefont {T.~J.}\ \bibnamefont
  {Kippenberg}}, \ and\ \bibinfo {author} {\bibfnamefont {F.}~\bibnamefont
  {Marquardt}},\ }\href@noop {} {\bibfield  {journal} {\bibinfo  {journal}
  {Reviews of Modern Physics}\ }\textbf {\bibinfo {volume} {86}},\ \bibinfo
  {pages} {1391} (\bibinfo {year} {2014}{\natexlab{b}})}\BibitemShut {NoStop}%
\bibitem [{\citenamefont {Lvovsky}\ and\ \citenamefont
  {Raymer}(2009)}]{tomRevLvovsky}%
  \BibitemOpen
  \bibfield  {author} {\bibinfo {author} {\bibfnamefont {A.~I.}\ \bibnamefont
  {Lvovsky}}\ and\ \bibinfo {author} {\bibfnamefont {M.~G.}\ \bibnamefont
  {Raymer}},\ }\href@noop {} {\bibfield  {journal} {\bibinfo  {journal} {Rev.
  Mod. Phys.}\ }\textbf {\bibinfo {volume} {81}},\ \bibinfo {pages} {299}
  (\bibinfo {year} {2009})}\BibitemShut {NoStop}%
\bibitem [{\citenamefont {Szorkovszky}\ \emph {et~al.}(2011)\citenamefont
  {Szorkovszky}, \citenamefont {Doherty}, \citenamefont {Harris},\ and\
  \citenamefont {Bowen}}]{PhysRevLett.107.213603}%
  \BibitemOpen
  \bibfield  {author} {\bibinfo {author} {\bibfnamefont {A.}~\bibnamefont
  {Szorkovszky}}, \bibinfo {author} {\bibfnamefont {A.~C.}\ \bibnamefont
  {Doherty}}, \bibinfo {author} {\bibfnamefont {G.~I.}\ \bibnamefont {Harris}},
  \ and\ \bibinfo {author} {\bibfnamefont {W.~P.}\ \bibnamefont {Bowen}},\
  }\href {\doibase 10.1103/PhysRevLett.107.213603} {\bibfield  {journal}
  {\bibinfo  {journal} {Phys. Rev. Lett.}\ }\textbf {\bibinfo {volume} {107}},\
  \bibinfo {pages} {213603} (\bibinfo {year} {2011})}\BibitemShut {NoStop}%
\bibitem [{\citenamefont {Doherty}\ \emph {et~al.}(2012)\citenamefont
  {Doherty}, \citenamefont {Szorkovszky}, \citenamefont {Harris},\ and\
  \citenamefont {Bowen}}]{sme}%
  \BibitemOpen
  \bibfield  {author} {\bibinfo {author} {\bibfnamefont {A.~C.}\ \bibnamefont
  {Doherty}}, \bibinfo {author} {\bibfnamefont {A.}~\bibnamefont
  {Szorkovszky}}, \bibinfo {author} {\bibfnamefont {G.~I.}\ \bibnamefont
  {Harris}}, \ and\ \bibinfo {author} {\bibfnamefont {W.~P.}\ \bibnamefont
  {Bowen}},\ }\href {\doibase 10.1098/rsta.2011.0531} {\bibfield  {journal}
  {\bibinfo  {journal} {Philos. Trans. Royal Soc. A}\ }\textbf {\bibinfo
  {volume} {370}},\ \bibinfo {pages} {5338} (\bibinfo {year}
  {2012})}\BibitemShut {NoStop}%
\bibitem [{\citenamefont {Dzhioev}\ and\ \citenamefont {Kosov}(2011)}]{Kosov2}%
  \BibitemOpen
  \bibfield  {author} {\bibinfo {author} {\bibfnamefont {A.~A.}\ \bibnamefont
  {Dzhioev}}\ and\ \bibinfo {author} {\bibfnamefont {D.}~\bibnamefont
  {Kosov}},\ }\href@noop {} {\bibfield  {journal} {\bibinfo  {journal} {J.
  Chem. Phys.}\ }\textbf {\bibinfo {volume} {134}},\ \bibinfo {pages} {044121}
  (\bibinfo {year} {2011})}\BibitemShut {NoStop}%
\bibitem [{\citenamefont {Berry}\ and\ \citenamefont
  {Wiseman}(2000)}]{berry2000optimal}%
  \BibitemOpen
  \bibfield  {author} {\bibinfo {author} {\bibfnamefont {D.~W.}\ \bibnamefont
  {Berry}}\ and\ \bibinfo {author} {\bibfnamefont {H.~M.}\ \bibnamefont
  {Wiseman}},\ }\href@noop {} {\bibfield  {journal} {\bibinfo  {journal}
  {Physical review letters}\ }\textbf {\bibinfo {volume} {85}},\ \bibinfo
  {pages} {5098} (\bibinfo {year} {2000})}\BibitemShut {NoStop}%
\bibitem [{\citenamefont {Mahler}\ \emph {et~al.}(2013)\citenamefont {Mahler},
  \citenamefont {Rozema}, \citenamefont {Darabi}, \citenamefont {Ferrie},
  \citenamefont {Blume-Kohout},\ and\ \citenamefont
  {Steinberg}}]{mahler2013adaptive}%
  \BibitemOpen
  \bibfield  {author} {\bibinfo {author} {\bibfnamefont {D.~H.}\ \bibnamefont
  {Mahler}}, \bibinfo {author} {\bibfnamefont {L.~A.}\ \bibnamefont {Rozema}},
  \bibinfo {author} {\bibfnamefont {A.}~\bibnamefont {Darabi}}, \bibinfo
  {author} {\bibfnamefont {C.}~\bibnamefont {Ferrie}}, \bibinfo {author}
  {\bibfnamefont {R.}~\bibnamefont {Blume-Kohout}}, \ and\ \bibinfo {author}
  {\bibfnamefont {A.}~\bibnamefont {Steinberg}},\ }\href@noop {} {\bibfield
  {journal} {\bibinfo  {journal} {Physical review letters}\ }\textbf {\bibinfo
  {volume} {111}},\ \bibinfo {pages} {183601} (\bibinfo {year}
  {2013})}\BibitemShut {NoStop}%
\bibitem [{\citenamefont {Welch}\ and\ \citenamefont
  {Bishop}(1995)}]{welch1995introduction}%
  \BibitemOpen
  \bibfield  {author} {\bibinfo {author} {\bibfnamefont {G.}~\bibnamefont
  {Welch}}\ and\ \bibinfo {author} {\bibfnamefont {G.}~\bibnamefont {Bishop}},\
  }\href@noop {} {\emph {\bibinfo {title} {An introduction to the Kalman
  filter}}}\ (\bibinfo  {publisher} {University of North Carolina, Department
  of Computer Science},\ \bibinfo {year} {1995})\BibitemShut {NoStop}%
\bibitem [{\citenamefont {Schumaker}\ and\ \citenamefont
  {Caves}(1985)}]{Caves}%
  \BibitemOpen
  \bibfield  {author} {\bibinfo {author} {\bibfnamefont {B.~L.}\ \bibnamefont
  {Schumaker}}\ and\ \bibinfo {author} {\bibfnamefont {C.~M.}\ \bibnamefont
  {Caves}},\ }\href@noop {} {\bibfield  {journal} {\bibinfo  {journal} {Phys.
  Rev. A}\ }\textbf {\bibinfo {volume} {31}},\ \bibinfo {pages} {3093}
  (\bibinfo {year} {1985})}\BibitemShut {NoStop}%
\bibitem [{\citenamefont {Louisell}(1973)}]{Louisell}%
  \BibitemOpen
  \bibfield  {author} {\bibinfo {author} {\bibfnamefont {W.~H.}\ \bibnamefont
  {Louisell}},\ }\href@noop {} {\emph {\bibinfo {title} {Quantum statistical
  properties of radiation}}},\ Vol.~\bibinfo {volume} {7}\ (\bibinfo
  {publisher} {Wiley New York},\ \bibinfo {year} {1973})\BibitemShut {NoStop}%
\bibitem [{\citenamefont {Attal}(2014)}]{attal2014quantum}%
  \BibitemOpen
  \bibfield  {author} {\bibinfo {author} {\bibfnamefont {S.}~\bibnamefont
  {Attal}},\ }\href@noop {} {\bibfield  {journal} {\bibinfo  {journal}
  {Institut Camille Jordan, University of Lyon}\ } (\bibinfo {year} {2014})},\
  \bibinfo {note} {online lecture notes: Quantum channels}\BibitemShut
  {NoStop}%
\end{thebibliography}%

\end{document}